\documentclass[12pt]{article}
\pdfoutput=1  %%%This is requested by JHEP to let them know to compile using pdflatex...
\usepackage{amsmath}
\usepackage{graphicx}
\usepackage{subfig}
\usepackage{slashed}
\newskip\zatskip \zatskip=0pt plus0pt minus0pt
\def\matth{\mathsurround=0pt}

\def\atversim#1#2{\lower0.7ex\vbox{\baselineskip\zatskip\lineskip\zatskip
  \lineskiplimit 0pt\ialign{$\matth#1\hfil##\hfil$\crcr#2\crcr\sim\crcr}}}

\begin{document} 
\def\sel{ \tilde{e}}
\def\smu{\tilde{\mu}}
\def\stau{\tilde{\tau}}
\def\LSP{\tilde{\chi}^0_1}
\def\Evis {E_{\mbox{\tiny vis}}}
\def\chip{\tilde{\chi}_1^+}
\def\chim{\tilde{\chi}_1^-}
\def\chipm{\tilde\chi_1^\pm}
\def\sneumu{\tilde{\nu}_\mu}
\def\neu2{\tilde \chi_2^0}
\def\neu1{\tilde \chi_1^0}
\def\jet{\mathrm{jet}}
\def\Emiss{E_{\mbox{\tiny miss}}}
\def\MET{\ifmmode E_T^{\mathrm{miss}} \else $E_T^{\mathrm{miss}}$\fi}
\def\MEFF{\ifmmode M_{\mathrm{eff}} \else $M_{\mathrm{eff}}$\fi}
\def\MMIN{\ifmmode M_{\mathrm{min}} \else $M_{\mathrm{min}}$\fi}
%%%%%%%%%%%%%%%%%%%%%%%%%%%%%%%%%%%%%%%%%%%
\def\Re{{\cal R \mskip-4mu \lower.1ex \hbox{\it e}\,}}
\def\Im{{\cal I \mskip-5mu \lower.1ex \hbox{\it m}\,}}
\def\ie{{\it i.e.}}
\def\eg{{\it e.g.}}
\def\etc{{\it etc}}
\def\etal{{\it et al.}}
\def\ibid{{\it ibid}.}
\def\sub#1{_{\lower.25ex\hbox{$\scriptstyle#1$}}}
\def\tev{\,{\rm TeV}}
\def\gev{\,{\rm GeV}}
\def\mev{\,{\rm MeV}}
\def\to{\rightarrow}
\def\slash{\not\!}
\def\subw{_{\rm w}}
\def\mh{\ifmmode m\sbl H \else $m\sbl H$\fi}
\def\mch{\ifmmode m_{H^\pm} \else $m_{H^\pm}$\fi}
\def\mt{\ifmmode m_t\else $m_t$\fi}
\def\mc{\ifmmode m_c\else $m_c$\fi}
\def\mz{\ifmmode M_Z\else $M_Z$\fi}
\def\mw{\ifmmode M_W\else $M_W$\fi}
\def\mws{\ifmmode M_W^2 \else $M_W^2$\fi}
\def\mhs{\ifmmode m_H^2 \else $m_H^2$\fi}   
\def\mzs{\ifmmode M_Z^2 \else $M_Z^2$\fi}
\def\mts{\ifmmode m_t^2 \else $m_t^2$\fi}
\def\mcs{\ifmmode m_c^2 \else $m_c^2$\fi}
\def\mchs{\ifmmode m_{H^\pm}^2 \else $m_{H^\pm}^2$\fi}
\def\ztwo{\ifmmode Z_2\else $Z_2$\fi}
\def\zone{\ifmmode Z_1\else $Z_1$\fi}
\def\mtwo{\ifmmode M_2\else $M_2$\fi}
\def\mone{\ifmmode M_1\else $M_1$\fi}
\def\tb{\ifmmode \tan\beta \else $\tan\beta$\fi}
\def\xw{\ifmmode x\subw\else $x\subw$\fi}
\def\ch{\ifmmode H^\pm \else $H^\pm$\fi}
\def\lum{\ifmmode {\cal L}\else ${\cal L}$\fi}
\def\inpb{\ifmmode {\rm pb}^{-1}\else ${\rm pb}^{-1}$\fi}
\def\infb{\ifmmode {\rm fb}^{-1}\else ${\rm fb}^{-1}$\fi}
\def\epem{\ifmmode e^+e^-\else $e^+e^-$\fi}
\def\ppb{\ifmmode \bar pp\else $\bar pp$\fi}
\def\pbp{\ifmmode ~^(\bar p^)p\else $~^(\bar p^)p$\fi}
\def\bsg{\ifmmode B\to X_s\gamma\else $B\to X_s\gamma$\fi}
\def\bsll{\ifmmode B\to X_s\ell^+\ell^-\else $B\to X_s\ell^+\ell^-$\fi}
\def\bstt{\ifmmode B\to X_s\tau^+\tau^-\else $B\to X_s\tau^+\tau^-$\fi}
\def\half{\textstyle{{1\over 2}}}
\def\elli{\ell^{i}}
\def\ellj{\ell^{j}}
\def\ellk{\ell^{k}} 

\def\matth{\mathsurround=0pt}
\def\undertext#1{$\underline{\smash{\vphantom{y}\hbox{#1}}}$}

\begin{titlepage}

\rightline{\vbox{\halign{&#\hfil\cr
&SLAC-PUB-14382\cr
&ANL-HEP-PR-11-13\cr
&NUHEP-TH/11-014\cr
&BONN-TH-2011-05\cr}}}
\vspace{1in}

\begin{center}

{\Large\bf
Supersymmetry Without Prejudice at the 7 TeV LHC}
\medskip

\normalsize
\large {John A. Conley}\\
       {\small\it Physikalisches Institut, Universit\"at Bonn, Nu\ss allee 12, 53113 Bonn, Germany}\\
\medskip
\large{James S. Gainer}\\
{\small\it High Energy Physics Division, Argonne National Laboratory,
Argonne, IL 60439 USA and \\
Department of Physics and Astronomy, Northwestern University, Evanston, IL 60208 USA}\\
\medskip
\large{JoAnne L. Hewett, My Phuong Le, Thomas G. Rizzo}\\
      {\small\it  SLAC National Accelerator Laboratory, 2575 Sand Hill Rd., Menlo Park, CA  94025, USA}\\
\end{center}

\begin{abstract}
% abstract
We investigate the model independent nature of the Supersymmetry search strategies
at the 7 TeV LHC.  To this end, we study the missing-transverse-energy-based searches
developed by the ATLAS Collaboration that were essentially designed for mSUGRA.  We simulate
the signals for $\sim 71$k models in the 19-dimensional parameter space of the pMSSM. These models
have been found to satisfy
existing experimental and theoretical constraints and provide insight into general features of
the MSSM without reference to a particular SUSY breaking scenario or any other assumptions at
the GUT scale. Using backgrounds generated by ATLAS, we find that imprecise knowledge of these
estimated backgrounds is a limiting factor in the potential discovery of these models and that some channels
become systematics-limited at larger luminosities.  As this systematic error is varied
between 20-100\%, roughly half to 90\% of this model sample is observable with significance
$S\geq 5$ for 1 fb$^{-1}$ of integrated luminosity.  We then examine the model characteristics
for the cases which cannot be discovered and find several contributing factors.  We find that
a blanket statement that squarks and gluinos are excluded with masses below a specific value
cannot be made.  We next explore possible modifications to the kinematic cuts in these analyses
that may improve the pMSSM model coverage.  Lastly, we examine the implications of a null search
at the 7 TeV LHC in terms of the degree of fine-tuning that would be present in this model set 
and for sparticle production at the 500 GeV and 1 TeV Linear Collider. 

\end{abstract}

\renewcommand{\thefootnote}{\arabic{footnote}} \end{titlepage}

%\maketitle

% introduction

\section {Introduction and Background}

The LHC has had an initial run at 7 TeV with both the ATLAS and CMS experiments collecting $\sim 35-45$ pb$^{-1}$ of useful data. Even with 
this low integrated luminosity these experiments have been able to extend searches far beyond the reach of the Tevatron for many new physics scenarios 
with, so far,  null results \cite {LaThuile,bigexpref}. 
This clearly demonstrates the power of increasing the center of mass energy in the search for new physics at hadron 
colliders. Starting soon, the LHC is to begin a longer run at 7 TeV and is expected to collect of order $1-7$ fb$^{-1}$ of data over the next 2 years. Such a 
data set will allow for a first exploration of the TeV mass scale, and if new strongly interacting particles exist in this kinematic regime, they
should be observed.

A well-motivated, and perhaps most popular, possibility for new physics that may be discovered during this coming LHC run is 
Supersymmetry (SUSY) \cite{mssmrev}.
Both ATLAS \cite{Aad:2009wy}  and CMS \cite{Ball:2007zza} 
have designed detailed searches for many of the SUSY partners of the Standard 
Model (SM) particles; these are mostly (but not exclusively) based on the assumption of mSUGRA/CMSSM-like soft breaking within the Minimal 
Supersymmetric Standard Model (MSSM) framework. This assumption greatly simplifies the exploration of the vast Supersymmetric parameter space.
While these searches are designed to well cover the parameter space of these SUSY-breaking 
scenarios it is important to ascertain their discovery
potential in a more general MSSM context. This is particularly worrisome in light of results from the Tevatron, where it has been realized
\cite{Alwall:2008ve,Berger:2008cq} that
relatively light gluinos may have escaped undetected.
The question then arises whether these specific scenarios and associated searches adequately describe the true breadth of the MSSM and its
possible collider signatures, and whether the LHC searches as presently designed could fail to observe sparticle production.  
This has prompted several studies of more
model independent search strategies as well as the development of simplified models \cite{Alwall:2008ag}.

Recently, we have addressed \cite{Conley:2010du} this question by
investigating the capability of the 14 TeV LHC to explore a more general MSSM model parameter space, i.e., that of the pMSSM 
(phenomenological MSSM) \cite{Berger:2008cq}, 
to be described below, from the 
point of view of the ATLAS detector.  In particular, we examined the performance of the planned ATLAS SUSY searches in exploring  
this more general MSSM scenario. This analysis provides insight into general features of the MSSM without reference to a particular SUSY breaking 
scenario or any other assumptions at the GUT scale.
We found that the ATLAS mSUGRA-inspired searches, based on missing $E_T$, did surprisingly well at covering 
the kinematically accessible portions of this model space; we also found that some interesting exceptions can arise in these more general models. 
Given the lower-energy run of the LHC over the next 2 years it behooves us to determine how well the corresponding mSUGRA motivated searches designed 
by ATLAS would perform under these conditions, since this is the situation presently before us. This is the goal of the present paper. 
We note that recently  there have been several mSUGRA-based studies evaluating the capability of the 7 TeV LHC run to probe that
parameter space \cite{bignewtheoryref}

As is well known, soft SUSY breaking within the MSSM in all generality leads to a theory with over $\sim 100$ a priori free parameters 
which prohibits a detailed study of this theory. A number of theoretically possible scenarios exist which describe the breaking of Supersymmetry; maybe 
even multiple mechanisms are simultaneously responsible. Practically speaking, there are two ways to approach reducing this large number of a priori unknown 
parameters to something more manageable. One approach is to consider only specific, well-motivated SUSY breaking scenarios, such as mSUGRA or others. 
This leads to a drastic reduction in the number of free parameters (to only $\sim 3-5$) so that detailed analyses of the
resulting parameter space can be easily achieved. A problem with performing such studies is that they may result in a bias as to the nature
of SUSY signals when searching for collider or other SUSY signatures. An alternative approach is to be less prejudicial and to instead follow a bottom-up 
analysis which we have employed in a number of recent works \cite{Conley:2010du,Cotta:2010ej}  
and will make use of here. By imposing a set of theoretically 
and experimentally well-motivated constraints on the general MSSM (to be described below), without making any reference to the specific mechanism of SUSY breaking, 
we arrive at a theory with 19  TeV-scale parameters.  This is known as the pMSSM, which is 
significantly more 
manageable than the full Supersymmetric parameter space, and yet allows for more breadth than is present in , \eg, mSUGRA.{\footnote {Even in such a case a 
full exploration of this large parameter space is at best difficult if not impossible with present computing 
power.}} These parameters will then completely define and describe all aspects of TeV-scale SUSY phenomenology. Such an approach 
has the advantage of being more general than any given (or given set of) specific SUSY breaking scenario(s) and allows one to be in some sense agnostic 
about the SUSY mass spectrum. 

To this end, we examine the \MET-based SUSY searches developed by the ATLAS collaboration for the 7 TeV LHC \cite{ATLAS:1278474}.  We simulate the pMSSM signal
for roughly 71k pMSSM models (hereafter `model' refers to a point in the 19-dimensional pMSSM parameter space) that we generated in our previous work 
\cite{Berger:2008cq}.  We employ SM backgrounds provided by the ATLAS Collaboration. In Section 2, we describe our SUSY model generation and LHC analysis 
procedure. It is important to note that we strictly adhere to the analyses as designed by ATLAS.  While numerous, and perhaps improved, SUSY collider
search techniques have been discussed in the literature \cite{Pape:2006ar}, it is not our present purpose here to discuss or employ them.
Section 3 contains our main results.  We find the systematic error in determining the SM background to SUSY production is a limiting factor
in the potential discovery of these models; in fact, some channels become systematics limited at larger luminosities.  In this section we
determine the fraction of our pMSSM model set that is discoverable at the 7 TeV LHC.  We then examine the model characteristics in some detail that
render some of the models undetectable.  We find that the observability of models depends on the precise details of the sparticle spectrum
and that a blanket statement of constraints on the mass of, say the gluino or squarks, cannot be made.  In this Section, we also explore potential
modifications to the kinematic cuts in these analyses that may improve model observability.  In Section 4, we examine the implications of a null SUSY
search during this run with respect to the degree of fine-tuning present in these models, as well as the implications for sparticle
production at a high energy $e^+e^-$ Linear Collider. 
A summary and our conclusions can be found in Section 5.

% review of model generation section
%\input{model_gen_rev.tex}

% procedure section

\section{Analysis Procedure for Inclusive SUSY Production at the LHC}
\label{analysisproc}

The purpose of this work is to explore how well
mSUGRA-inspired inclusive SUSY searches (in particular the
set proposed by the ATLAS collaboration~\cite{ATLAS:1278474})
apply to the larger and much more general
pMSSM parameter space for the 7 TeV run of the LHC.  This is similar in spirit to
to Ref. \cite{Conley:2010du}, which explored this question for the more powerful 14 TeV LHC. 
The pMSSM model sample that we
study was generated in Ref. \cite{Berger:2008cq}; here, we briefly review the procedure employed to generate
this sample.  We then describe our procedure for
generating the signal events, comparing to background, and determining the statistical criteria 
for discovery.  We will show that we faithfully reproduce the ATLAS results in each analysis
channel for their benchmark SUSY model.

As stated above, we study the 19-dimensional parameter space of the pMSSM.  
This set of parameters was arrived at \cite{Djouadi:2002ze} by imposing the following set of requirements 
onto the general R-Parity 
conserving MSSM: ($i$) the soft parameters are taken to be real so that 
there are no new CP-violating sources beyond those in the usual CKM matrix; ($ii$) Minimal Flavor Violation(MFV) \cite{D'Ambrosio:2002ex} is assumed to be 
valid at the TeV scale; ($iii$) the first two generations of sfermions having the same quantum numbers are taken to be degenerate and to have 
negligible Yukawa couplings and ($iv$) the lightest neutralino is assumed to be the Lightest Supersymmetric Particle(LSP) and is a stable 
thermal WIMP. Most of these assumptions are applied in order to avoid issues associated with flavor physics constraints.
With these conditions, the remaining 19 free soft-breaking parameters are given by the three gaugino masses, $M_{i=1-3}$, the ten sfermion 
masses $m_{\tilde f}$, the three $A$-terms for the third generation fermions ($A_{b,t,\tau}$), and the usual Higgs sector parameters $\mu$, $M_A$ 
and $\tan \beta$. 

To generate the specific pMSSM parameter points that we study below (hereafter referred to as
our set of models), we performed numerical scans over 
the space formed by these 19 parameters. This required both a selection 
of the parameter range intervals as well as an assumption about the nature of the scan prior for how points are chosen within these intervals. These 
issues are both described in detail in our previous works \cite{Berger:2008cq,Conley:2010du,Cotta:2010ej}
and the interested reader should refer to them directly. Here, we simply note 
that two scans were performed: one employing a flat prior beginning with $10^7$ points and one
with a log prior employing $2\times 10^6$ points. The main 
distinctions between these two scans directly relevant to our analysis here are that ($a$)  all SUSY mass parameters were 
restricted to be $\leq 1$ TeV for the flat prior case, while for the log case the upper limit on mass parameters was raised to 3 TeV, and (b) the 
choice of the log prior generally 
leads to a more compressed sparticle spectrum than does the flat prior case.  Note that the
restriction on the upper limit for the mass parameters ensures relatively large production cross
sections at the LHC for the case of the flat prior model sample.

Once these points were generated, we demanded that they be consistent with a large
number of both theoretical and experimental constraints in order to ensure that the model 
sets are valid to study.    We mention the most important of these restrictions 
here{\footnote {For full details, see Ref. \cite{Berger:2008cq}}}: 
($i$) The spectrum is required to be tachyon free, color and charge breaking minima must be avoided,  
a bounded Higgs potential must be obtained and electroweak symmetry breaking must be consistent. ($ii$) We impose a number of flavor and electroweak 
constraints arising from $g-2$, $b\to s\gamma$, $B\to \tau \nu$, $B_S \to \mu^+\mu^-$, meson--anti-meson mixing, the invisible width of the $Z$ and 
$\Delta \rho$. ($iii$) We demand that the LSP contribution to the dark matter density not
exceed the upper bound determined by WMAP; note that the LSP is not required to saturate the
measured relic density, leaving room for the existence of other dark matter candidates.
Constraints from dark matter direct detection searches are also applied. ($iv$) We then include the restrictions imposed from the numerous direct searches for both 
the SUSY particles themselves as well as the extended SUSY Higgs sector at LEP. Here, some care was required as some of these searches needed to be 
re-evaluated in some detail due to particular SUSY model-dependent assumptions present in the
analysis which we needed to remove. ($v$) Finally, the null results from a number 
of Tevatron searches are imposed. In addition to the Higgs searches, the most restrictive searches
were found to be those hunting for stable charged 
particles \cite{Abe:1992vr} and those looking for an excess of multijet plus MET events \cite{:2007ww}. 
We note that in the latter case, the search strategies were designed for kinematics expected in mSUGRA-like models.  We thus were forced to simulate them 
in some detail, at the level of fast Monte Carlo, for our full model set. 
At the end of this analysis chain, $\sim 68.4k$ models from the flat prior set survived this set of constraints, 
as well as a corresponding set of $\sim 2.9k$ log prior models. These are the models that we will consider in our following analysis. 

We now turn our attention to the analysis procedure that we followed in generating and analyzing
the signal events from sparticle production at the 7 TeV LHC.  Throughout our analysis,
we adhere to the search strategies developed by ATLAS \cite{ATLAS:1278474} as closely as possible.  In this
reference, ATLAS
considers 10 MET search
channels, including selections where the minimum number of jets is 2, 3, or 4
and the number of leptons is 0, 1, or 2.  In the dilepton case, opposite-sign
(OSDL) and same-sign (SSDL) pairs are considered separately, with SSDL only being considered
in association with 2 jets.  Note that flavor tagged final states are not considered here. 
We consider 85 SUSY production processes that contribute to these 10 signatures.

Accurate estimates of the SM
backgrounds for the various channels are crucial to the validity of this study.  
We obtained details of the background
distributions presented in Ref.~\cite{ATLAS:1278474} directly from the ATLAS SUSY
Group~{\cite{thanks}}.  These backgrounds were produced
with state-of-the-art Monte Carlo event generators and the full
ATLAS detector simulation.  Employing these ATLAS computed backgrounds in our analysis
allows us to concentrate on
generating and analyzing signal events for each of the $\sim$ 71k parameter space points
in our pMSSM model sample.

\subsection{Generation of the Signal Events}

The steps involved in the generation of the signal events are very similar to
those detailed in Ref.~\cite{Conley:2010du}.  Here we will briefly summarize
the procedure and point out any differences in the present analysis.
For the generation and analysis of events for a single model, the workflow is:
\begin{enumerate}
  \item 
    The spectrum and decay table was generated with a
    modified~\cite{Conley:2010du} version of SUSY-HIT~\cite{Djouadi:2006bz}.
  \item The NLO cross sections for the 85 distinct SUSY production processes
    considered were computed using
    Prospino2.1~\cite{Beenakker:1996ch}
    and the CTEQ6.6M parton distribution functions \cite{Nadolsky:2008zw}
    (which were also used in the event generation).
  \item
    Using PYTHIA 6.418, events were generated, fragmented, showered, and hadronized
    for each of 85 SUSY production processes, with each
    process being weighted by its K-factor.
  \item
    Detector effects were simulated using an ATLAS-tuned version of the fast
    detector simulation PGS4 \cite{PGS} with the default isolation cuts removed.
  \item
    The simulated events were then analyzed using the analysis cuts for the 10
    ATLAS analyses listed above, as well as the isolation cuts described in Ref.~\cite{ATLAS:1278474}.
\end{enumerate}
We note that as in our previous work, a subset (about 1\%) of the models suffered serious
enough errors that the Pythia event generation halted.  These ``PYSTOP''
models are excluded from our results.

\subsection{Analysis Cuts}

For the reader's convenience, we provide here the full set of kinematic
cuts for each analysis channel, summarizing the information given in Section 4
of Ref.~\cite{ATLAS:1278474}.  All channels have a missing energy cut of
$\MET>80$ GeV, and all analyses except the SSDL channel have a transverse sphericity
cut of $S_T>0.2$.    Table~\ref{jet-cuts} summarizes the cuts for the searches with $n=2,3,4$ jets
which are
independent of the choice of lepton channel except for SSDL, which will be
described below. 

\begin{table}
\centering
  %\centering
\begin{tabular}{|l|c|c|c|}
  \hline
  Number of jets & $\geq 2$ & $\geq3$ & $\geq 4$ \\
  \hline
  Leading jet $p_T$ (GeV) & $> 180$ & $> 100$ & $> 100$ \\
  \hline
  Other jets $p_T$ (GeV) &  $> 50$ & $> 40$ & $> 40$ \\
  \hline
  min. $\Delta\phi(\mathrm{jet}_i,\MET)$ & $0.2,0.2$ & $0.2,0.2,0.2$ &
  $0.2,0.2,0$\\
  \hline
  $\MET>f\times\MEFF$ & $f=0.3$ & $f=0.25$ & $f=0.2$ \\
  \hline
\end{tabular}

  \caption{The kinematic cuts employed in the event selection: the cut
    on the $p_T$ of the leading jet, the $p_T$ of the other
    selected jets, the azimuthal angle between the selected jets and the
    missing transverse energy, and the missing energy as a fraction of the
    effective mass.}
  \label{jet-cuts}
\end{table}

We complete the description of the kinematic cuts by specifying the additional cuts that are
specific to the various lepton channels.  For the case with zero leptons, events
are rejected that have at least one lepton with $p_T>20$ GeV.  For the
one-lepton channels, one lepton with $p_T>20$ GeV is required, no additional
leptons with $p_T>10$ GeV are allowed, and the transverse mass $M_T$ of the
selected lepton and the missing energy vector must satisfy $M_T>100$ GeV. (The
definition of $M_T$ can be found in Ref.~\cite{Aad:2009wy}.)  For the OSDL
channel, exactly two leptons with $p_T>10$ GeV are required, and they must
have opposite charge.

As mentioned above, the SSDL channel has distinct cuts; unlike all the
other analyses, the three different jet selection options
specified in Table~\ref{jet-cuts} are not employed.  Instead, two jets with
$p_T>80$ GeV are required.  In addition, two leptons with $p_T>20$ GeV, same
charge, and invariant mass $m_{\ell\ell}>5$ GeV must be present, and there is a veto
on additional leptons with
$p_T>10$ GeV.  The transverse mass of the leading lepton
and the missing energy vector must satisfy $M_T>80$ GeV.

Lastly, when performing our statistical analysis, the \MEFF\ cut is optimized for each channel, and for each pMSSM model, in steps of 400 GeV
\cite{ATLAS:1278474}.

\subsection{Statistical Procedure}

To compute the significance of the signal for each search channel, 
we follow the statistical procedure described in detail in
Refs.~\cite{Conley:2010du,Aad:2009wy}, which is that employed by the ATLAS collaboration.
The probability that the expected background fluctuates to the number of observed events
is computed assuming that the systematic error on the background is Gaussian and the
statistical error is Poissonian. A significance of $S\geq 5$ is required for the
observation of a signal. As mentioned above, 
the \MEFF\ cut is optimized for each channel, and for each pMSSM model, in steps of 400 GeV
\cite{ATLAS:1278474}.  As was discussed in Ref.~\cite{Conley:2010du}, and 
will be further demonstrated here, the accuracy of the background estimation
has a profound impact on the signal significance and the resulting search reach.  In
order to quantify this, we will present results assuming a 20, 50, and 100\%
systematic error on the background.  We will also consider integrated
luminosities of 0.1, 1, and 10 fb$^{-1}$.

\subsection{Comparison with ATLAS Benchmark Models}

The ATLAS SUSY group has published signal rates for a single mSUGRA benchmark point
(SU4) in their study of Supersymmetry at the 7 TeV LHC \cite{ATLAS:1278474}.  
It is imperative for us to check the results of our analysis
against these published results for this benchmark model before proceeding to
apply our analysis to the pMSSM model set.

\begin{figure}
\centerline{
  \includegraphics[width=0.45\textwidth]{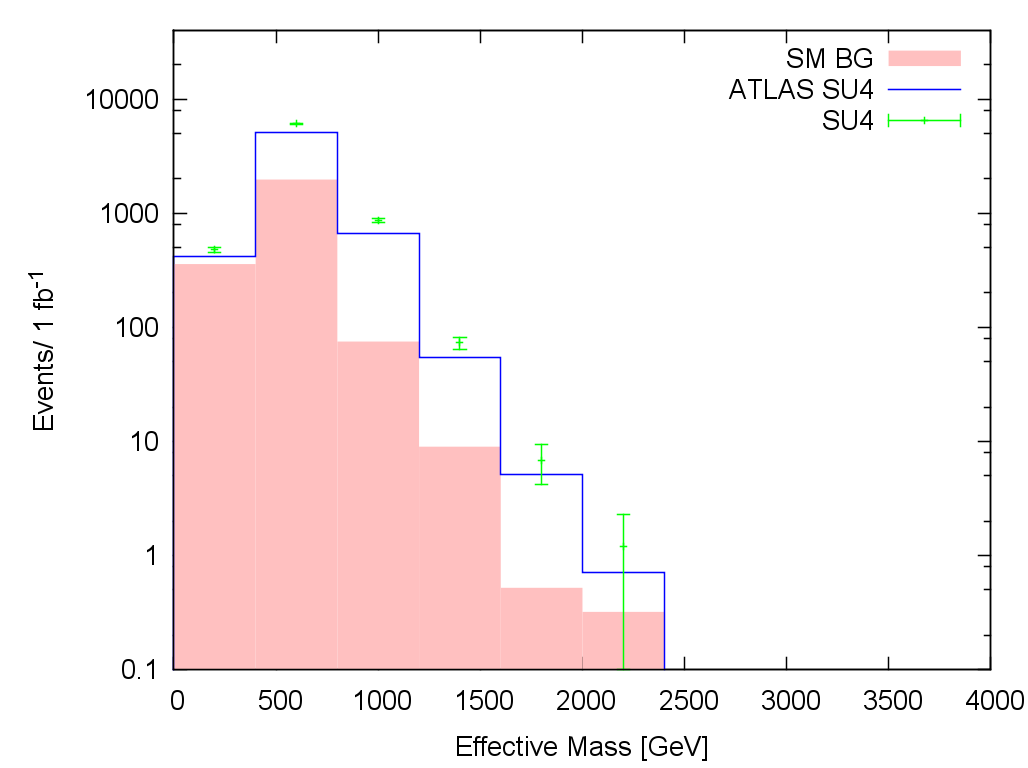}
  \includegraphics[width=0.45\textwidth]{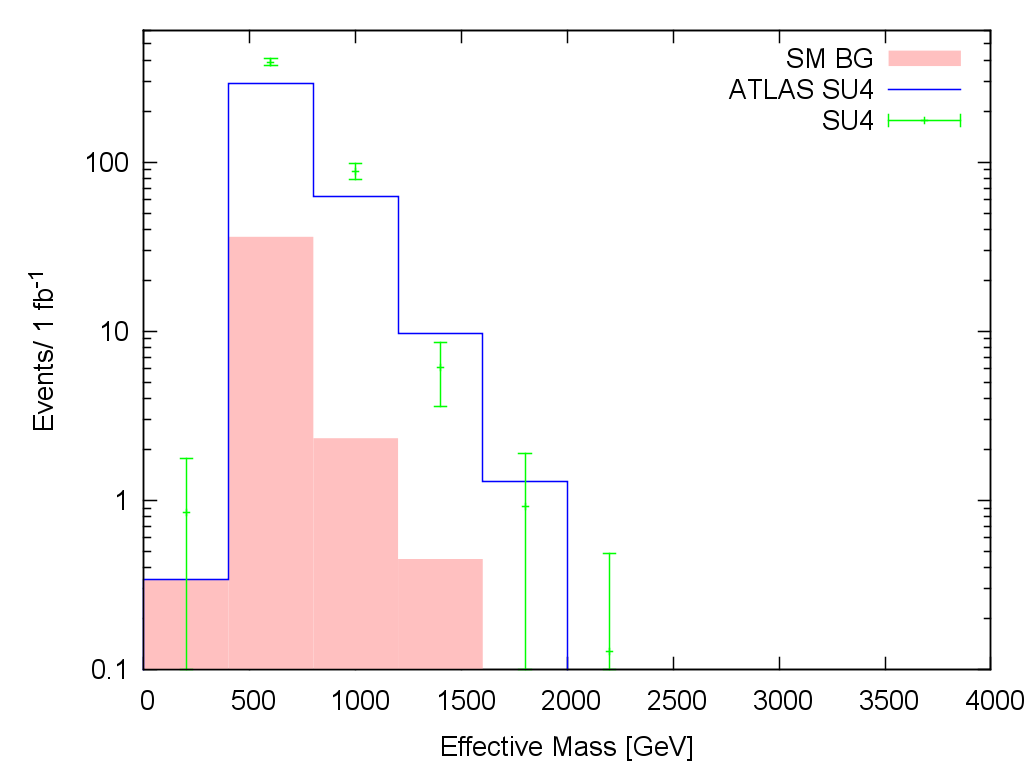}}
  \caption{The \MEFF\ distribution for the 4 jet, 0(1) lepton analysis on the
    left(right) for the SU4 benchmark model.  The green data points represent our
    analysis, while the blue line is the result from the ATLAS study
    \cite{ATLAS:1278474}.  The red shaded area represents the SM background.}
  \label{4j-benchmark}
\end{figure}

\begin{figure}
\centerline{
  \includegraphics[width=0.45\textwidth]{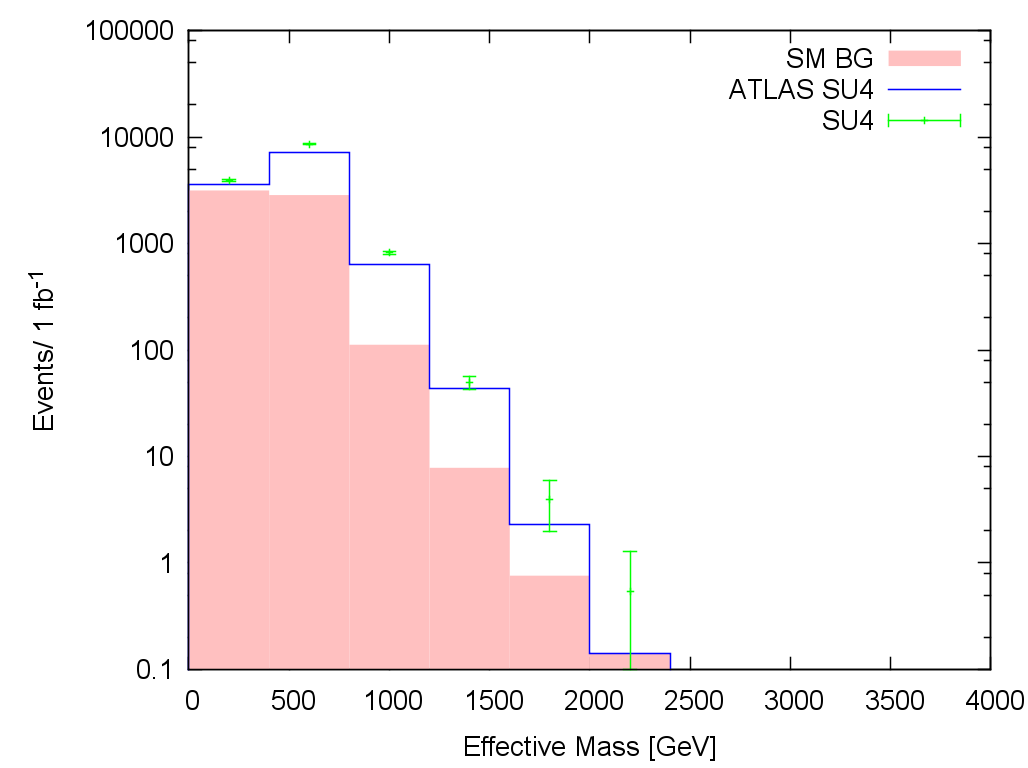}
  \includegraphics[width=0.45\textwidth]{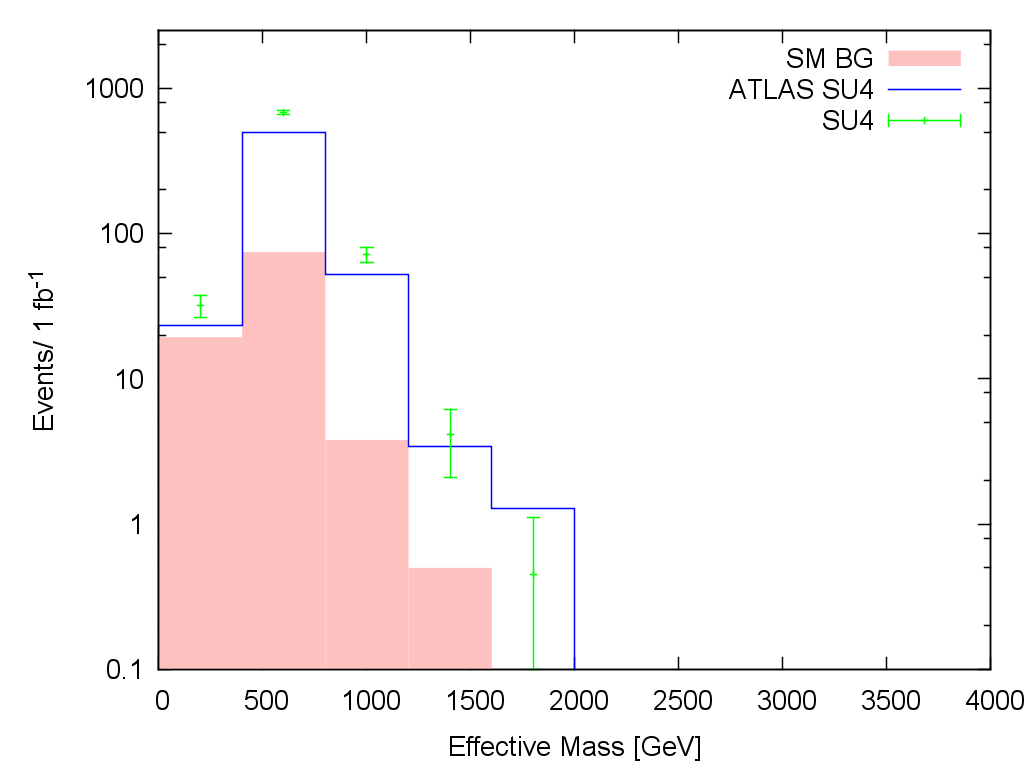}}
  \caption{The same as Figure~\ref{4j-benchmark}, except for the 3 jet, 0(1)
    lepton analysis on the left(right).}
  \label{3j-benchmark}
\end{figure}

\begin{figure}
\centerline{
  \includegraphics[width=0.45\textwidth]{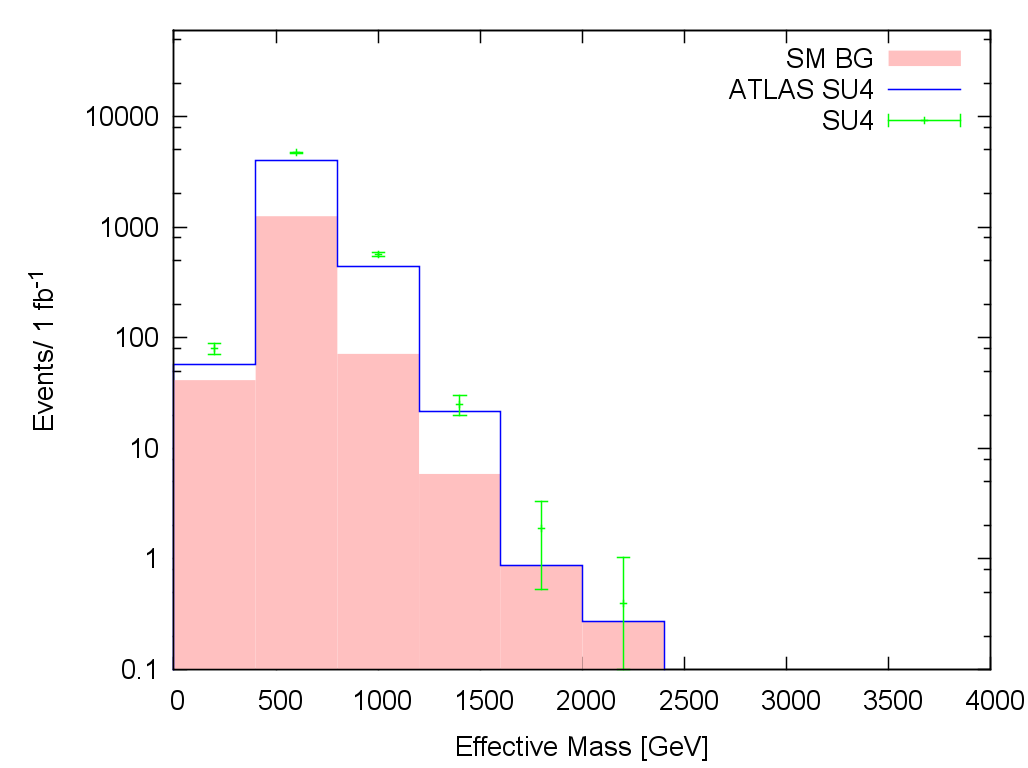}
  \includegraphics[width=0.45\textwidth]{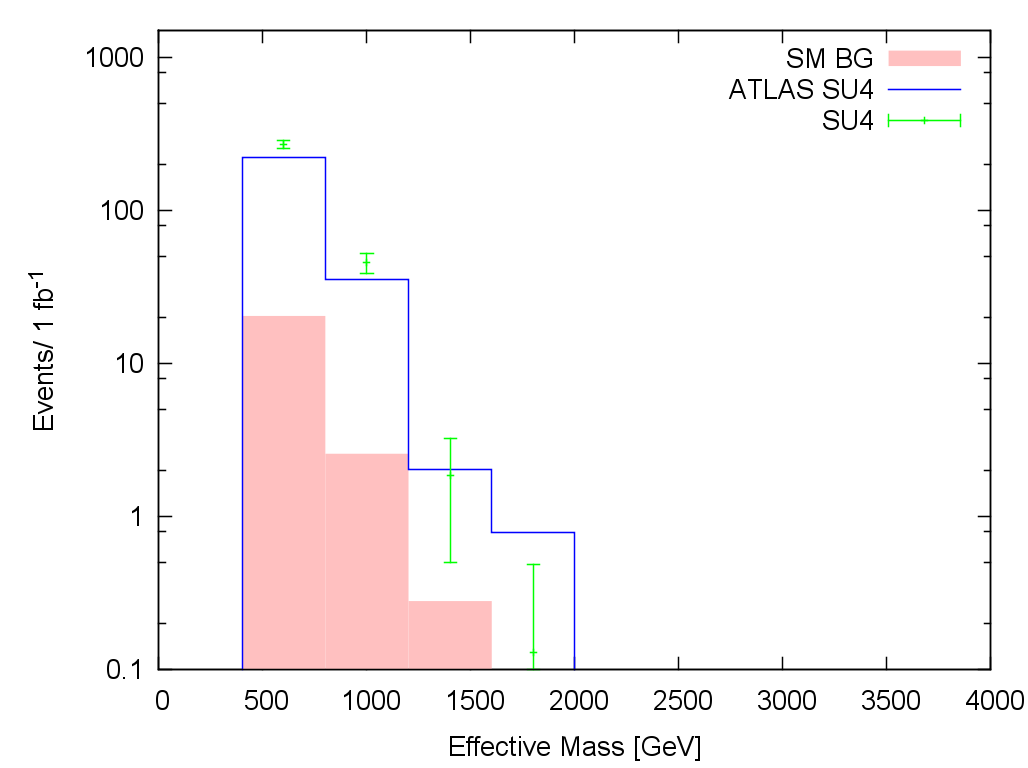}}
  \caption{The same as Figure~\ref{4j-benchmark}, except for the 2 jet, 0(1)
    lepton analysis on the left(right).}
  \label{2j-benchmark}
\end{figure}

\begin{figure}
\centerline{
  \includegraphics[width=0.45\textwidth]{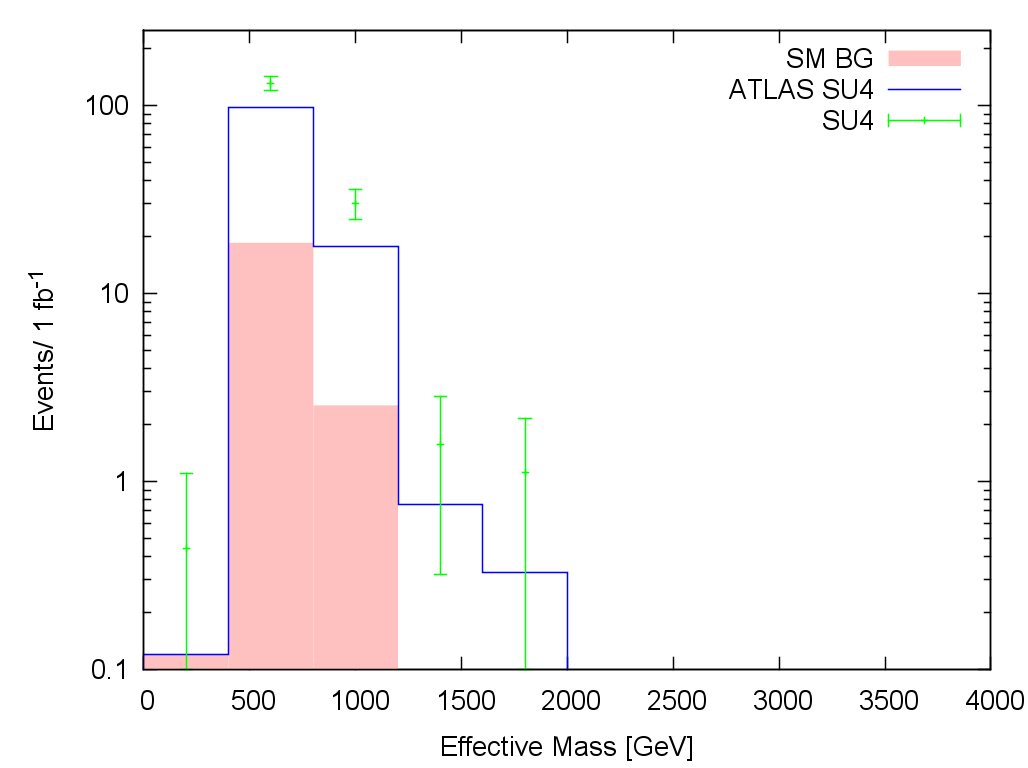}
  \includegraphics[width=0.45\textwidth]{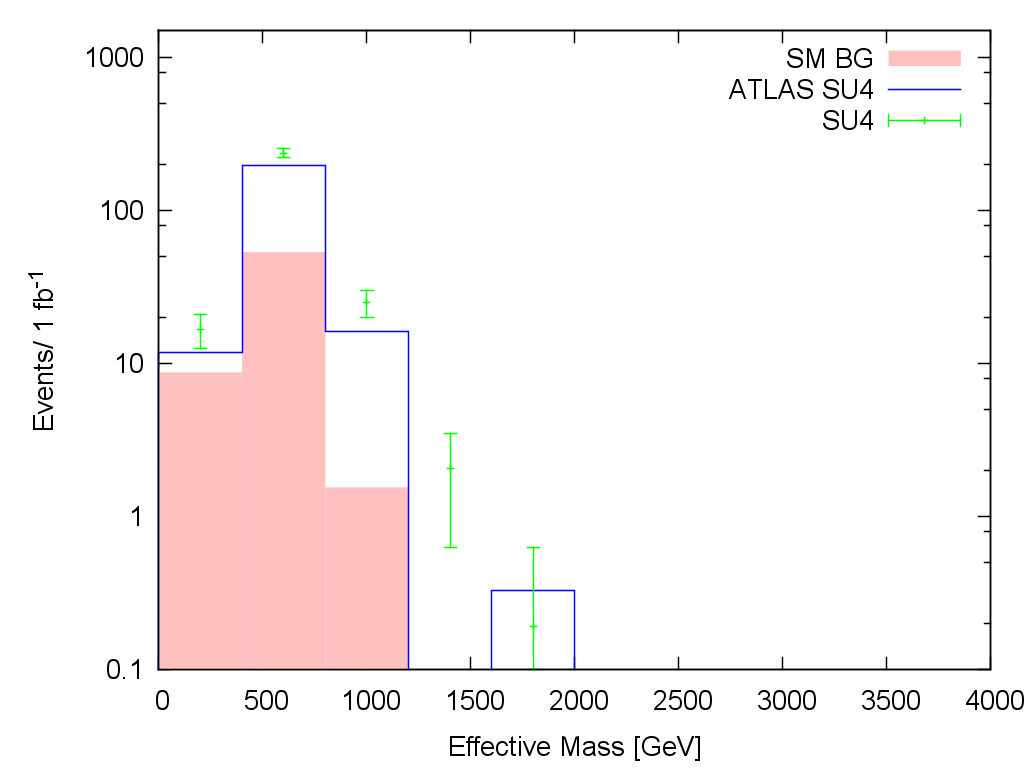}}
  \caption{The same as Figure~\ref{4j-benchmark}, except for the 4(3) jet, OSDL
    lepton analysis on the left(right).}
  \label{OSDL1-benchmark}
\end{figure}

\begin{figure}
\centerline{
  \includegraphics[width=0.45\textwidth]{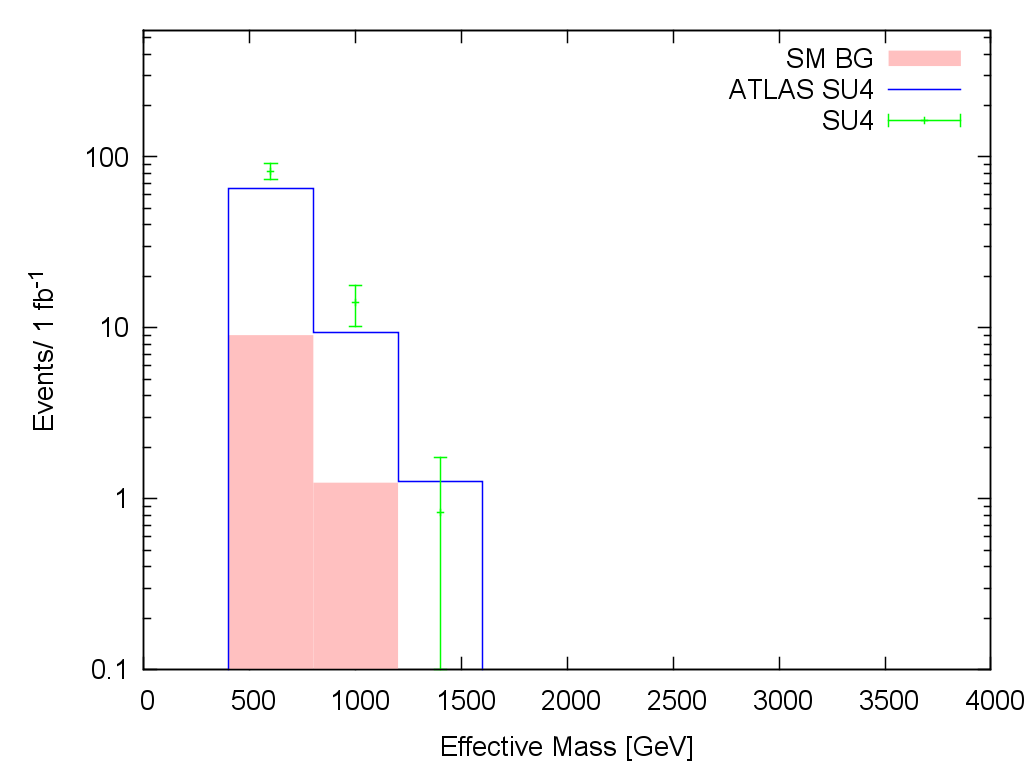}
  \includegraphics[width=0.45\textwidth]{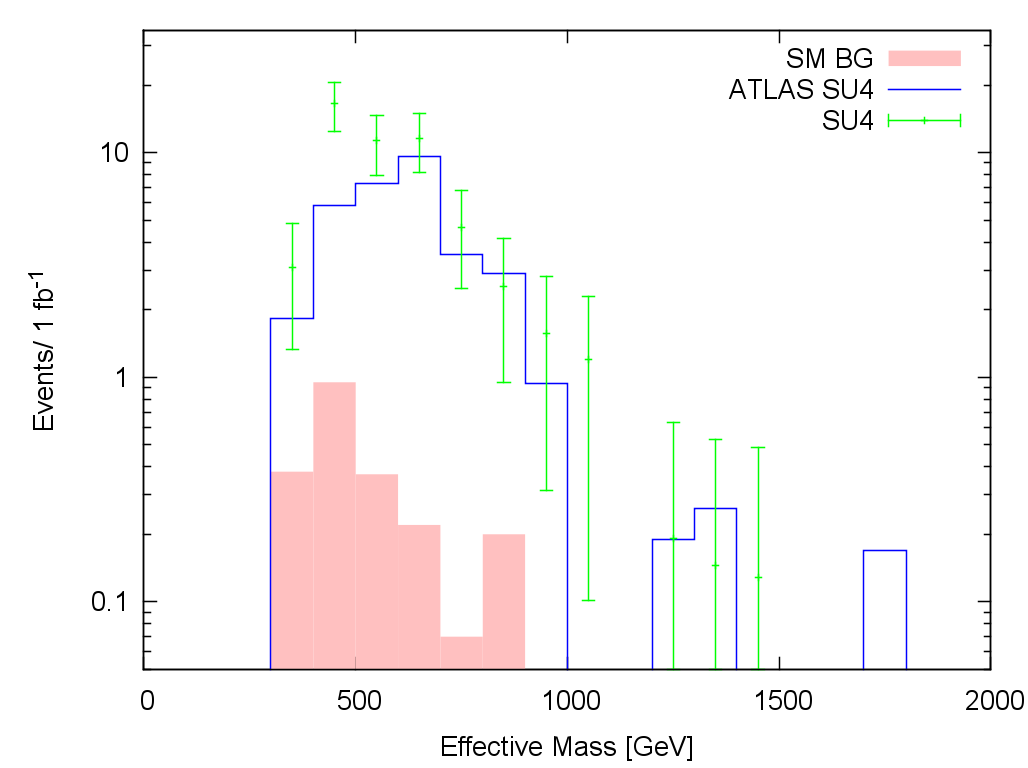}}
  \caption{The same as Figure~\ref{4j-benchmark}, except for the 2 jet, OSDL(SSDL)
    analysis on the left(right).}
  \label{2jl-benchmark}
\end{figure}

Here it is important to remind the reader that our SUSY signal generation, as 
described in detail in Ref.~\cite{Conley:2010du}, differs slightly  
from the procedure employed by
ATLAS.  In particular, the numerical programs used to compute the SUSY spectrum and decay tables,
as well as event generation, are different.  Furthermore, we use, by
necessity, a fast detector simulation as opposed to the ATLAS full GEANT-based
simulation.  Therefore, a small degree of discrepancy can be expected.  
The comparisons shown in Figures \ref{4j-benchmark}-\ref{2jl-benchmark}, however, indicate
that we are indeed able to faithfully reproduce the results
obtained by ATLAS for this benchmark model for all of the various inclusive analyses.

\clearpage

%\section{pMSSM Results}

\section{Results of the 7 TeV Analysis}

In this Section, we relate the results of our study on the effectiveness 
of the ATLAS 7 TeV \MET\ analyses in detecting our pMSSM model 
sample.  We first discuss the impact of the size of the background systematic errors
on SUSY searches, and then turn to the discovery coverage of the pMSSM.  We examine the characteristics
which cause some models to be undetectable as well as study the effects of modifying
the ATLAS SUSY analysis cuts.  We remind the reader
that our sample of $\sim 70$k models is not intended to be a full description of the
19-dimensional pMSSM parameter space.  However, the sample does contain numerous models
which exhibit properties that are quite different than those expected in mSUGRA and thus
provides insight into general features of the full MSSM.

\subsection{Influence of Background Systematic Errors}

\begin{figure}
\centerline{
\includegraphics[width=10.6cm]{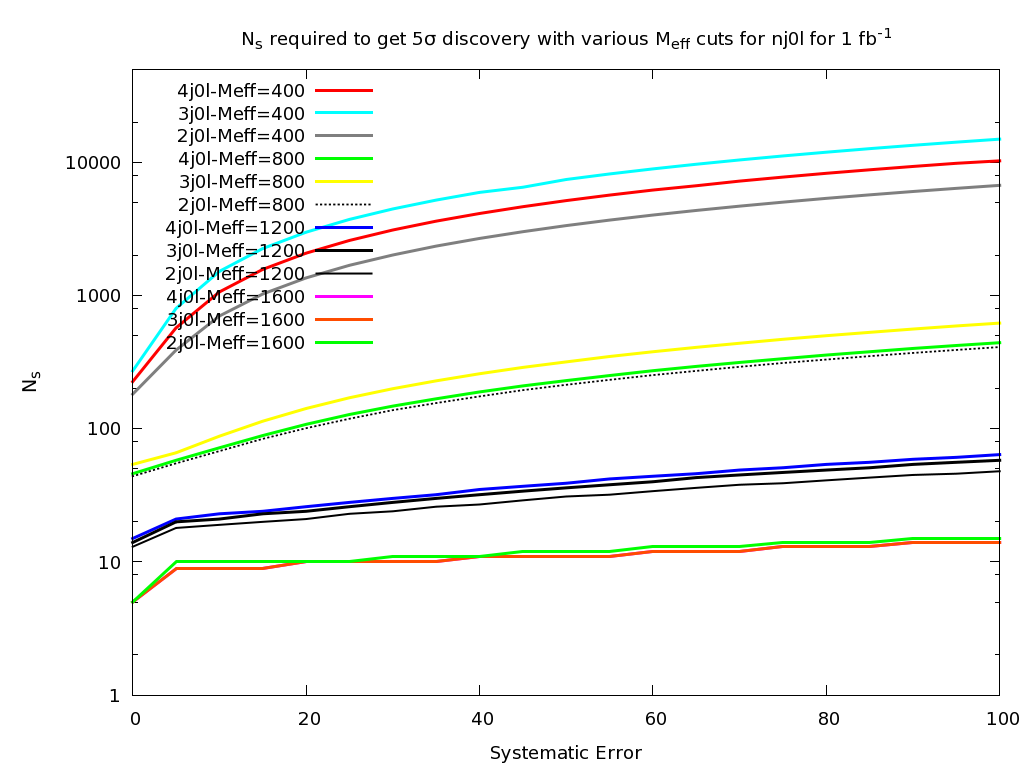}}
\vspace*{-0.3cm}
\centerline{
\includegraphics[width=10.5cm]{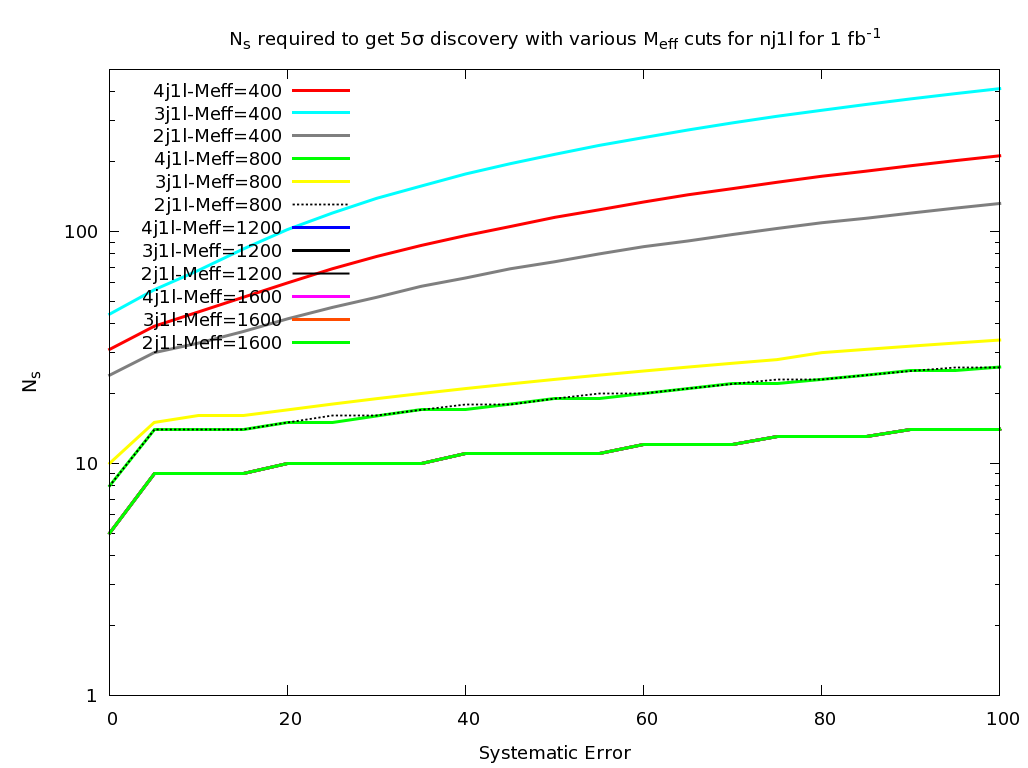}}
%\vspace*{0.1cm}
    \caption{Number of events required to reach the $S=5$ level of discovery as a function of the fractional 
      systematic error in the SM background for the 
      ATLAS \MET\ searches for various values of the \MEFF\ cut. The results for the nj0l and nj1l searches are shown in the top and bottom panels, 
      respectively.  The curves are color coded from top to bottom as indicated in
the legend.  For higher values of the \MEFF\ cut, we see that the curves are essentially
indistinguishable, lying on top of one another.}
\label{system1}
\end{figure}

\begin{figure}
\centerline{
\includegraphics[width=10.6cm]{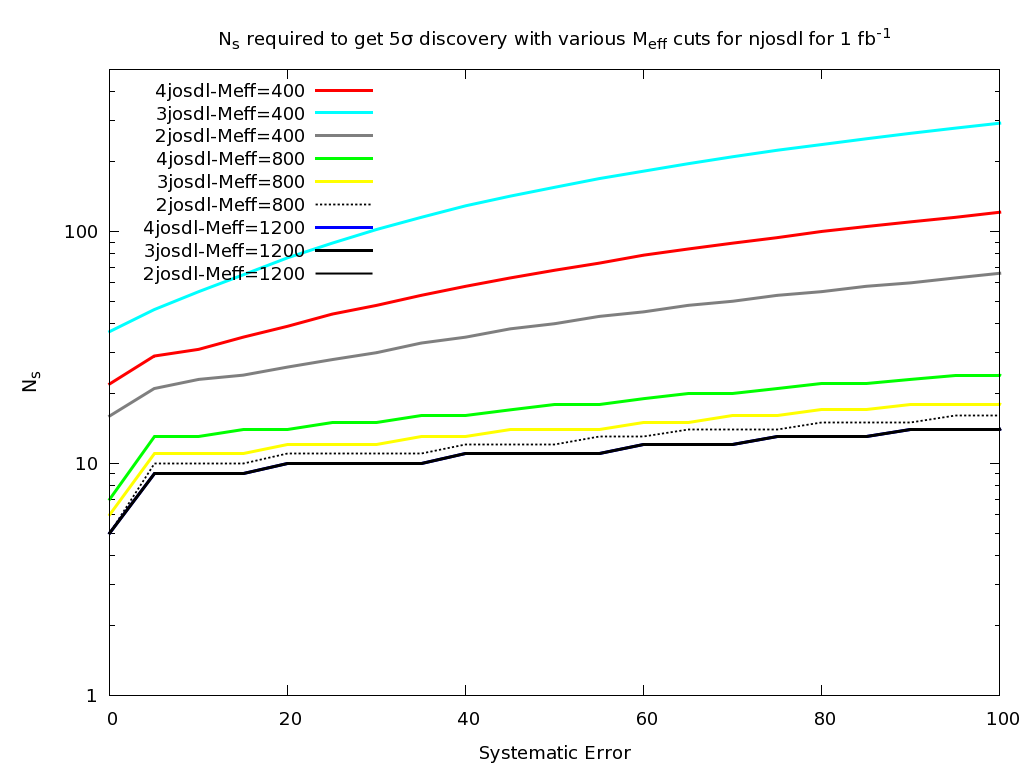}}
\vspace*{-0.3cm}
\centerline{
\includegraphics[width=10.5cm]{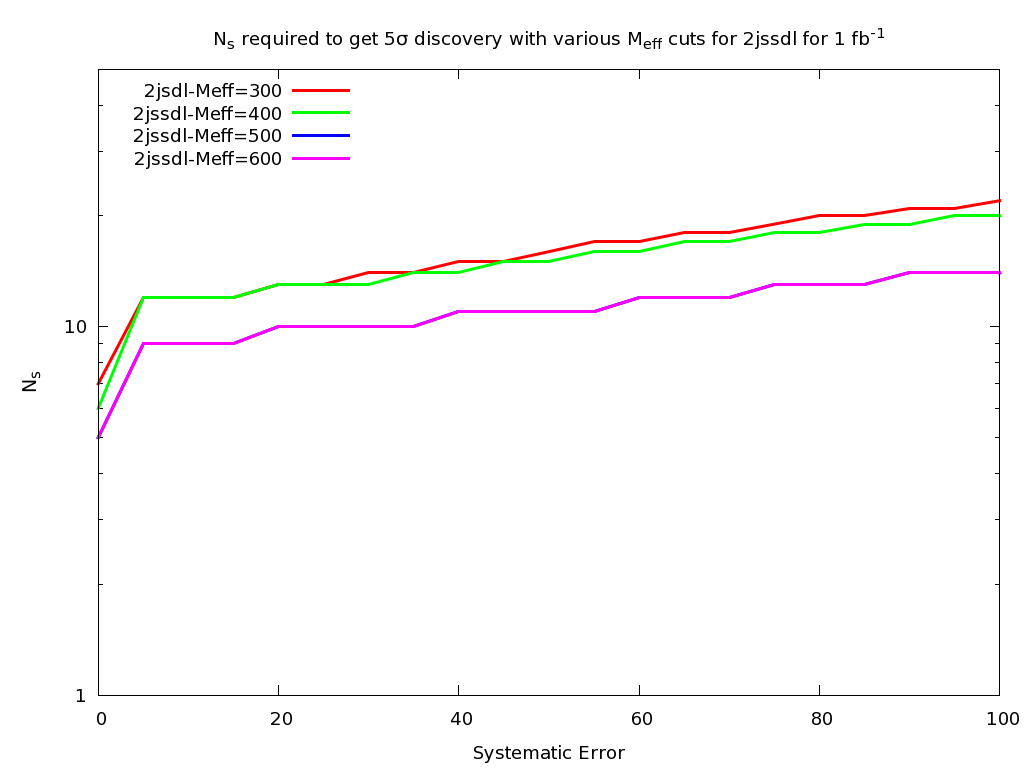}}
%\vspace*{0.1cm}
%    \includegraphics[width=0.70\columnwidth]{figs/Ns5sigReq_njosdl.png} \\
%    \includegraphics[width=0.70\columnwidth]{figs/Ns5sigReq_2jssdl.png}
    \caption{Same as the previous figure but now for the njSSDL(top) and 2jSSDL(bottom) search channels.}
    \label{system2}
\end{figure}

As mentioned above, the size of the SM background systematic errors plays an important role in the ability of the ATLAS \MET\ 
searches to discover Supersymmetry, including
the pMSSM. 
This is not surprising as the number of signal events necessary to reach $S=5$ critically depends upon both the size of the estimated background itself 
as well 
as the background uncertainty. For a fixed systematic uncertainty, search channels with large backgrounds clearly require a large number of 
signal events in order to claim a discovery. To 
get a feel for this in the case of the ATLAS \MET\ analyses studied here,  we  
determine the necessary
number of signal events to reach the $S=5$ level in each analysis as function of the fractional background uncertainty. We remind the reader 
that the SM backgrounds for each channel were supplied to us by the ATLAS SUSY working group~\cite{thanks}.
In performing these calculations we 
exactly follow the discussion as given by ATLAS in Ref.~\cite{Aad:2009wy}. Our results are
displayed in Figs.~\ref{system1} and ~\ref{system2} for the ten ATLAS \MET\ channels assuming 1 fb$^{-1}$ of integrated luminosity.  Here, we 
see the number of signal events that are required to obtain the discovery criterion of $S=5$ for various values of the final \MEFF\ cut.  
In the case of the nj0l channel, which has the largest SM
background, we note that the required number of signal events is  
particularly large and is quite sensitive to the value of the \MEFF\ cut.  
Note that as the systematic error increases, the number of
required signal events can rise drastically, in some cases by an order of magnitude or more.
In particular, the difference between a reasonably low 20\% systematic error and taking a 0\%
error (i.e., ignoring this effect) is substantial
and theoretical analyses that do not include this error are thus wildly optimistic.
 
We will use these numerical results  
in our subsequent analyses of the pMSSM model coverage in these \MET-based searches in the next subsection.  They indicate the importance of 
reducing background systematic errors in 
order to increase the coverage of new physics parameter spaces. 

\clearpage

\subsection{pMSSM Model Coverage}

\begin{figure}
\centerline{
\includegraphics[width=8cm,angle=90]{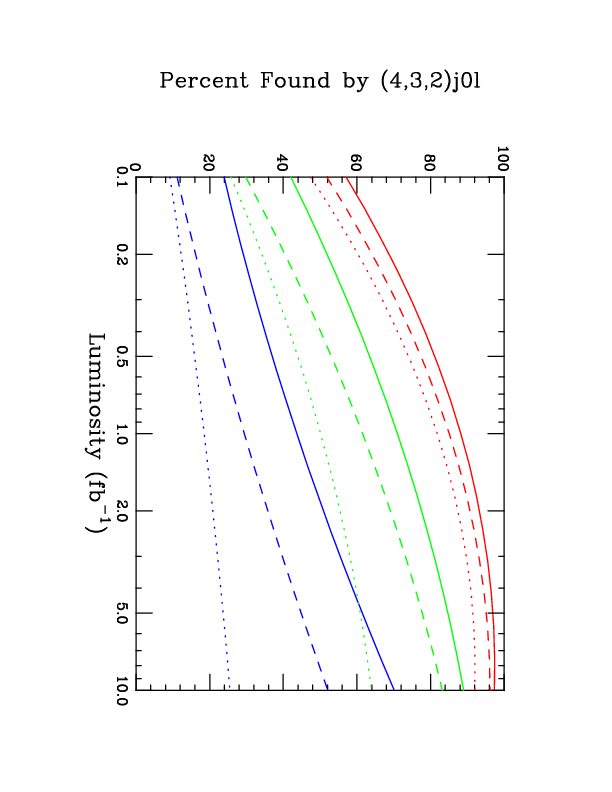}
\hspace*{-1.0cm}
\includegraphics[width=8cm,angle=90]{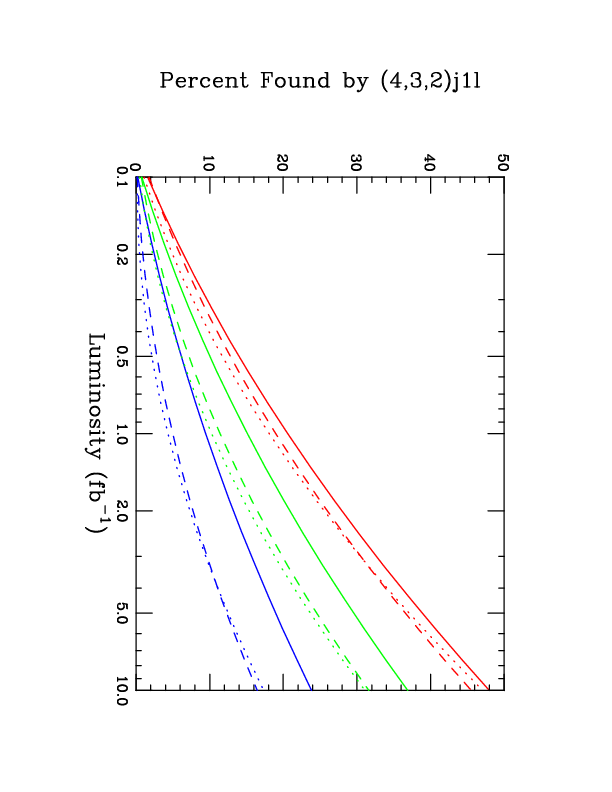}}
\vspace*{-0.5cm}
\centerline{
\includegraphics[width=8cm,angle=90]{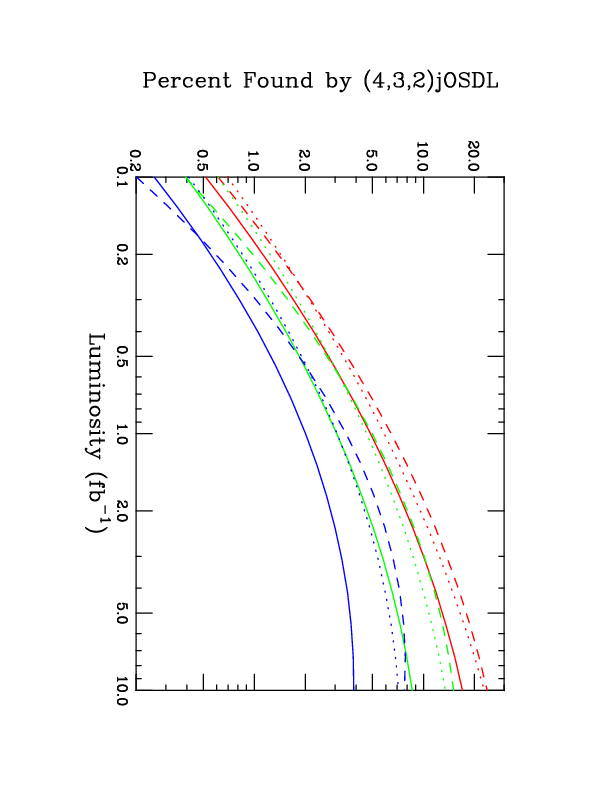}
\hspace*{-1.0cm}
\includegraphics[width=8cm,angle=90]{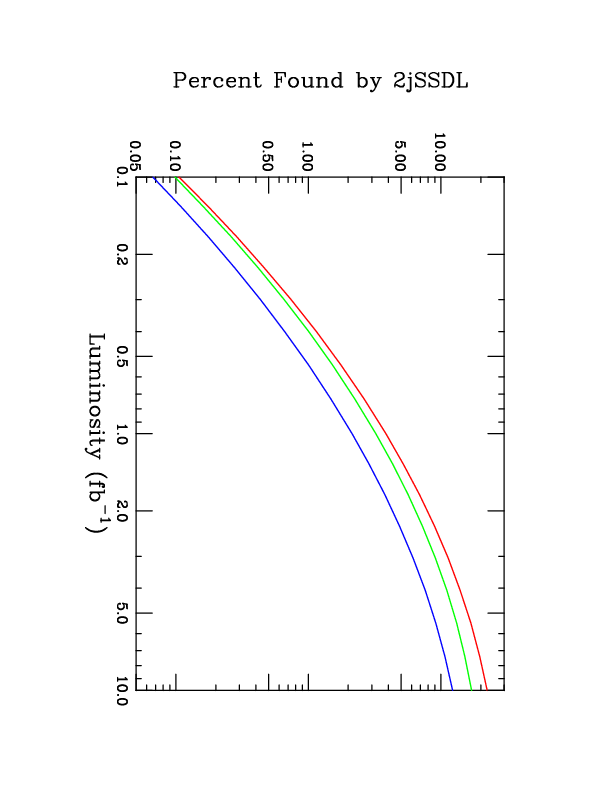}}
    \caption{Fraction of flat prior pMSSM model set that can be observed with $S\geq 5$ in the 
nj0l(top-left), nj1l(top-right), njOSDL(bottom-left), and 2jSSDL(bottom) search channels as a function of the integrated 
     luminosity. The solid(dashed, dotted) curves in each case correspond to n=4(3,2), respectively for the nj0l, nj1l, njOSDL channels. The red(green, blue) curves 
     correspond to background systematic uncertainties of 20(50, 100)\%, respectively.}
    \label{res1flat}
\end{figure}

\begin{figure}
    \centerline{
\includegraphics[width=8cm,angle=90]{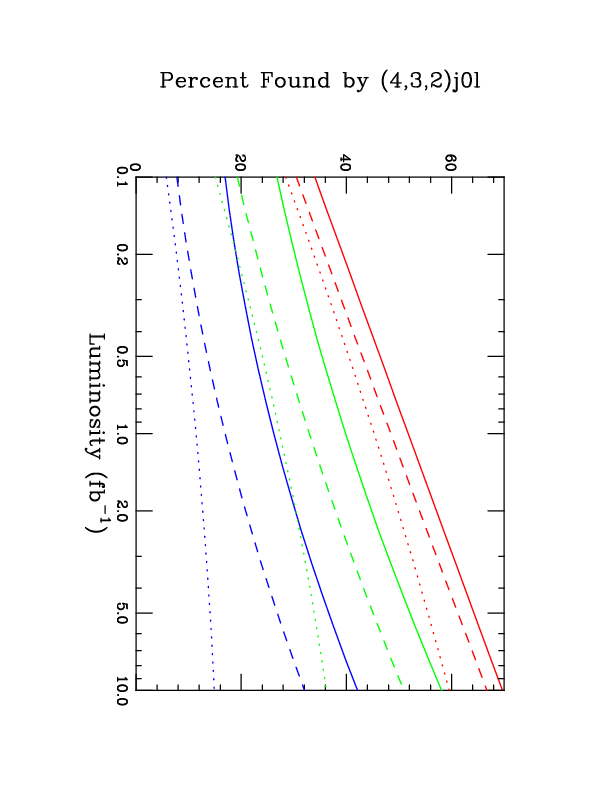}
\hspace*{-1.0cm}
\includegraphics[width=8cm,angle=90]{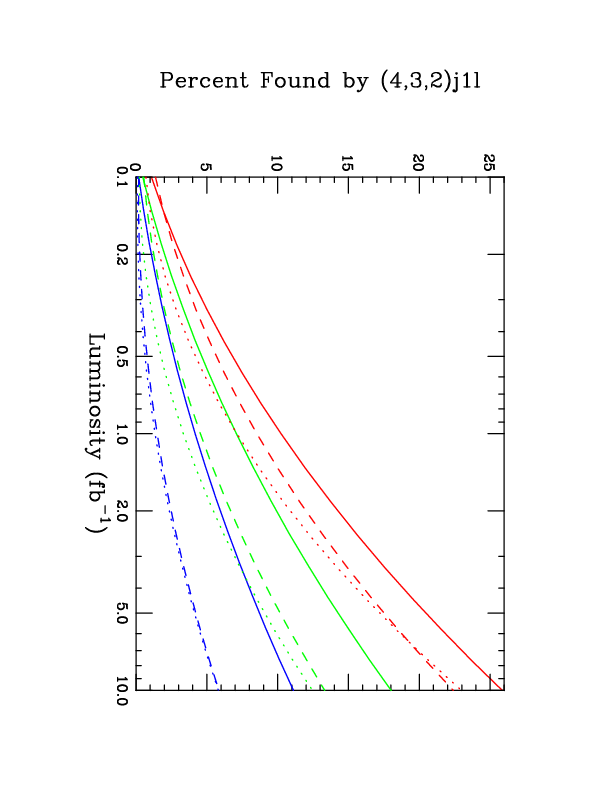}}
\vspace*{-0.5cm}
\centerline{
\includegraphics[width=8cm,angle=90]{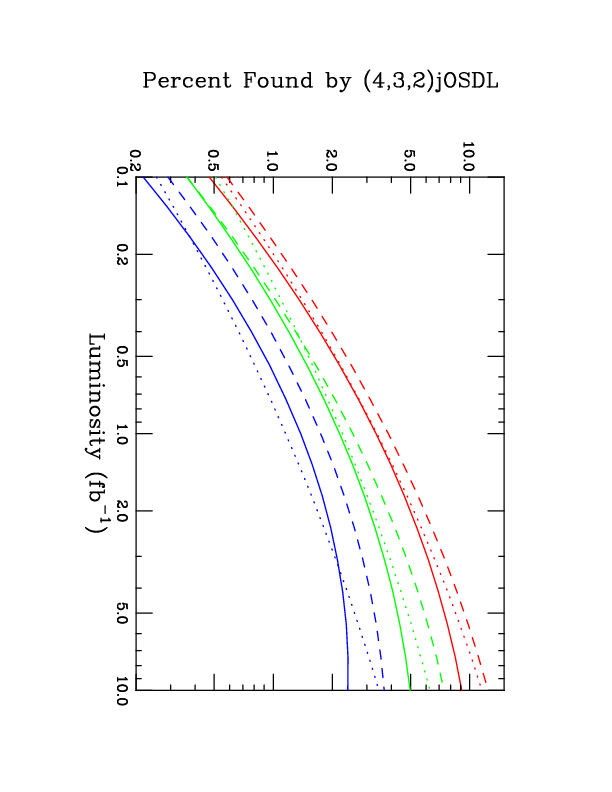}
\hspace*{-1.0cm}
\includegraphics[width=8cm,angle=90]{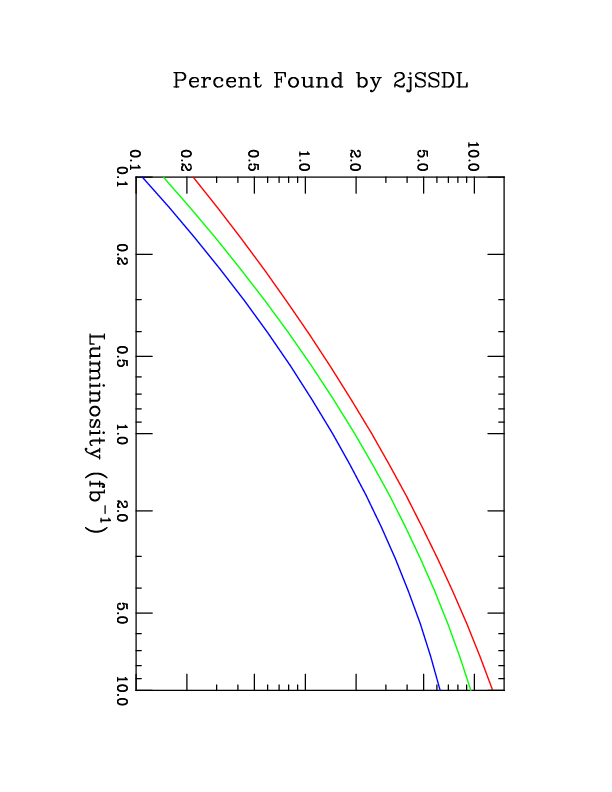}}
    \caption{Fraction of log prior pMSSM model set that can be observed with $S\geq 5$ in the 
nj0l(top-left), nj1l(top-right), njOSDL(bottom-left), and 2jSSDL(bottom-right) search channels as a function of the integrated 
     luminosity. The solid(dashed, dotted) curves in each case correspond to n=4(3,2), respectively for the nj0l, nj1l, njOSDL channels. The red(green, blue) curves 
     correspond to background systematic uncertainties of 20(50, 100)\%, respectively.}
    \label{res1log}
\end{figure}

We now run each of our pMSSM models through the analysis chain described above.  The first
question we address is how well do the various search analyses cover the pMSSM model sample,
or, more precisely, what fraction of these models can be discovered (or not) by these
searches. Further, we 
also determine which of the analyses provide the best model discovery capabilities. Clearly the answers to these questions will be highly sensitive to 
the assumed values of both the integrated luminosity and the estimated SM background uncertainty. Figure~\ref{res1flat} shows the fraction 
of the pMSSM models that can be discovered with $S\geq 5$ 
in each of the ATLAS \MET\ channels for the flat prior model set as a function of the integrated luminosity assuming three different choices for the 
background systematic error. The corresponding results obtained in the case of the log prior model set can be found in Fig.~\ref{res1log}. 
Here, we again emphasize that our pMSSM sample is not
meant to provide full coverage of the 19-dimensional parameter space (such coverage would be
computationally prohibited).  However both the very large number of models in our pMSSM sample, 
and the distinct characteristics they possess, make this sample an ideal testbed for
this set of mSUGRA designed search strategies.  Our fractional results based on our pMSSM
model set are thus indicative of the behavior of the MSSM under these search routines.

These figures reveal a number of interesting results: ($i$) The size of the background systematic errors makes a significant impact on model coverage for all 
search channels and integrated luminosities. For the searches with significant SM backgrounds, i.e., the nj0l and nj1l channels, 
variation in the background uncertainty 
leads to substantial changes in the capability to observe the model sample.  
The search least affected by systematics is 2jSSDL since the backgrounds in this case are quite small. This behavior confirms 
the results of the previous subsection. ($ii$) The model coverage in almost all cases is significantly better for the flat prior model set 
than for the log prior sample. This, too, is not very surprising since the masses of the sparticles in the log prior case extend out to much 
larger values and the sparticle spectrum is 
generally more compressed in this set \cite{Berger:2008cq}. The latter leads to softer jets and leptons in the corresponding cascade 
decays which have a more difficult time passing the analysis cuts. ($iii$) For all values of the background systematic error, the nj0l channels yield the best  
model space coverage with 4j0l affording the best discovery opportunity. In fact, we see that
the channels which require more jets to be present have a better chance of being observed for
the nj0l and nj1l searches.  For the flat prior sample, the 4j0l analysis with low background systematics is observed to cover a very large 
fraction of the model set by itself once significant integrated luminosities are obtained.
($iv$) As the number of leptons required to be present in the final state increases, 
the model coverage is found to decrease significantly, especially for smaller values of the integrated luminosity. 
This is due to the fact that the branching fractions for leptons to appear in squark and gluino induced cascade 
decays are generally not very large in our model sample, 
as we have seen in our earlier work \cite{Conley:2010du}. 
($v$) Independently of the specific \MET\ search, as the background systematic errors become large,
the pMSSM model coverage is seen to increase more slowly with the 
integrated luminosity. Some of the search channels nearly saturate at high luminosity due to the large background uncertainties and thus become systematics dominated.

As discussed above, the final step in the ATLAS \MET\ analyses is to apply a cut on \MEFF,
where the particular value of \MEFF\ that is chosen  (in units of 400 GeV) is the one 
that maximizes the signal significance given the SM background and its corresponding uncertainty. This choice not only depends upon the particular channel but 
also on the amount of integrated luminosity. It is important to note that if this cut is taken
to be {\it too} large when maximizing the signal, then the analysis will be 
very sensitive to the detailed shape in the tails of both the signal and expected background distributions, especially with higher luminosities. This happens 
when there are very few events with large values of \MEFF, e.g., 2 TeV. In this situation, small fluctuations in the SM background and/or SUSY signal 
expectations due to limited Monte Carlo statistics can lead to inconsistencies in whether a given model is observable in a specific analysis or even whether or not it 
would be detected overall.  This effect only occurs in the case of the search analyses designed
for the 7 TeV run, as the \MEFF\ cut applied in the planned 14 TeV ATLAS analyses was fixed at relatively low values.  
 
\begin{figure}
\centerline{
  \includegraphics[width=0.45\columnwidth]{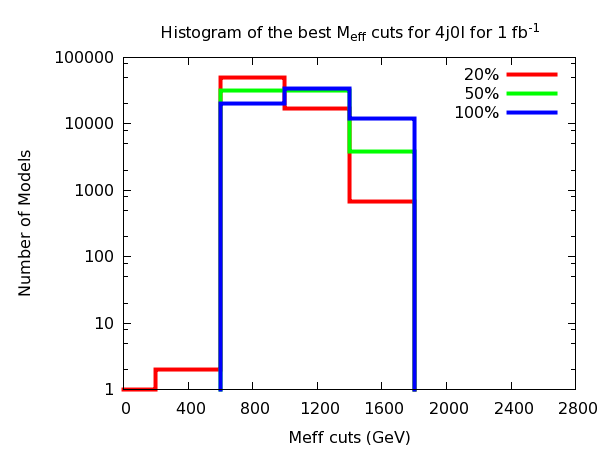}}
\centerline{
  \includegraphics[width=0.45\columnwidth]{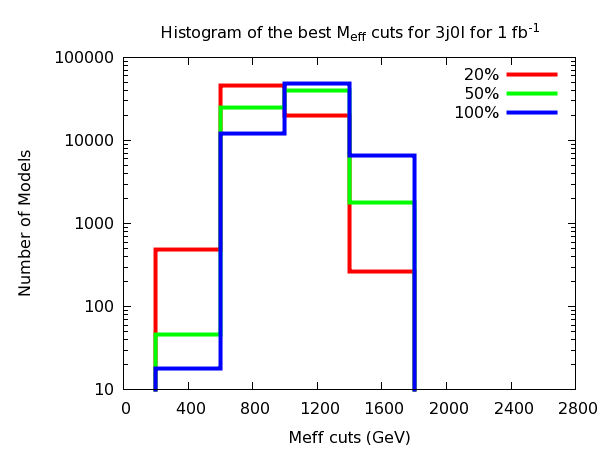}
  \includegraphics[width=0.45\columnwidth]{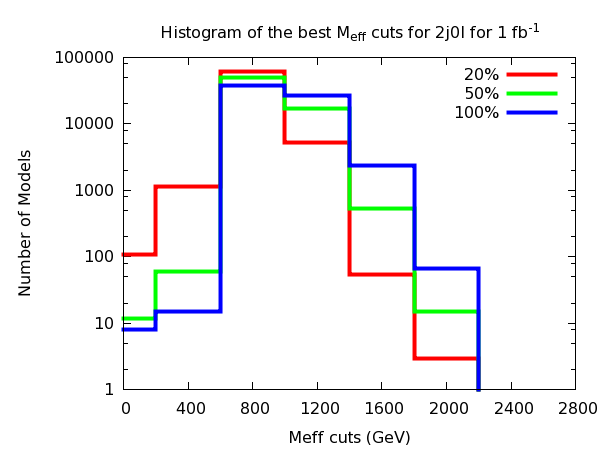}}
  \caption{Optimized \MEFF\ cut for the flat prior model set for the 4j0l(top),  3j0l(bottom left) and 2j0l analyses(bottom right). 
    The red(green, blue) histograms correspond to background systematic uncertainties of 20(50, 100)\%, respectively. An integrated luminosity of 1 fb$^{-1}$ has 
    been assumed in these figures for purposes of demonstration.}
  \label{meffcut}
\end{figure}

Figure~\ref{meffcut} shows the optimized value for the \MEFF\ cut for the nj0l analyses, as an example, for the flat prior model sample.   
The three things we see here are: ($i$) for large background systematic errors, a harder
\MEFF\ cut is required to optimize the search, ($ii$) as the number of required jets in
the final state decreases, the strength of the cut can be reduced.
Both of these results are also found to hold for the nj1l and njOSDL searches although the 
\MEFF\ cut itself turns out to be less important as the number of required leptons in the final state increases. ($iii$) Given the warning about distribution 
tails in the discussion above, it is a welcome result to see that in the majority of cases only a moderately strong \MEFF\ cut is required to optimize the signal 
significance. Note that there are some cases where the \MEFF\ cut does not contribute very much to increase the significance of the SUSY signal; this 
happens in particular for scenarios where the background is low and the effect of systematic uncertainties is not very significant.  

It is interesting to evaluate the fraction of models that can be discovered in multiple analyses. This is important to consider as, first, it is valuable to verify 
the discovery of new physics in more than one channel, and second, the availability of multiple discovery channels admits for the possibility of further 
studies that will allow for, e.g., the extraction of sparticle masses. To be specific, Tables~\ref{n_analyses_20}, \ref{n_analyses_50} and \ref{n_analyses_100} 
show the fraction of pMSSM models which are observed in (exactly) $n$ channels assuming a SM background systematic error of 20, 50, and 100\%, respectively.  Note that 
the distribution shifts towards more models being observed in multiple channels as the integrated luminosity increases and the background systematic 
error decreases, as expected..

\begin{table}
\centering
  \begin{tabular}{ | c || c | c | c | c | c| c | }
  \hline
  \# anl. & Flat $\mathcal{L}_{0.1}$ & Flat $\mathcal{L}_{1}$ & Flat $\mathcal{L}_{10}$ & 
Log $\mathcal{L}_{0.1}$ & Log $\mathcal{L}_{1}$ & Log $\mathcal{L}_{10}$ \\ \hline \hline
  0  &  38.172 &        7.5501 &        0.9965 &        63.64 &         43.988 &        22.92 \\ \hline
  1  &  9.2928 &        4.1988 &        0.90862 &       5.376 &         4.8674 &        5.8482 \\ \hline
  2  &  8.7432 &        4.6665 &        1.6102 &        3.6687 &        5.6665 &        6.0298 \\ \hline
  3  &  41.836 &        59.878 &        39.573 &        26.008 &        34.907 &        35.38 \\ \hline
  4  &  0.65686 &       4.9257 &        7.9422 &        0.25427 &       2.2158 &        6.4657 \\ \hline
  5  &  0.53472 &       4.2629 &        6.7163 &        0.47221 &       2.0341 &        4.8311 \\ \hline
  6  &  0.54366 &       8.5391 &        13.494 &        0.32692 &       3.0875 &        6.5383 \\ \hline
  7  &  0.067026 &      2.5217 &        8.9044 &        0.21794 &       1.453 &         4.1773 \\ \hline
  8  &  0.062558 &      1.2288 &        5.6364 &        0.036324 &      0.72648 &       2.2884 \\ \hline
  9  &  0.077452 &      1.2958 &        6.548 &         0 &     0.58118 &       2.9422 \\ \hline
  10  &  0.013405 &     0.93241 &       7.6711 &        0 &     0.47221 &       2.579 \\ \hline
\end{tabular}

  \caption{The fraction of models that are observed in (exactly) $n$ \MET\ search channels assuming a SM background systematic error of 20\%.}
\label{n_analyses_20}
\end{table}

\begin{table}
\centering
  \begin{tabular}{ | c || c | c | c | c | c| c | }
  \hline
  \# anl. & Flat $\mathcal{L}_{0.1}$ & Flat $\mathcal{L}_{1}$ & Flat $\mathcal{L}_{10}$ & Log 
$\mathcal{L}_{0.1}$ & Log $\mathcal{L}_{1}$ & Log $\mathcal{L}_{10}$ \\ \hline \hline
  0  &  54.756 &        21.772 &        4.8782 &        71.558 &        55.903 &        32.546 \\ \hline
  1  &  14.143 &        10.547 &        4.847 &         8.1729 &        7.3011 &        9.8801 \\ \hline
  2  &  7.8435 &        11.453 &        9.959 &         5.0854 &        7.1195 &        12.532 \\ \hline
  3  &  22.552 &        42.949 &        40.705 &        14.857 &        24.228 &        28.478 \\ \hline
  4  &  0.29938 &       4.1407 &        8.3533 &        0.18162 &       1.7436 &        4.5768 \\ \hline
  5  &  0.15788 &       3.1562 &        7.619 &         0 &     1.3803 &        3.4871 \\ \hline
  6  &  0.1415 &        3.3036 &        9.1487 &        0.072648 &      1.0534 &        3.4871 \\ \hline
  7  &  0.061068 &      1.4075 &        6.049 &         0.036324 &      0.79913 &       1.9615 \\ \hline
  8  &  0.031279 &      0.58536 &       3.6166 &        0.036324 &      0.32692 &       1.4166 \\ \hline
  9  &  0.013405 &      0.43493 &       2.9716 &        0 &     0.036324 &      1.235 \\ \hline
  10  &  0.0014895 &    0.25172 &       1.853 &         0 &     0.10897 &       0.39956 \\ \hline
\end{tabular}

  \caption{Same as the previous Table but now assuming a SM background systematic error of 50\%.}
\label{n_analyses_50}
\end{table}

\begin{table}
\centering
  \begin{tabular}{ | c || c | c | c | c | c| c | }
  \hline
  \# anl. & Flat $\mathcal{L}_{0.1}$ & Flat $\mathcal{L}_{1}$ & Flat $\mathcal{L}_{10}$
& Log $\mathcal{L}_{0.1}$ & Log $\mathcal{L}_{1}$ & Log $\mathcal{L}_{10}$ \\ \hline \hline
  0  &  74.112 &        47.23 &         17.635 &        81.911 &        69.016 &47.875 \\ \hline
  1  &  13.894 &        16.834 &        15.996 &        9.8438 &        10.825 &14.094 \\ \hline
  2  &  4.4759 &        13.331 &        21.917 &        3.1602 &        7.7007 &15.91 \\ \hline
  3  &  7.3282 &        18.166 &        26.186 &        4.9401 &        10.607 &15.365 \\ \hline
  4  &  0.10575 &       1.8827 &        6.478 &         0.036324 &      0.79913 &2.6153 \\ \hline
  5  &  0.037237 &      1.0322 &        4.7174 &        0.036324 &      0.32692 &1.7436 \\ \hline
  6  &  0.023832 &      0.7075 &        3.3008 &        0 &     0.32692 &       1.0171 \\ \hline
  7  &  0.019363 &      0.57345 &       2.1703 &        0.072648 &      0.36324 &0.87178 \\ \hline
  8  &  0.0029789 &     0.16831 &       1.0025 &        0 &     0.036324 &      0.36324\\ \hline
  9  &  0.0014895 &     0.064047 &      0.46474 &       0 &     0 &     0.10897 \\ \hline
  10  &  0 &    0.011916 &      0.13257 &       0 &     0 &     0.036324 \\ \hline
\end{tabular}

  \caption{Same as the previous Table but now assuming a SM background systematic error of 100\%.}
\label{n_analyses_100}
\end{table}

\begin{figure}
\centerline{
\includegraphics[width=10.5cm,angle=90]{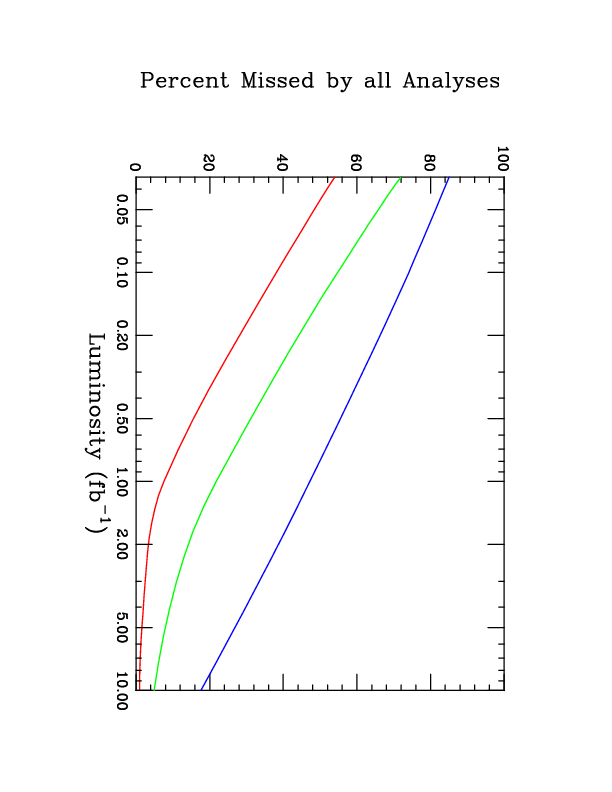}}
\vspace*{-1.5cm}
\centerline{
\includegraphics[width=10.5cm,angle=90]{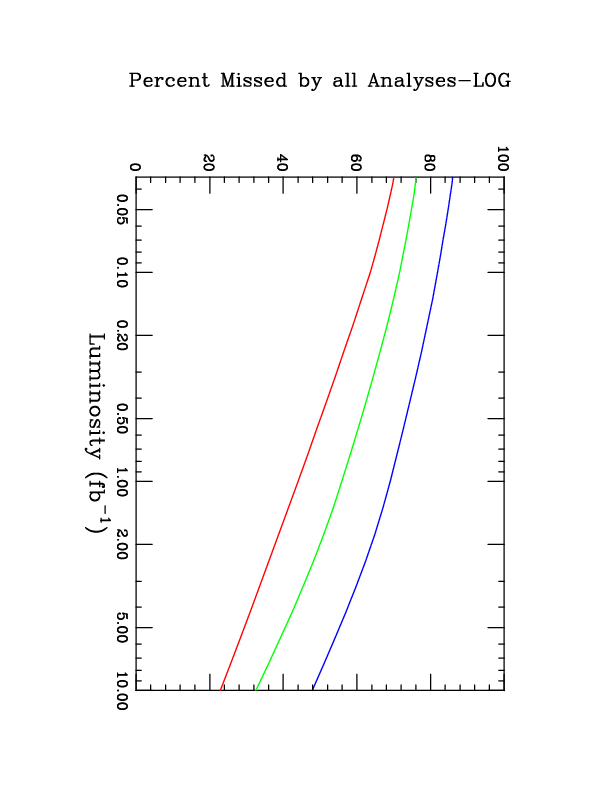}}
\vspace*{0.1cm}
    \caption{Fraction of flat prior(top) or log prior(bottom) pMSSM model sets which are 
undetected  after combining all of the ATLAS \MET\ search analyses. 
     The red(green, blue) curves correspond to background systematic uncertainties of 20(50, 100)\%, respectively.}
\label{search}
\end{figure}

By combining our results for these \MET\ searches, we can also determine the fraction of pMSSM models that are undetected in {\it all} 
of the 7 TeV search analyses designed by ATLAS; 
this corresponds to the case of $n=0$ in these Tables. Figure~\ref{search} presents the fraction of pMSSM models which are undetected in all
of the search channels 
as a function of integrated luminosity for both the flat and log prior sets. In the flat prior case we see that as the integrated 
luminosity increases, the model coverage substantially improves, and approaches (or exceeds) $\sim 95\%$ for 10 fb$^{-1}$ with 
SM background uncertainties of 50\% or less. In the log prior case, the improvement in pMSSM 
model coverage as the luminosity increases is much more gradual as we expected. However, even in this case, 
at high integrated luminosities 
substantial model coverage is seen to be obtainable at 7 TeV. 

This figure also shows the important playoff between increasing the integrated luminosity and decreasing the SM background systematic error in terms of pMSSM 
model coverage. (Of course, increased luminosity often results in decreased systematic errors,
up to a point.)
For example, it is interesting to compare the effectiveness of the analyses for the flat prior set assuming $\delta B=100\%$ 
and $\lum=5(10)$~fb$^{-1}$ with other values. 
Figure~\ref{search} shows that taking $\delta B=50\%$ and $\lum=0.65(1.4)$~fb$^{-1}$ or 
$\delta B=20\%$ with $\lum=0.20(0.39)$~fb$^{-1}$ produces essentially identical model 
coverage. This demonstrates that small reductions in the SM background uncertainty can be worth a significant amount of increased integrated luminosity in terms 
of pMSSM model coverage.

Another very important message to take home from this figure is that for the integrated luminosity collected during the 2010 LHC run at 7 TeV  
($\sim 35$ pb$^{-1}$ for ATLAS), some significant fraction of these pMSSM models should already have been observed at the $S=5$ level. 
Explicitly, for a background systematic 
error of 20(50,100)\% we find that 46(28,15)\% of the flat prior model set should have been discovered; in the log prior case the corresponding results are found 
to be 30(24,14)\%. Since these are {\it discovery} results, an even greater portion of the pMSSM model sample would be expected to 
be {\it excluded} by these analyses. 
This shows the incredible power of going from Tevatron energies up to the 7 TeV LHC in performing searches for massive objects, such as 
SUSY sparticles, even when only small amount of integrated luminosity is available.

\clearpage

\subsection{Why are Models not Detected by the ATLAS \MET\ Searches?}

Here, we investigate the main reasons why some pMSSM models are not discoverable in
the ATLAS \MET\ searches at 7 TeV. 
We addressed this question in some detail in our earlier work for the case of the analyses designed for the 14 TeV 
LHC \cite{Conley:2010du}, so our discussion here will not be as extensive. Clearly, many of our previous results will carry over qualitatively into the present 
7 TeV analysis. 

\begin{figure}
\centerline{
\includegraphics[width=8.0cm]{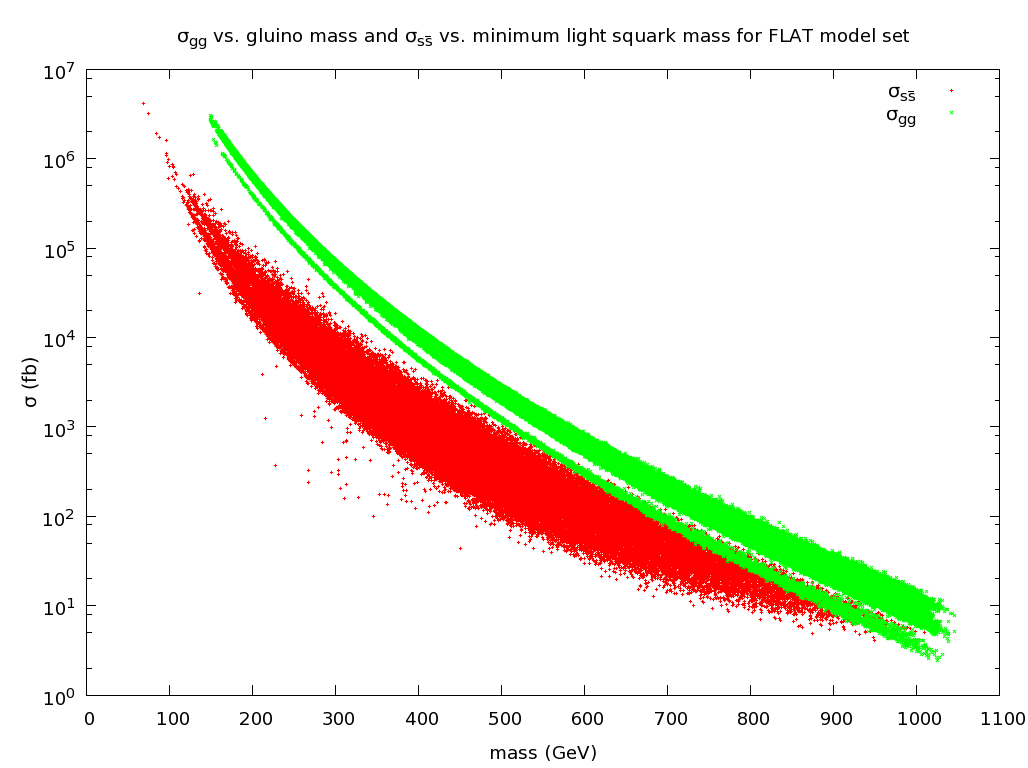}
\hspace*{0.4cm}
\includegraphics[width=8.0cm]{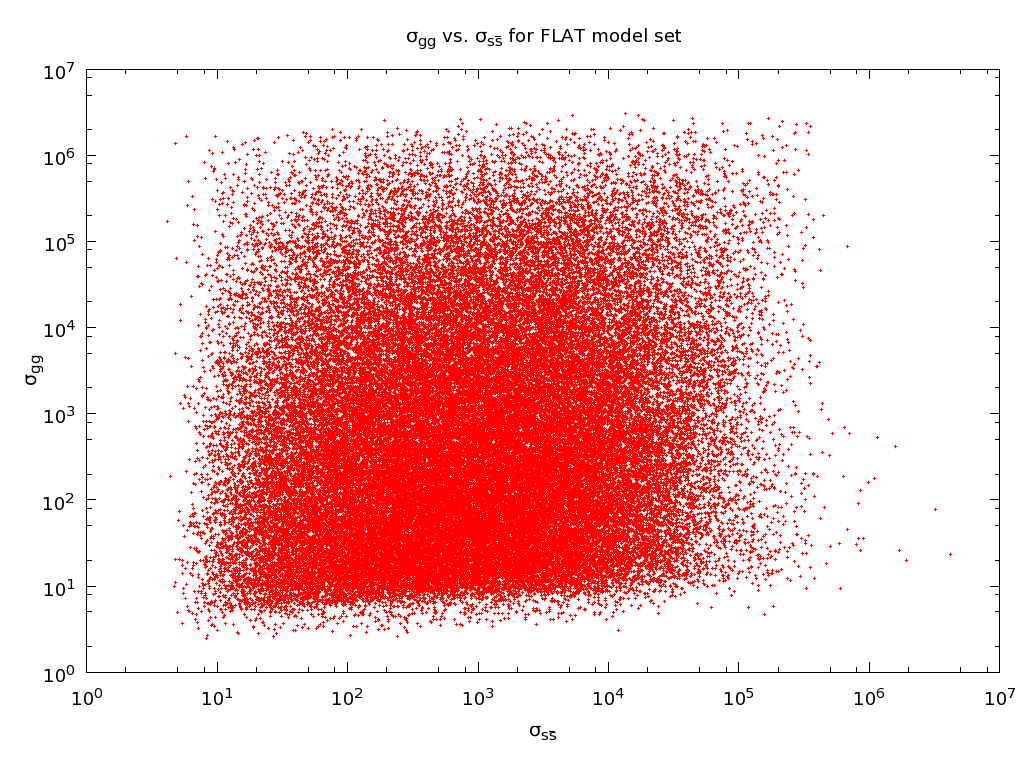}}
%\vspace*{-0.4cm}
\centerline{
\includegraphics[width=8.0cm]{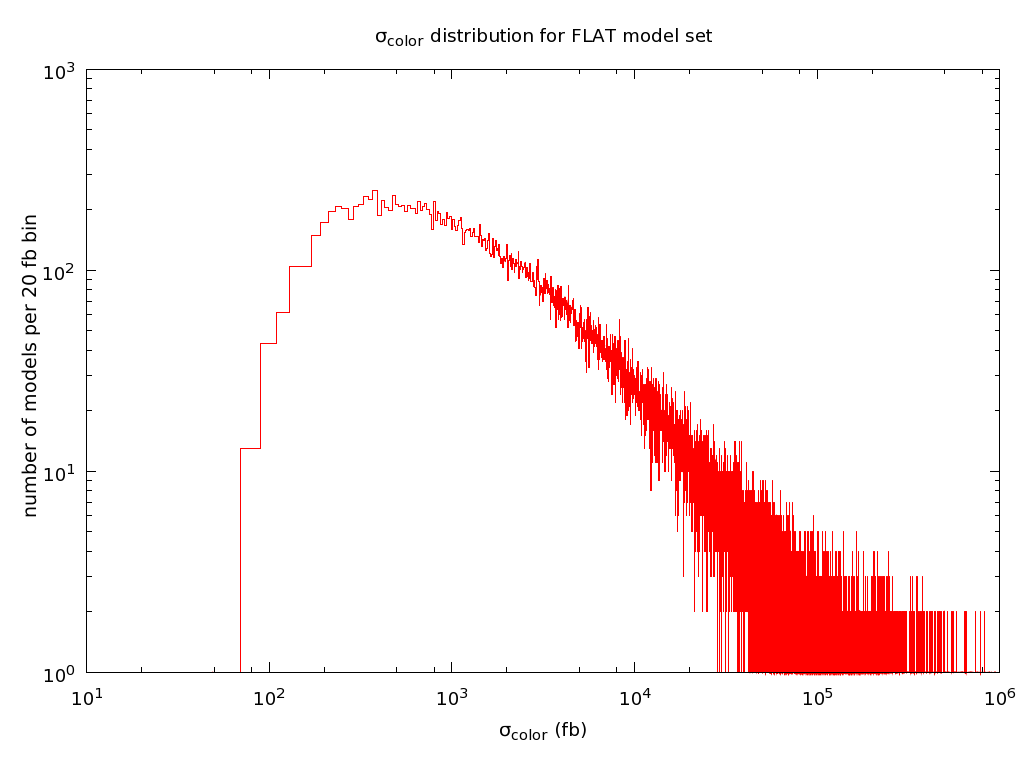}
\hspace*{0.4cm}
\includegraphics[width=8.0cm]{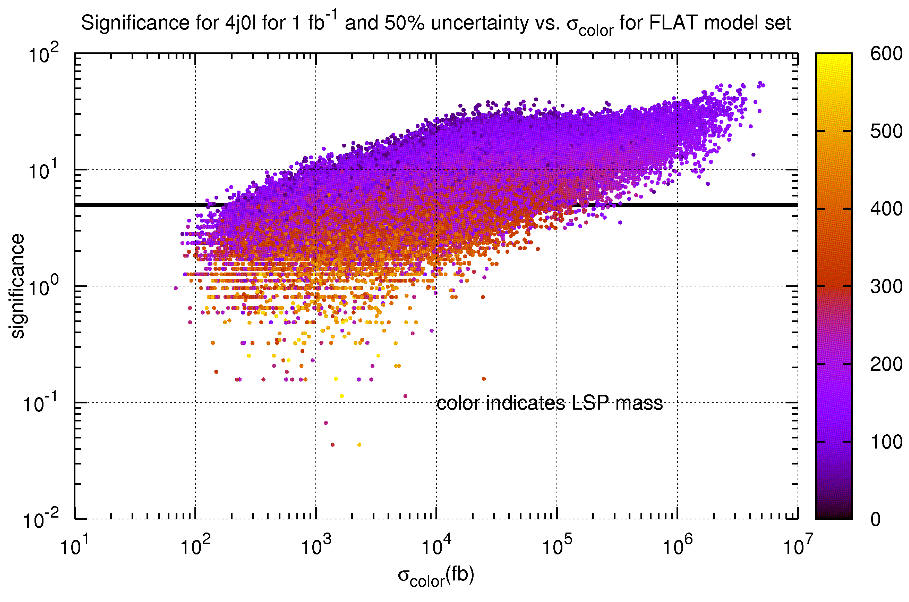}}
\vspace*{0.9cm}
    \caption{(Top left ) NLO first generation squark and gluino pair production cross sections
at $\sqrt s=7$ TeV as a function of their masses for the flat prior model set. The green(red) points represent the
gluino(squark) cross sections. 
(Top right) Correlation of the squark pair and gluino pair NLO cross sections in the flat prior set.  Each point represents one model. (Bottom left) 
Total NLO QCD production cross section 
distribution for the flat model set. (Bottom right) Search significance of the 4j0l analysis as a function of the total NLO QCD production cross section assuming 
$\lum=1$ fb$^{-1}$ and $\delta B=50\%$. The solid line highlights the $S=5$ discovery level and the color code reflects the mass of the LSP.}
 \label{sig-spread}
\end{figure}

There are multiple explanations as to why some pMSSM models are undetected by the \MET\ searches, the most obvious one being small production 
cross sections for the colored 
sparticles that initiate the familiar SUSY cascades. As an example, we note that for our pMSSM models in the flat prior set, the cross sections for the production 
of gluino and squark pairs are found to cover an enormous range of several orders of magnitude as can be seen in Fig.~\ref{sig-spread} (recall 
that the upper limit on sparticle masses in
our flat model set is $\sim 1$ TeV). Here we see that the 
large (or small) values of the gluino pair cross section is completely uncorrelated with the corresponding values for first generation squarks 
within a particular model. Furthermore, 
by summing over all of the QCD production channels involving gluinos and/or first generation squarks (i.e., $\tilde g \tilde g$, $\tilde g \tilde q$, 
$\tilde q \tilde q$ and $\tilde q \tilde q^*$) we obtain an approximate handle on the total overall rate for SUSY production which we see ranges over four orders of 
magnitude. Note that for any particular value of the squark or gluino mass, the corresponding production cross section itself can vary by up to an order of magnitude 
or more depending upon the remainder of the pMSSM model spectrum.  

\begin{figure}
\centerline{
\includegraphics[width=8.0cm]{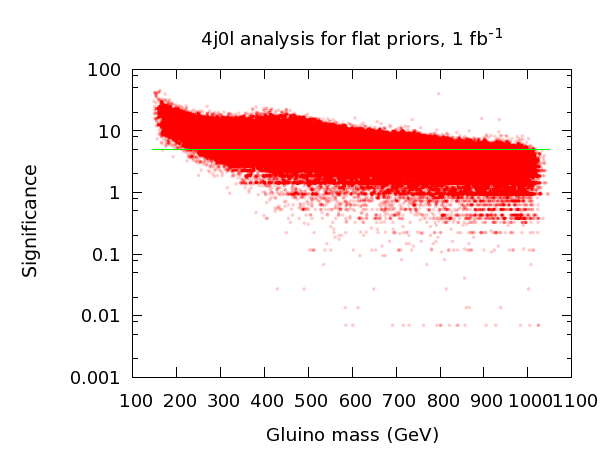}
\hspace*{0.0cm}
\includegraphics[width=8.0cm]{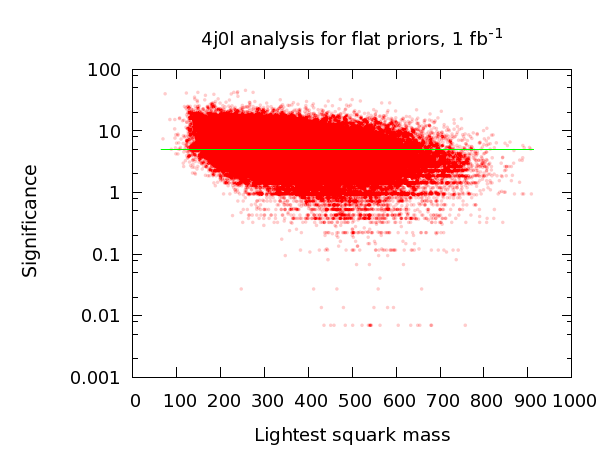}}
\vspace*{0.5cm}
\caption{Significance of the 4j0l search for the flat prior model set as a function of the gluino(left) and lightest 
  squark(right) masses. $\lum=1$ fb$^{-1}$ and 
  $\delta B=50\%$ have been assumed.}
\label{masses}
\end{figure}

While it is certainly clear from this figure that some models have too small a cross section
to be discovered, the bottom-right panel indicates that this cannot be the entire explanation. 
Here we show the search significance, $S$, of the 4j0l channel 
(as it is the most powerful channel in terms of discovery capability) as a function of the total NLO QCD production 
cross section assuming $\lum=1$~fb$^{-1}$ and $\delta B=50\%$. Here we observe that ($i$)  there are models with cross sections $\sim 20$ pb which are {\it missed} by 
this analysis, while ($ii$) there are models with cross sections $\sim 100$ fb which are discovered. ($iii$) For any given value of the cross 
section, the range of the significance is large and can be up to two orders of magnitude or more.
This validates the claim that there are reasons other than small production cross sections that
render models unobservable by these \MET\ analyses. ($iv$) For any given gluino mass there is a strong correlation of the signal significance with 
the mass of the LSP. Clearly when these two masses are close the average $p_T$ of the jets will be softer and this will make it more difficult to pass analysis cuts. 
Also if the LSP mass is large then that implies even larger squark and 
gluino masses that will result on average in smaller production cross sections. In the case of the nj0l analyses, 
a larger number of signal events is required for discovery due to the sizeable 
SM backgrounds and hence such models will be missed by these analyses. Visibility will then 
require the production of leptons with significant $p_T$ in cascade decays in order 
to pass the lower background nj1l, njOSDL and  2jSSDL searches. Unfortunately, lepton 
branching fractions are low in these cascades in our model sample (since, e.g., sleptons are heavy) 
and thus some models will be missed entirely. Of course, models that are lepton-rich will  
automatically fail all the nj0l analyses since they veto events with high $p_T$ leptons, but will
be picked up by the searches containing leptons.

\begin{figure}
\centerline{
\includegraphics[width=8.0cm]{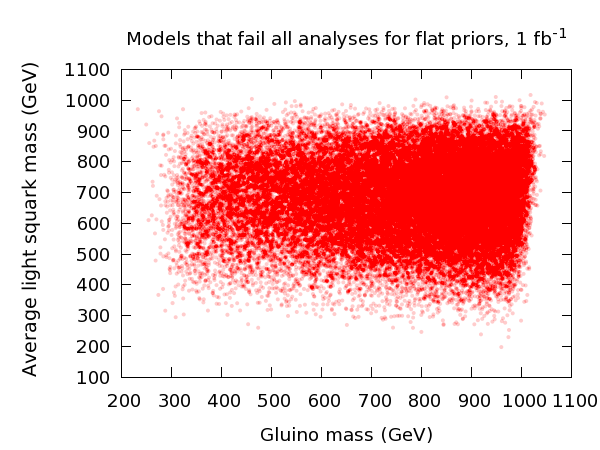}}
%\vspace*{-1.0cm}
\centerline{
\includegraphics[width=8.0cm]{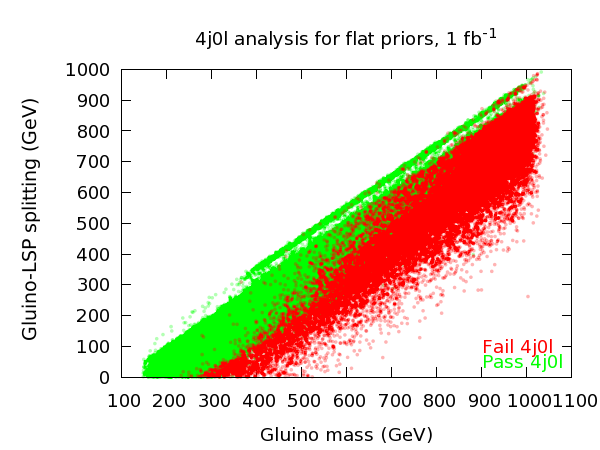}
\hspace*{0.0cm}
\includegraphics[width=8.0cm]{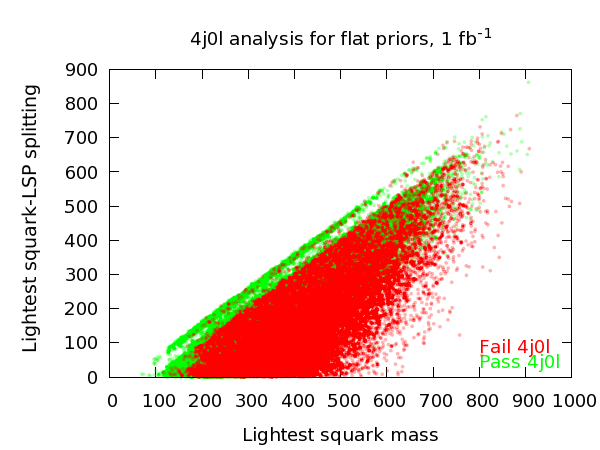}}
\vspace*{0.9cm}
\caption{(Top) Flat prior models that are unobservable in all of the 
  \MET-based search analyses in the average light squark -gluino mass plane. (Bottom) Flat prior models that pass(green) or fail(red) 
  the 4j0l analysis in the gluino(lightest squark) mass vs gluino-LSP(lightest squark-LSP) mass splitting plane in the left(right) panel. $\lum=1$ fb$^{-1}$ 
  and $\delta B=50\%$ have again been assumed.}
\label{splitting}
\end{figure}

Though the production cross section for SUSY particles is reasonably well correlated with their masses, we can ask more directly if larger sparticle masses lead to 
their non-observation in these searches. Figure~\ref{masses} shows that, indeed, 
models with lighter squarks or gluinos tend to lead to signals with greater significance in the 
4j0l channel. However, as we can also see from this Figure, this is not true 
universally, e.g., there are many models with gluino (lightest squark) masses below 300 (200) 
GeV that have $S<5$, while conversely there are models with 1 TeV gluinos that have $S>5$.
We see that for any given squark or gluino mass the value of $S$ can vary 
significantly. The top panel of Fig.~\ref{splitting} displays this property 
even more strongly where we see that these results hold even when all of the \MET\ searches are combined. 
This Figure shows the set of flat prior models that are unobservable in all the 
search channels in average light squark mass--gluino mass plane. Note that there are a significant number of 
these models which contain light squarks and gluinos. Thus while the masses of the colored sparticles do play an important role in model observability clearly there 
are additional important factors.

\begin{figure}
\centerline{
\includegraphics[width=12.0cm]{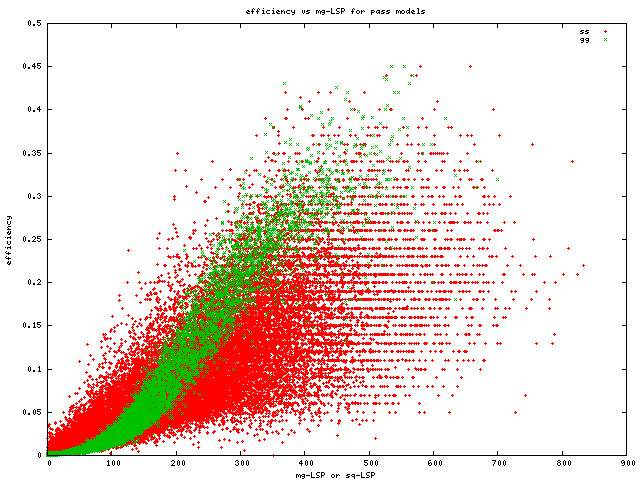}}
\vspace*{0.9cm}
\caption{The efficiency for passing the 4j0l analysis cuts as a function of the (green)gluino- or (red)squark-LSP mass splitting.  
Here, efficiency is defined as the fraction of generated signal events that pass the analysis cuts.}
\label{splitting2}
\end{figure}

In the lower panels of Fig.~\ref{splitting} we see that the mass splitting between squarks and/or gluinos and the LSP can play an important role in determining 
model observability as was first noted in Ref.~\cite{Alwall:2008ve} and was seen explicitly in our earlier work on the generation of 
the pMSSM models \cite{Berger:2008cq} 
and the 14 TeV ATLAS SUSY analyses \cite{Conley:2010du}. The obvious reasoning here is that as the degeneracy in the spectrum 
increases and mass splittings become smaller, the values of, e.g., the $p_T$ of the jets, will be
reduced so that it will be more difficult to satisfy any of the analysis 
cuts. These figures show this result explicitly. Note, however, that, e.g., in the case of light gluinos with small gluino-LSP splittings there are many models which 
are still discoverable in the 4j0l channel. The reason for this is that while the 
efficiency\footnote{Here, efficiency is defined as the fraction of generated signal events
that pass the analysis cuts.} 
for passing the 4j0l analysis cuts may be quite low for small mass splittings, as seen in Fig.~\ref{splitting2},  the cross section to produce the lighter 
gluinos/squarks is very large and more than compensates for these low efficiencies, especially if there are any additional hard jets in the event from ISR.  However, 
there are numerous unobservable models that have larger raw sparticle production cross sections than observable models with somewhat similar spectra; the difference 
then being in their respective abilities to pass the necessary analysis cuts.

Another cause for models being undetected is the occurrence of detector-stable sparticles at the end of 
gluino or squark induced decay chains instead of the LSP \cite{Conley:2010du}. 
This happens with reasonable frequency in both the 7 and 14 TeV analyses.
In such cases, the amount of \MET\ that is produced is substantially decreased which reduces the capability 
of the relevant models to pass any \MET\ analysis requirements. Most commonly, these sparticles are actually long-lived charginos that are reasonably degenerate 
with the LSP in wino- or Higgsino-like LSP scenarios.  In these cases, searches for long-lived sparticles, as discussed below, will be a important supplement 
to the conventional \MET\ searches. Of course, a loss of the \MET\ signature can happen in other ways. For example, if 
the initial squark or gluino produces a very long decay chain then the particles produced at the end of such a chain will be somewhat soft. In some cases this may 
lead to the inability to pass the necessary $p_T$ and/or \MET\ requirements for the various
\MET-based searches and the model will not be observed. Such long decay chains were 
shown to occur with a reasonable frequency in our earlier work \cite{Conley:2010du} and can be a contributor to models failing to pass the analyses requirements.

There are, of course, other reasons that prevent models from being discovered. As noted above, subtleties in any sector of the sparticle spectrum can make 
a significant difference as to whether a given model is observed by various analyses. Here, we will discuss a couple of examples where this occurs. As in 
Ref. {\cite{Conley:2010du}}, the approach we follow is to compare a model which fails to be
observed in all search channels to one with a similar spectrum (dubbed a `sister' model) that is detected in at 
least one channel and then examine the difference between them. For this study, we concentrate on the more difficult cases by taking the flat prior model sample and 
assume $\lum=10$ fb$^{-1}$ and $\delta B=20\%$; this leaves only $\sim$ 670 models that are not observed. 
In order to avoid the statistical issues associated with the tails of the 
\MEFF\ distribution discussed above, and to further reduce this model set to a more manageable size, we will only 
consider models whose optimized \MEFF\ value is $\leq 800$ 
GeV. We note that a large fraction of models in this set have relatively heavy LSPs with masses in excess of 400 GeV. 
We now discuss two brief examples of these comparisons.

\begin{figure}
\centerline{
\includegraphics[width=8.0cm]{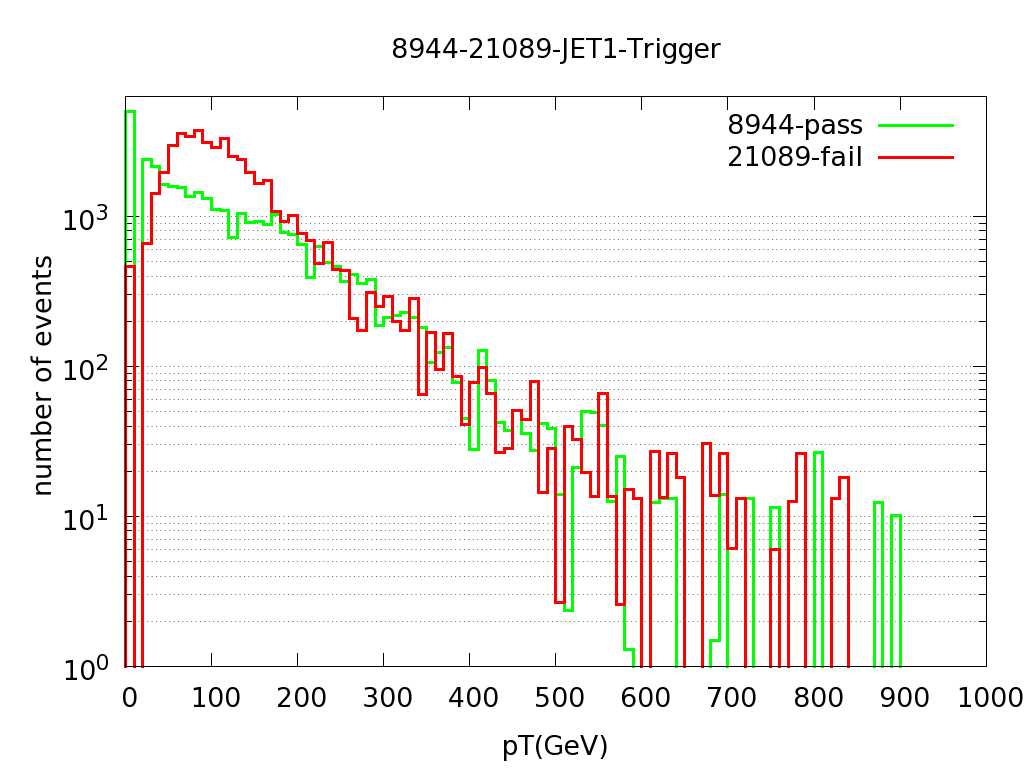}
\hspace*{0.4cm}
\includegraphics[width=8.0cm]{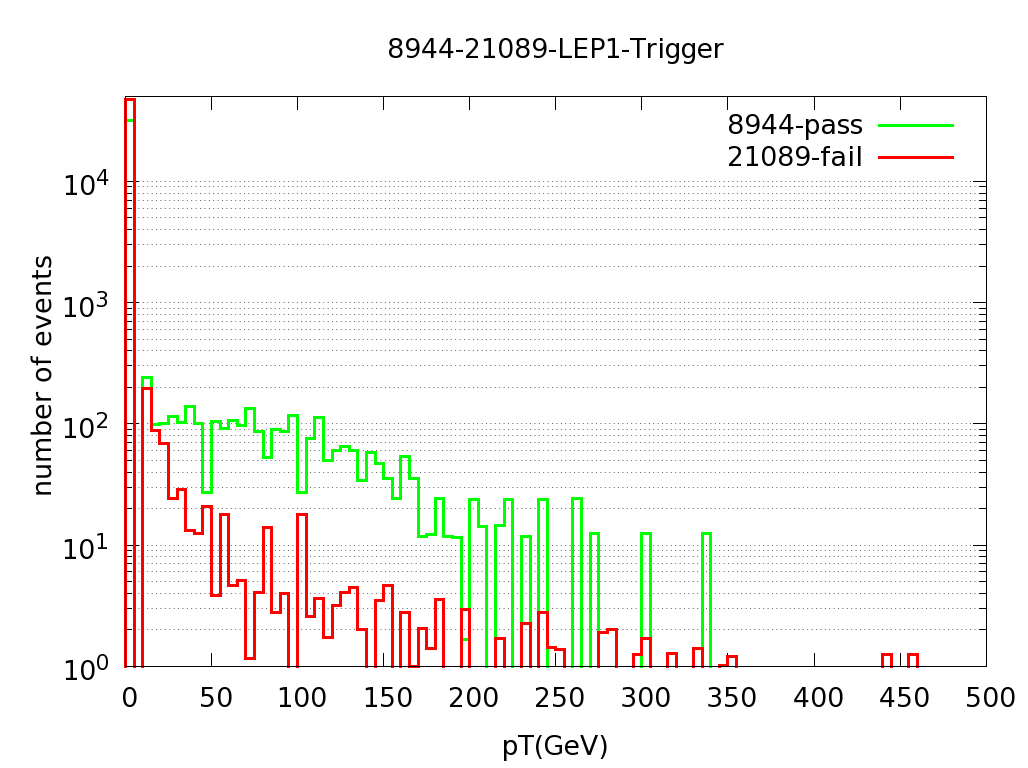}}
%\vspace*{-1.0cm}
\centerline{
\includegraphics[width=8.0cm]{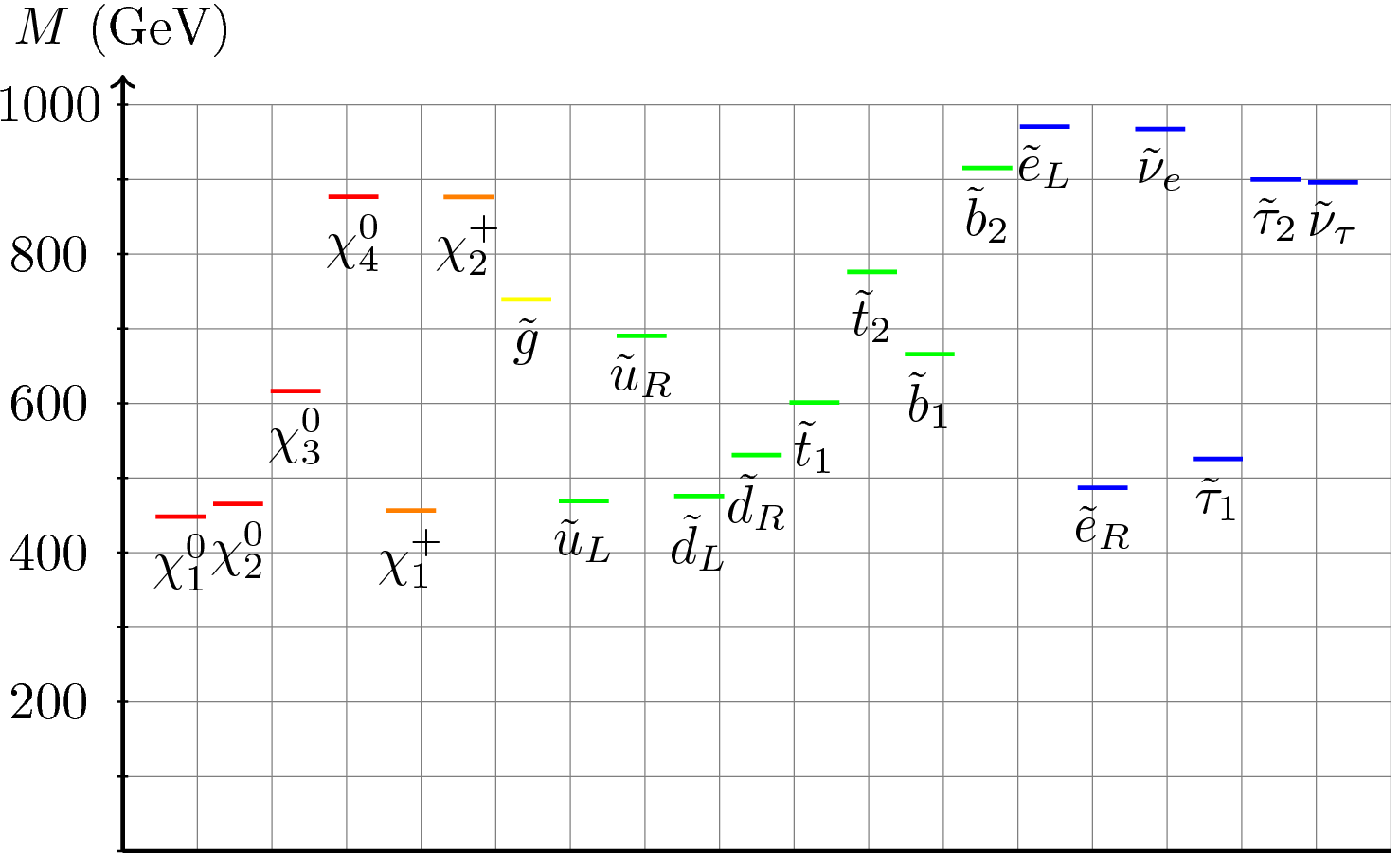}
\hspace*{0.4cm}
\includegraphics[width=8.0cm]{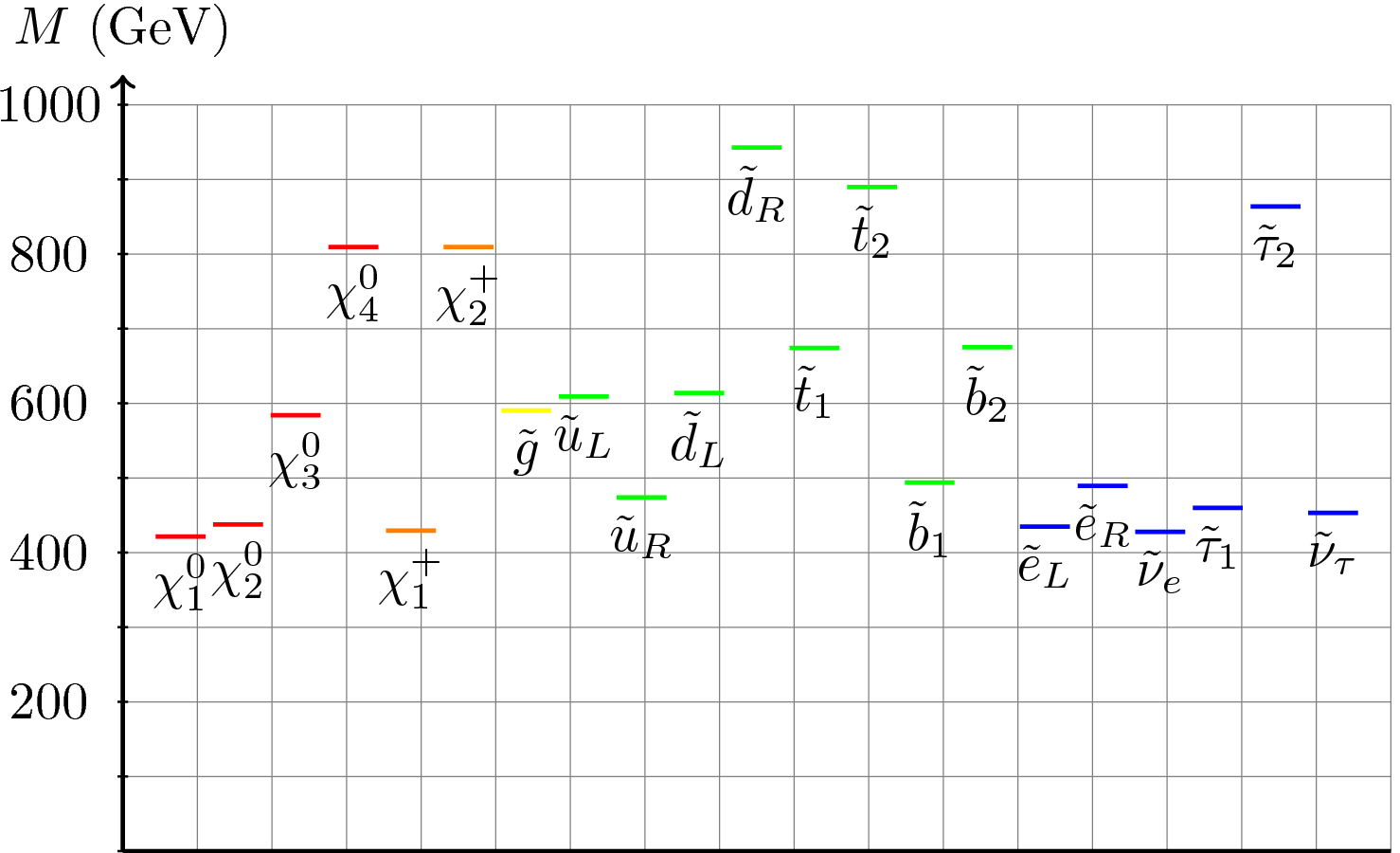}}
\vspace*{0.9cm}
\caption{The top panels compare the leading jet (left) and leading lepton(right) $p_T$ spectra for models 8944 and 21089. The bottom panels show the sparticle spectra for 
these two models.}
\label{case1}
\end{figure}

Figure~\ref{case1} shows a comparison of two similar models, 8944 (observed in the 3,4jOSDL channels) 
and 21089 (missed by all analyses), that have comparable total colored sparticle 
production rates (3.4 and 4.6 pb, respectively).  Both of these models are not observed in the 
nj0l searches since the lighter squarks are too close in mass to the LSP to produce hard 
jets. The gaugino sectors of these two models are quite similar (with the LSP and $\tilde \chi_2^0$ being Higgsino-like and $\tilde \chi_3^0$ being 
bino-like), while their colored sparticle spectra are somewhat different.  In either model the decay of $\tilde \chi_3^0$ allows for the OSDL production through an 
intermediate slepton which has sufficient $p_T$ to pass the analysis requirements. However, while 8944 has a $\tilde u_R$ with mass below that of the gluino
(which is light enough to give a reasonable cross section), allowing 
for the decay into $\tilde \chi_3^0$, only $\tilde d_R$ is (sufficiently) heavier than  
the $\tilde \chi_3^0$ 
in model 21089. In this case, $\tilde d_R$ is also much more massive than the gluino, through which it will dominantly decay, and so it will 
not have a large enough 
branching fraction into $\tilde \chi_3^0$ to produce the OSDL signature.

\begin{figure}
\centerline{
\includegraphics[width=8.0cm]{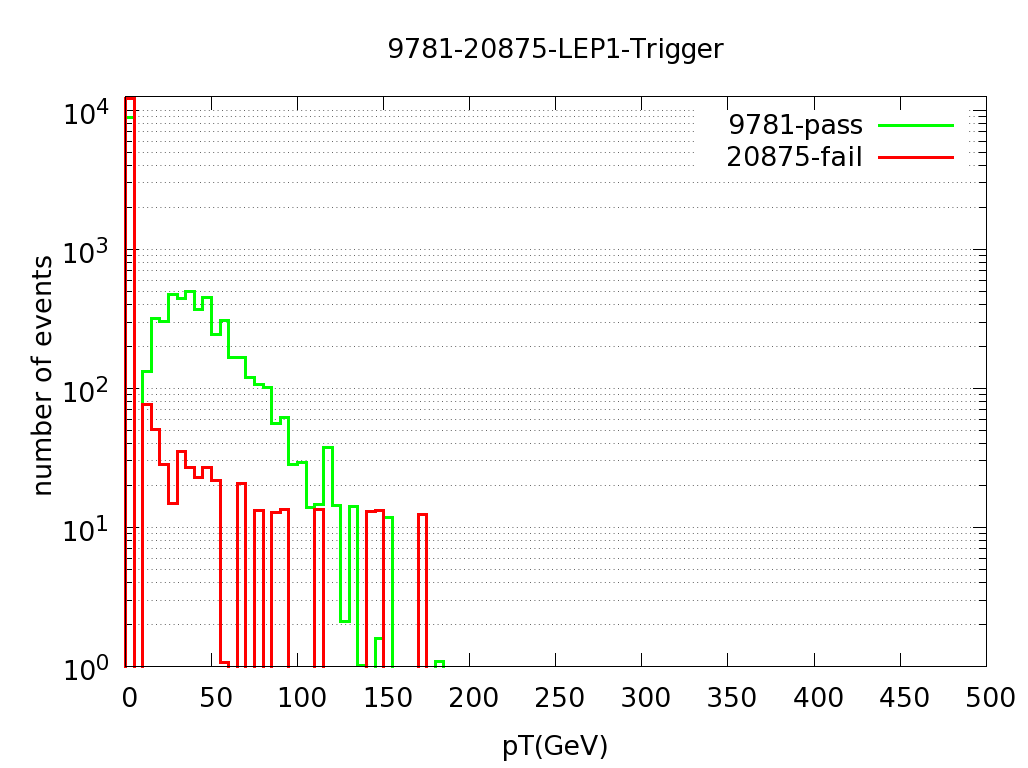}
\hspace*{0.4cm}
\includegraphics[width=8.0cm]{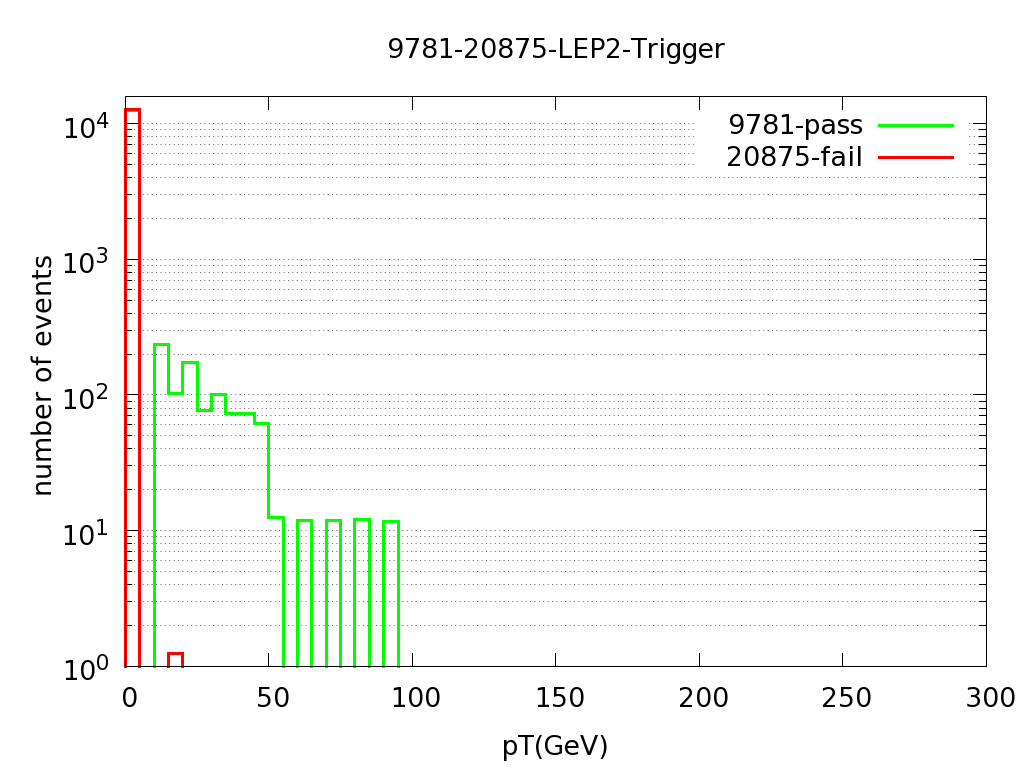}}
%\vspace*{-1.0cm}
\centerline{
\includegraphics[width=8.0cm]{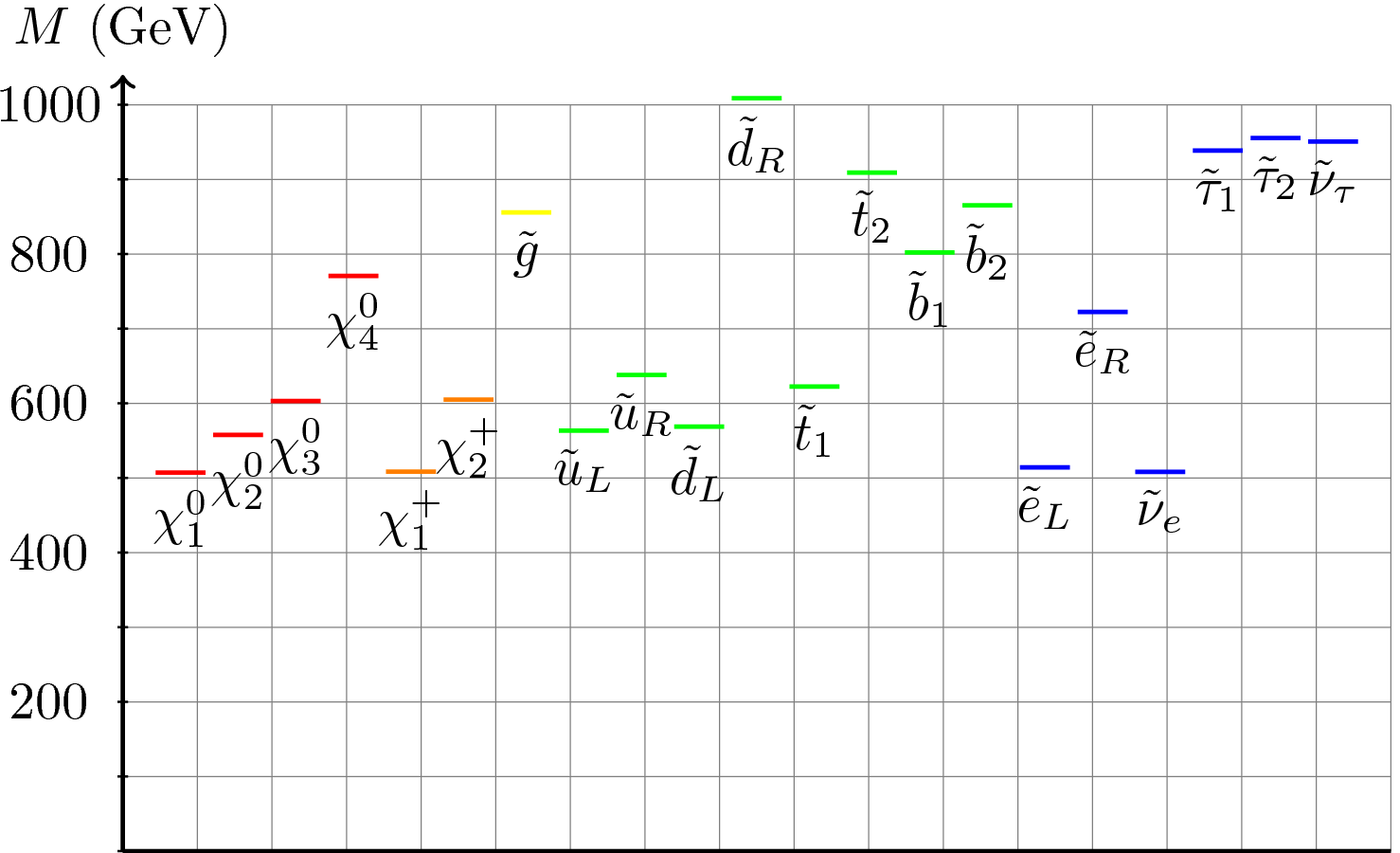}
\hspace*{0.4cm}
\includegraphics[width=8.0cm]{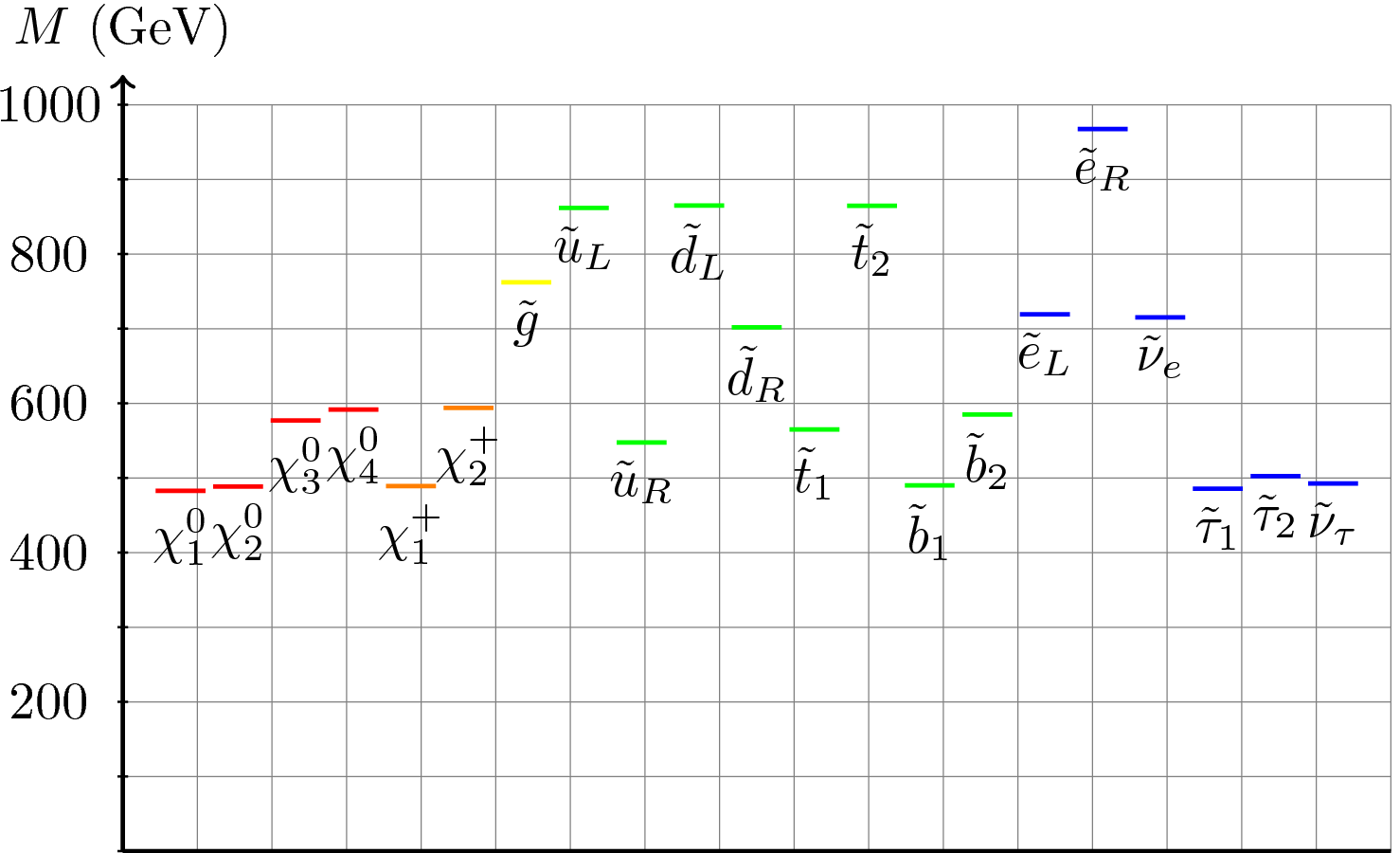}}
\vspace*{0.9cm}
    \caption{The top panels compare the leading(left) and secondary(right) lepton $p_T$ spectra for models 9781 and 20875. 
The bottom panels show the sparticle spectra for 
these two models.}
\label{case2}
\end{figure}

Figure~\ref{case2} compares models 9781 (discovered in the 2jSSDL channel) and 20875 (completely missed) that have total colored sparticle production cross sections of 
1.3 and 1.1 pb, respectively, but yet produce too few hard jet plus \MET\ events to be found in the nj0l channels due to spectrum compression. Model 
9781 is quite interesting as the charginos and lightest three neutralinos are highly mixed combinations of winos and Higgsinos. In this model, 
$\tilde u_R$ (which 
is relatively light) decays to $j+\tilde \chi_2^0$ with a $\sim 98\%$ branching fraction. 
Since $\tilde \chi_2^0$ has a large bino content, it decays 
$\sim 95\%$ of the time 
through sleptons which subsequently decay directly to the LSP with a branching fraction of $\sim 45\%$. Thus model 9781 can easily populate the leptonic final state and 
since the neutralinos are Majorana fermions, the 2jSSDL final state becomes accessible, a channel with an 
extremely small background. On 
the other hand, model 20875 does not allow for the generation of a typical leptonic signal. This
is because the 
$\tilde e,\tilde \mu$ are quite heavy so that the neutralinos only allow for decay to $\tau$
leptons via an intermediate on-shell $\tilde \tau$. This is 
correlated with the 
relative lightness of the $\tilde \tau$ as well as the Higgsino-like LSP. In addition, 
the lightest squarks decay primarily directly to the LSP which will not 
produce any high-$p_T$ leptons. 
    
These two examples demonstrate that the full sparticle spectrum may conspire to render a
model undiscoverable in the \MET-based analysis suite, even if the colored sparticle
cross section is large.  Discovery is not based on
the value of the squark and/or gluino masses alone and blanket limits that claim 
$m_{\tilde q,\tilde g}$ are ruled out below some value cannot be set.

%\clearpage

\subsection{Detector Stable Sparticles in Cascades}

\begin{figure}
\centerline{
\includegraphics[width=10.5cm,angle=90]{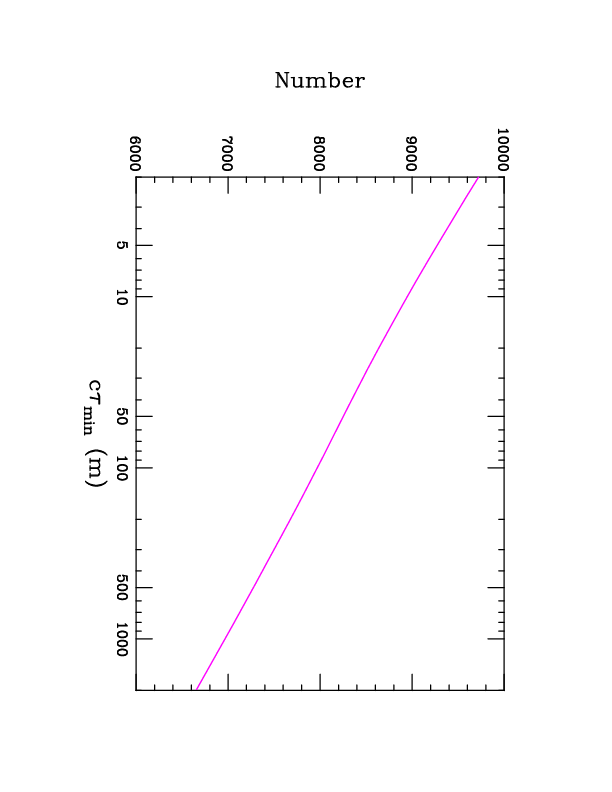}}
\vspace*{-0.5cm}
\centerline{
\includegraphics[width=10.5cm]{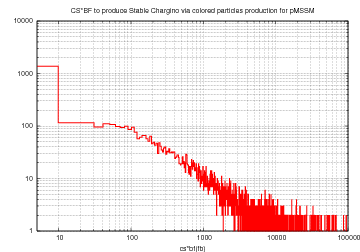}}
\vspace*{0.9cm}
    \caption{(Top) Number of models from both the log and flat prior sets combined having a charged sparticle with an unboosted decay length above a given value.
     (Bottom) Distribution of the estimated cross section times branching fraction for the production of detector-stable charginos in cascade decays in the 
     flat prior model set.}
   \label{stables}
\end{figure}

As we mentioned above, one of the reasons that some pMSSM models may not be observed in the ATLAS \MET\ analyses is that squark and gluino cascade 
decays can sometimes lead to a final state with low \MET.  In many cases this is due to the 
existence of long-lived sparticles, usually charginos, which appear with sizable 
branching fractions in such cascade chains and are essentially detector stable. Particles that  decay outside the detector when produced at the LHC would 
provide a dramatic signal of new physics (see~\cite{Fairbairn:2006gg} and references therein). 
In fact, data from the LHC are already extending the mass limits on such detector-stable 
sparticles \cite{Khachatryan:2011ts}. Since the inclusive \MET\ analyses discussed above do not consider such sparticles, we will briefly sketch the $7$ TeV 
discovery prospects for detector-stable sparticles in our model sets. As discussed in our earlier work \cite{Conley:2010du}, the existence of such long-lived 
states is relatively common in our pMSSM model sample as can be seen in 
Fig~\ref{stables}.\footnote {Recall that we will define a particle to be detector-stable if 
its unboosted decay length in at least 20m. Note that typical values of $\gamma \beta$ for long-lived particles resulting from cascade decays are in the 2-3 range. 
Further note that the dependence of the number of detector-stable sparticles of various species in this model set on the value of the decay width 
$\Gamma_{\text{stable}}$ that is assumed is discussed in detail in~\cite{Conley:2010du}.}

\begin{table}
\begin{center}
\begin{tabular}{|l|c|c|c|} \hline\hline
Sparticle & LHC Reach $100$ pb$^{-1}$ GeV & LHC Reach $1$ fb$^{-1}$
& LHC Reach $10$ fb$^{-1}$ \\
\hline \hline
 $\tilde\chi^+$ (Wino-like)    & $206$ GeV & $264$ GeV & $334$ GeV \\
 $\tilde\chi^+$ (Higgsino-like)& $153$ GeV & $204$ GeV & $267$ GeV \\
 $\tilde\tau$                  & $79$ GeV  & $109$ GeV & $146$ GeV \\
 $\tilde t$                    & $294$ GeV & $363$ GeV & $441$ GeV \\
 $\tilde g$                    & $563$ GeV & $654$ GeV & $751$ GeV \\
\hline\hline
\end{tabular}
\end{center}
\caption{Approximate $7$ TeV LHC search reaches for detector-stable sparticles
of the given species with $100$pb$^{-1}$, $1$ fb$^{-1}$, and
$10$ fb$^{-1}$ \cite{Raklev:2009mg}. }
\label{reach}
\end{table}

Our estimation of the $7$ TeV LHC mass reach for each long-lived sparticle is shown in Table~\ref{reach} 
for the specified integrated luminosities, assuming direct pair 
production. These results are deduced from 
Figure 1 of~\cite{Raklev:2009mg} by taking the geometric mean of the search reach for a $5$ and $10$ TeV LHC at the specified luminosities and
are interpolated to the luminosities considered here where necessary. These results are somewhat conservative as only 
detector-stable sparticle  production in the hard process is considered;  additional detector-stable sparticles could be produced through cascade decays as discussed 
below. 

The number of detector-stable sparticles of various species in our model sample is shown in Table~\ref{stable-results}. This Table also shows the number of 
detector-stable sparticles which will not be discovered at LHC with $100$ pb$^{-1}$ and $1$ fb$^{-1}$ of integrated luminosity, using the 
approximate mass reaches presented in Table~\ref{reach}.  We 
assume here that the mass reach is roughly generation-independent and that it is the same for, e.g., stops and sbottoms.  This assumption is reasonable except 
where there could be significant $t$-channel production for the first or (to a lesser extent) second generation, for instance in the case of up or down squarks.

\begin{table}
\begin{tabular}{|l|c|c|c|c| } \hline\hline
Sparticle & In Model Set & LHC Reach $100$ pb$^{-1}$ & LHC Reach $1$ fb$^{-1}$
& LHC Reach $10$ fb$^{-1}$ \\ \hline \hline
 $\tilde\chi_1^+$   & $8642$ & $8623$ & $3471$ & $1024$ \\
 $\tilde\tau_1$     & $179$  & $179$  & $174$  & $129$ \\
 $\tilde t_1$       & $66$   & $20$   & $9$    & $1$ \\
 $\tilde c_R$       & $49$   & $10$   & $4$    & $1$ \\
 $\tilde\mu_R$      & $17$   & $17$   & $17$   & $11$ \\
 $\tilde b_1$       & $11$   & $0$    & $0$    & $0$\\
 $\tilde c_L$       & $8$    & $0$    & $0$    & $0$ \\
 $\tilde s_R$       & $8$    & $3$    & $0$    & $0$ \\
 $\tilde g$         & $5$    & $0$    & $0$    & $0$ \\
 \hline\hline
\end{tabular}
\caption{The second column from the left gives the number of
  detector-stable sparticles
  of various types in our model set.  The next two columns show the number of
  such sparticles that will not be discovered after
  $100$ pb$^{-1}$, $1$ fb$^{-1}$, and $10$ fb$^{-1}$ at the $7$ TeV LHC,
  following~\cite{Raklev:2009mg}.}
\label{stable-results}
\end{table}

We now focus on the specific case of long-lived charginos, which are by far the most common long-lived sparticles in our model sets. If the 
production cross section for colored sparticles (times relevant branching fractions into
charginos) are sufficiently large, these stable charginos should be found in searches for (effectively) 
stable charged particles occurring at the end of a cascade decay chain. In Fig.~\ref{stables} we display the 
estimated value for $\sigma B$ for the production of detector-stable charginos in cascade decays in our flat prior set. 
(Note that this does 
not include the direct contribution arising from direct chargino pair production.) 
Here, we see that roughly $\sim 84\%$ of models with detector-stable charginos lead to $\sigma B$ 
values in excess of 10 fb at 7 TeV and so we expect them to be observable in the upcoming run of the LHC. 
In this estimation, we assumed the largest contribution to the production 
cross section arises from the production of gluino and light squark species. Using the decay tables generated for each model with detector-stable 
charginos, we calculated the branching fraction for the gluino and light squarks to produce a stable chargino 
at the end of each possible decay chain and then weighted 
them by their corresponding production cross-sections. Note that mass information for neither the mother particles or the daughter chargino is used to 
indicate how likely it is that
the chargino will pass the trigger criteria for detection. As we used PYTHIA to compute the LO cross-section for the light squarks, we do not separately generate 
the production for the various light squark species. Thus in order to obtain our estimate, we make the assumption that the overall cross 
section is 100\% dominated 
by that arising from the lightest squark.  This assumption will break down when the squarks are nearly degenerate, especially when their 
branching fractions to charginos 
are vastly different due to the complexities in the gaugino sector. Many of our models have a large production cross-section for colored sparticles, but the relevant 
branching fractions to charginos can be simultaneously quite small. This can result in very small overall production rates for stable charginos well below $\sim 10$ fb.

\subsection{SUSY Mass Scale From \MEFF}

\begin{figure}
    \includegraphics[width=0.47\columnwidth]{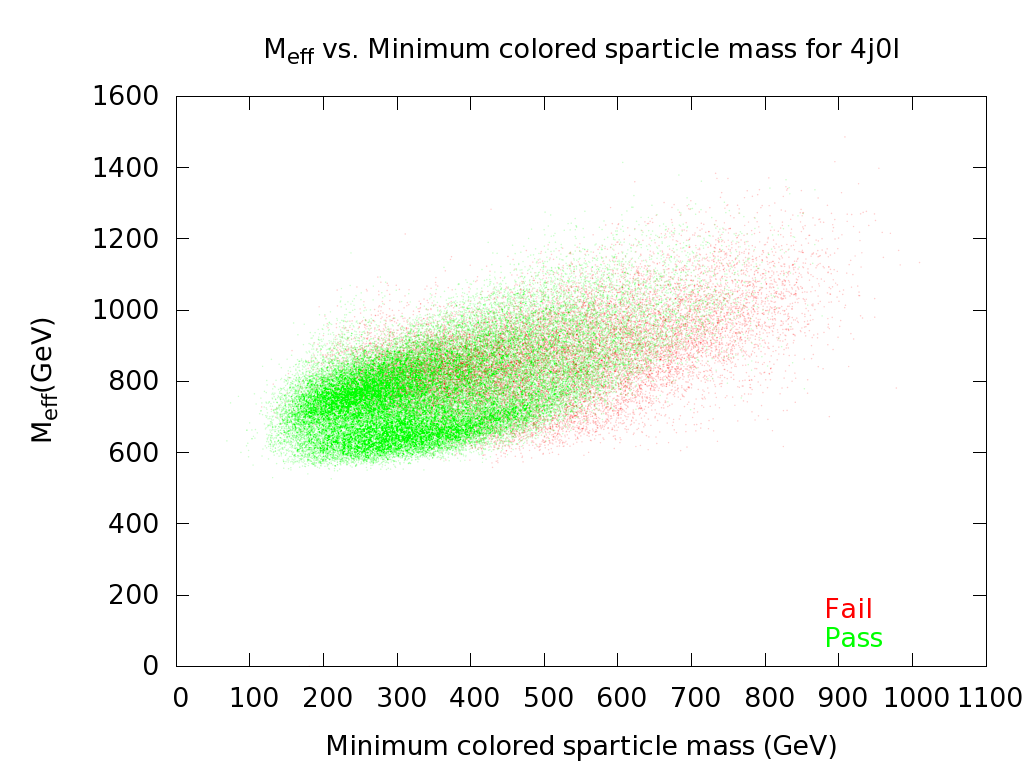} 
    \includegraphics[width=0.47\columnwidth]{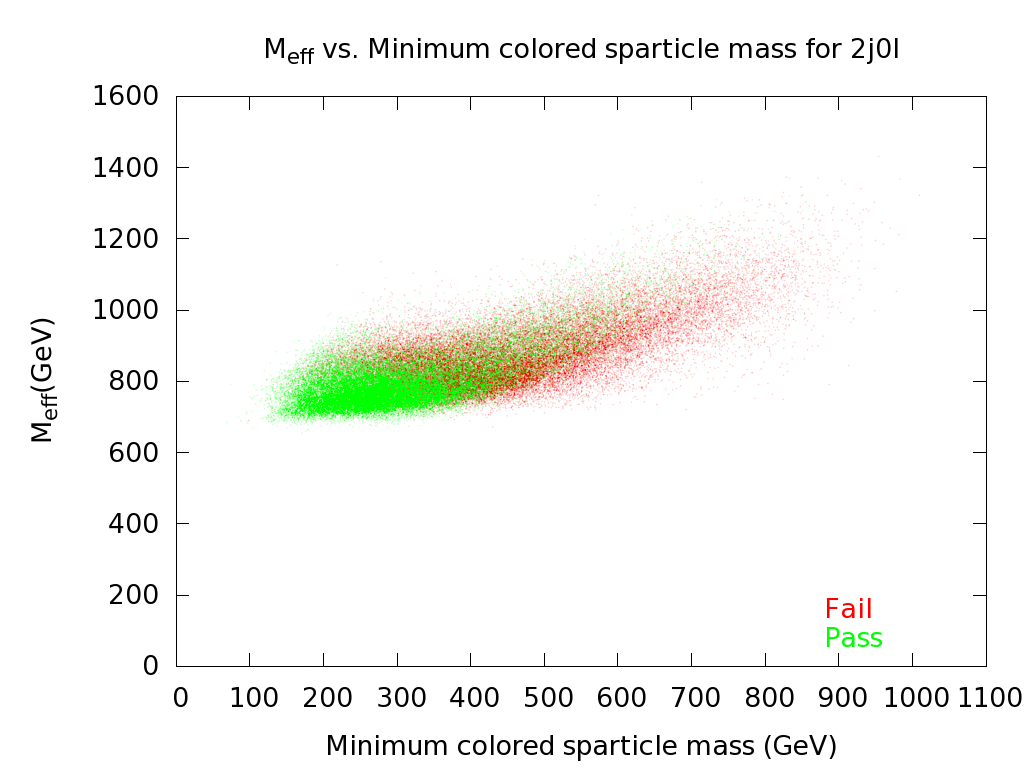}
    \caption{Correlation between the value of \MEFF\ and the mass of the lightest colored sparticle for the 7 TeV 4j0l(left) and 2j0l(right) ATLAS \MET\ channels.  
     The red(green) points correspond to flat prior models which are missed(found) in these two search analyses.}
    \label{relationship}
\end{figure}

In our earlier work on the 14 TeV ATLAS \MET\ analyses, we demonstrated that the relationship between \MEFF\ and the mass of the lightest colored sparticle  
found in mSUGRA, i.e., $\MEFF\simeq 1.5 m_{\text{LCP}}$
(where LCP stands for Lightest Colored Particle), proposed long ago \cite{Hinchliffe:1996iu} does not necessarily hold in the pMSSM. This possible relationship 
is important as it might be used to get the first handle on the overall mass scale of the sparticle spectrum.  Here, we briefly note that this  
result remains valid for the 7 TeV ATLAS \MET\ analyses as can be seen in Fig.~\ref{relationship}. For both the 4j0l and 2j0l channels we see explicitly that 
the values of \MEFF\ lie mostly above the expected value of $1.5m_{\text{LCP}}$, especially in the low sparticle mass region. However, for lightest colored sparticle 
masses in excess of $\simeq 550-600$ GeV we see that, indeed, the relationship $\MEFF\simeq 1.5 m_{\text{LCP}}$ provides a fairly good estimate in both of these 
\MET\ searches. 

\subsection{Modifying ATLAS SUSY Analysis Cuts}

Given the properties of the various sparticles in our model sets, we can try to determine whether the canonical cuts employed in the ATLAS \MET-based search 
analyses can be strengthened to reduce SM backgrounds without any significant loss in the coverage of our pMSSM model space. This is certainly a non-trivial 
issue and the structure of our analysis, being based on the fixed ATLAS \MET\ analyses cuts, is not directly set up to obtain completely definitive answers. However, it 
is possible to make some reasonable estimates based on the information that we do have available. We will concentrate on the three nj0l analyses as they 
generally provide the greatest pMSSM model coverage and have large statistics. The most important kinematic quantities for these searches are the requirements 
on the leading jet $p_T$ and the required amount of \MET. Here, we make use of the average values of the distributions in these quantities for 
our pMSSM model sample, as well as the corresponding fitted width of the part of the distribution below
this average value for pre-selected events. This information then provides us with 
an estimate of where these two kinematic distributions `turn on,' which we take to be the average value minus this
width, on the low energy/momentum side below their peak average values.

\begin{figure}
\centerline{
\includegraphics[width=8.0cm]{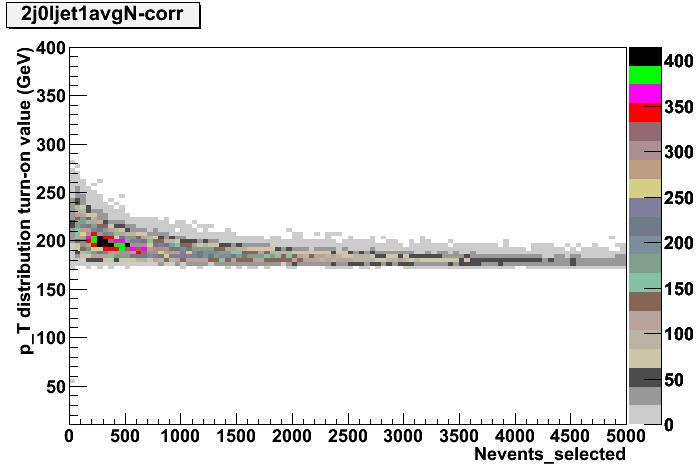}}
%\vspace*{-0.4cm}
\centerline{
\includegraphics[width=8.0cm]{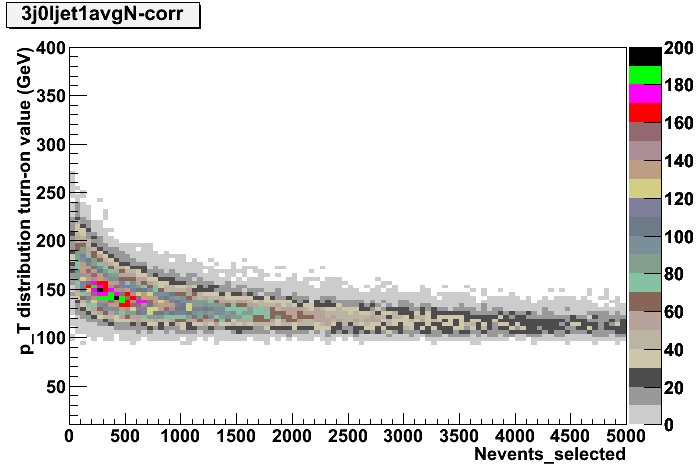}
\hspace*{0.4cm}
\includegraphics[width=8.0cm]{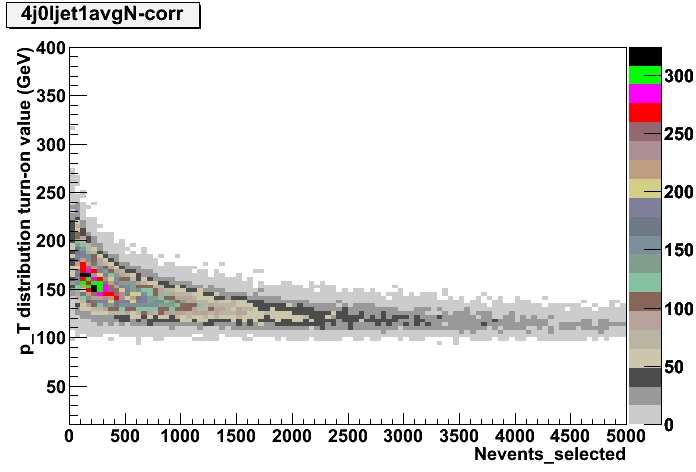}}
\vspace*{0.9cm}
    \caption{Approximate location of the lower edge of the leading jet $p_T$ distribution as a function of the number of preselected events employing the flat 
prior model set for the 2j0l(top), 3j0l(lower left) and 4j0l(lower right) ATLAS search analyses as discussed in the text. The color code reflects the number of models. }
 \label{jet1}
\end{figure}

First consider the cut on the leading jet, $p_{T_1}$, for the 2(3,4)j0l analyses; ATLAS chooses the value for this cut to be 180(100,100) GeV, respectively. 
Fig.~\ref{jet1} shows the distribution of the `turn-on' $p_T$ values for these three ATLAS analyses obtained from analyzing the flat prior model set. For the 
2j0l analysis, we see that the lower edge of the `turn-on' values lies somewhat below the cut value of 180 GeV from which we can conclude that this cut is already 
reasonably hard and cannot be increased without a loss of model coverage. However, for both the 3j0l and 4j0l analyses we instead observe that the `turn-on' values 
lie above those of the ATLAS cuts by $\sim 20$ GeV suggesting that that the $p_{T_1}$ cut in these two channels may be raised without impacting coverage rates for 
the pMSSM. 

\begin{figure}
\centerline{
\includegraphics[width=8.0cm]{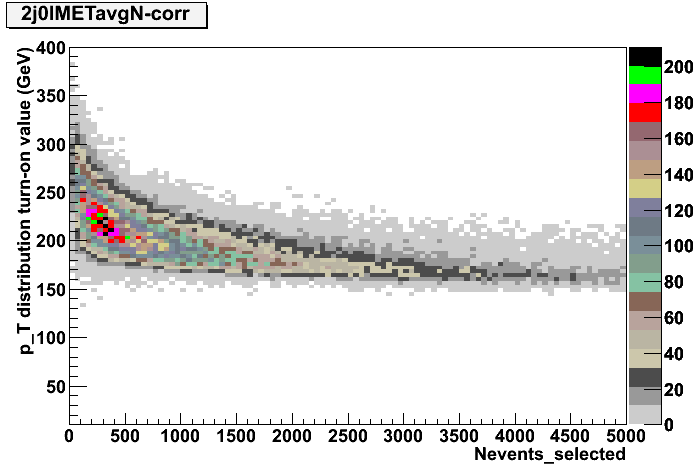}}
%\vspace*{-0.4cm}
\centerline{
\includegraphics[width=8.0cm]{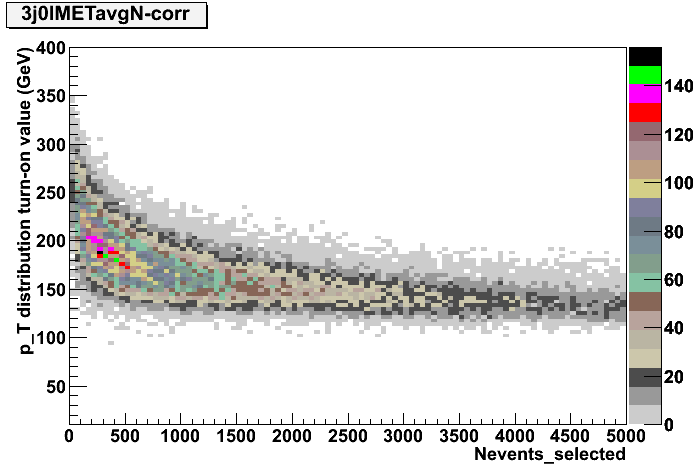}
\hspace*{0.4cm}
\includegraphics[width=8.0cm]{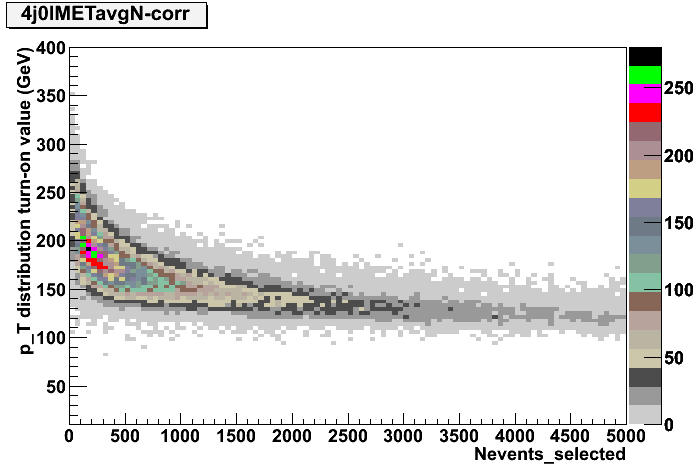}}
\vspace*{0.9cm}
    \caption{Same as in the previous figure but now for the \MET\ distribution.}
 \label{MET}
\end{figure}

Similarly, the nominal \MET\ cut imposed by ATLAS for the nj0l analyses is 80 GeV. However, there is an additional subsequent cut imposed by ATLAS based on the 
value of \MEFF, i.e., $\MET \geq f\,\MEFF$\, where $f=0.30(0.25,0.20)$ for the $n=2(3,4)$ analyses. Fig.~\ref{MET} shows the distributions for the `turn-on' 
values of \MET\ for these three channels employing the flat prior model set. Here we see that the lower edge of this distribution occurs at $\sim 160(130,120)$ GeV 
for $n=2(3,4)$, respectively. This is suggestive that the nominal \MET\ cut made by ATLAS may be increased for these three analyses without losing significant 
pMSSM model coverage.

\section{Implications of the 7 TeV Run}

In this section, we explore some implications of a null search for Supersymmetry at the 7 TeV LHC.  We examine the degree of fine-tuning that would
be placed on our pMSSM model sample and we discuss the resulting expectations for sparticle production at a 500 GeV and 1 TeV Linear Collider.

\subsection{Fine-tuning in the Undiscovered pMSSM Models}

As has recently been discussed in the mSUGRA/CMSSM context \cite{Strumia:2011dv,Cassel:2011tg}, it is apparent that if SUSY signatures are not discovered at the 
7 TeV LHC as the integrated luminosity accumulates it is likely that the SUSY parameter space must become more fine-tuned, and hence more problematic as a solution 
to the hierarchy problem. Since we know which models in our sample are discoverable (or not) by the ATLAS \MET\ search analyses, 
we can ask whether this same result also holds in the 
case of our pMSSM model sets.\footnote{Note that the amount of fine-tuning in both our model sets was examined in some detail in our earlier 
work \cite{Berger:2008cq}} 

\begin{figure}
\centerline{
\includegraphics[width=7.5cm]{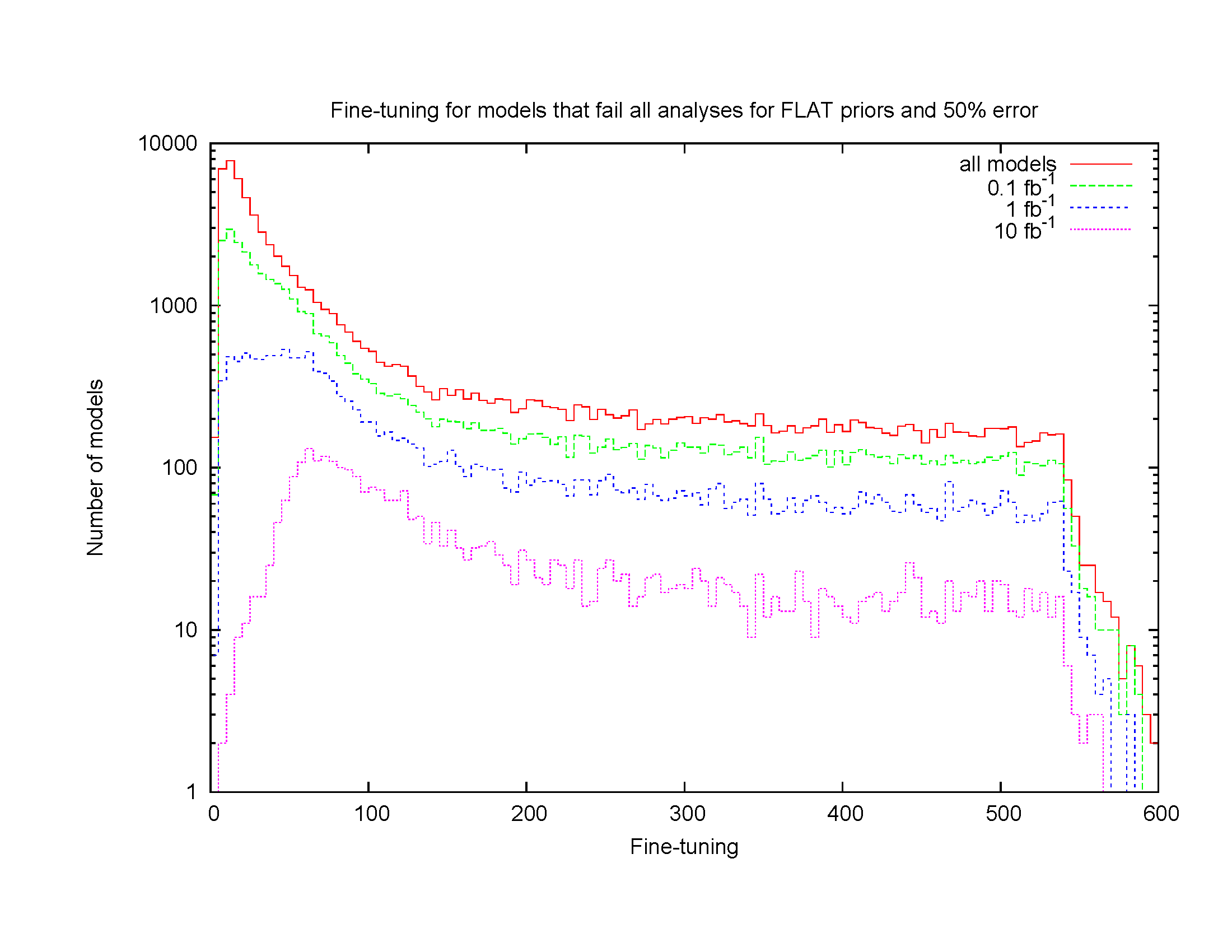}
\hspace*{-0.5cm}
\includegraphics[width=7.5cm]{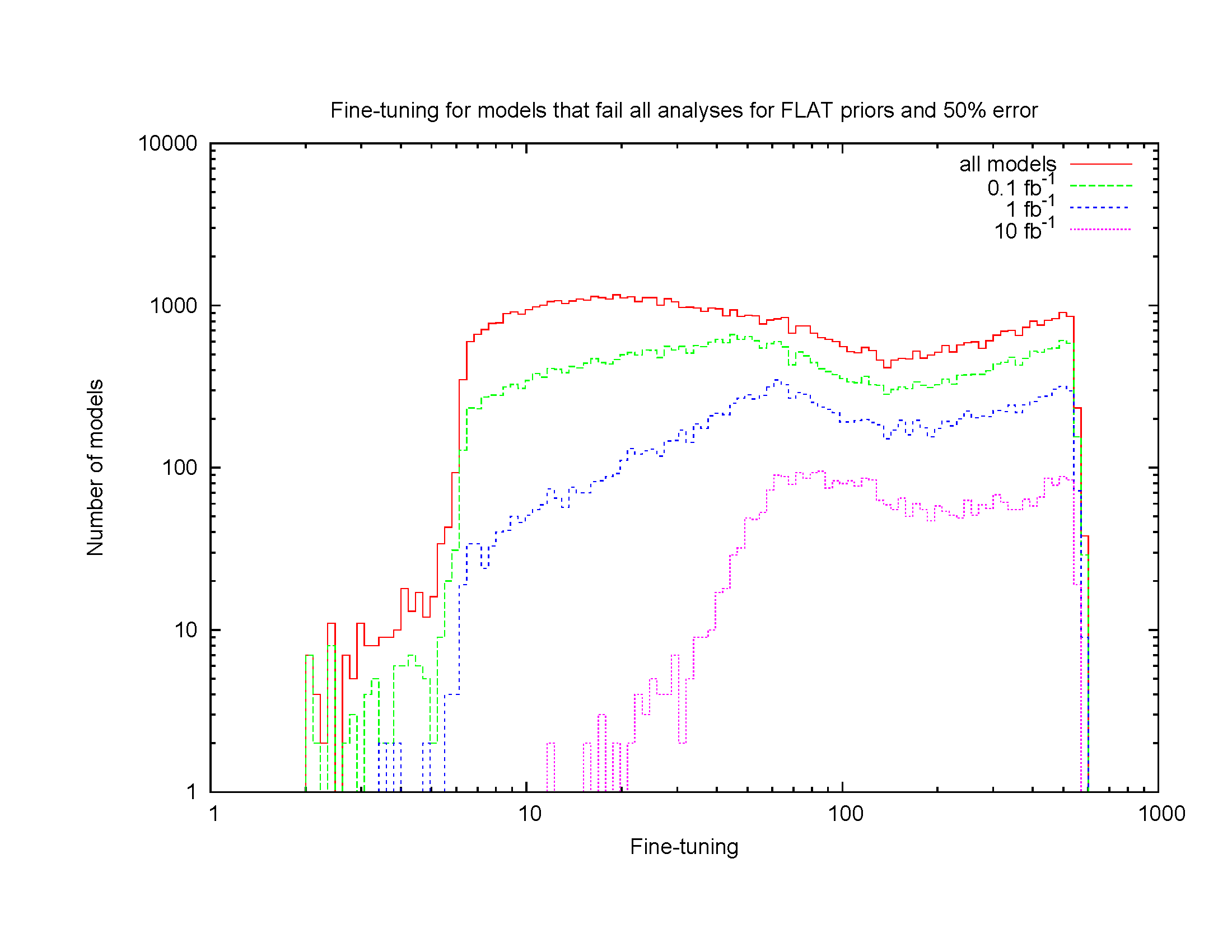}}
\vspace*{-0.3cm}
\centerline{
\includegraphics[width=7.5cm]{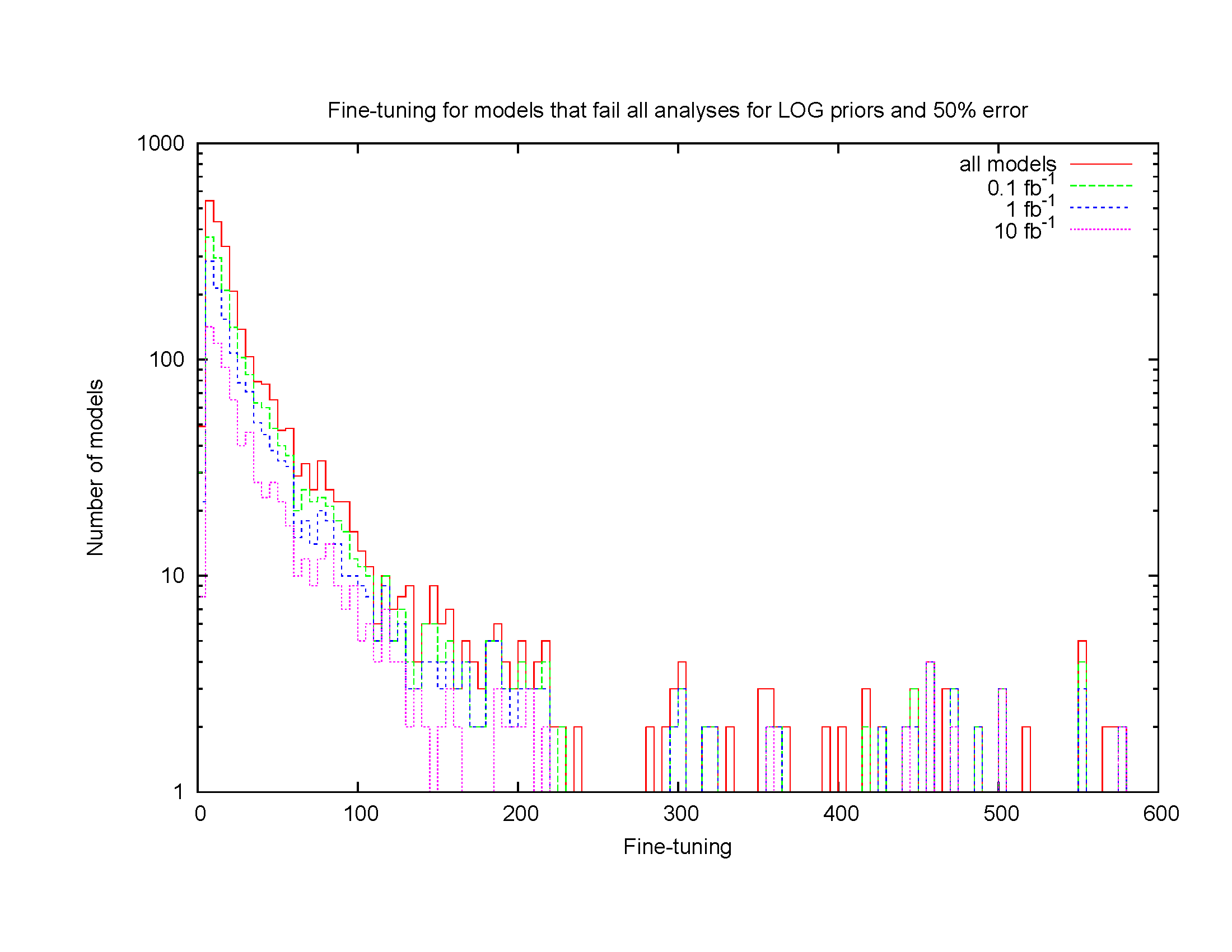}
\hspace*{-0.5cm}
\includegraphics[width=7.5cm]{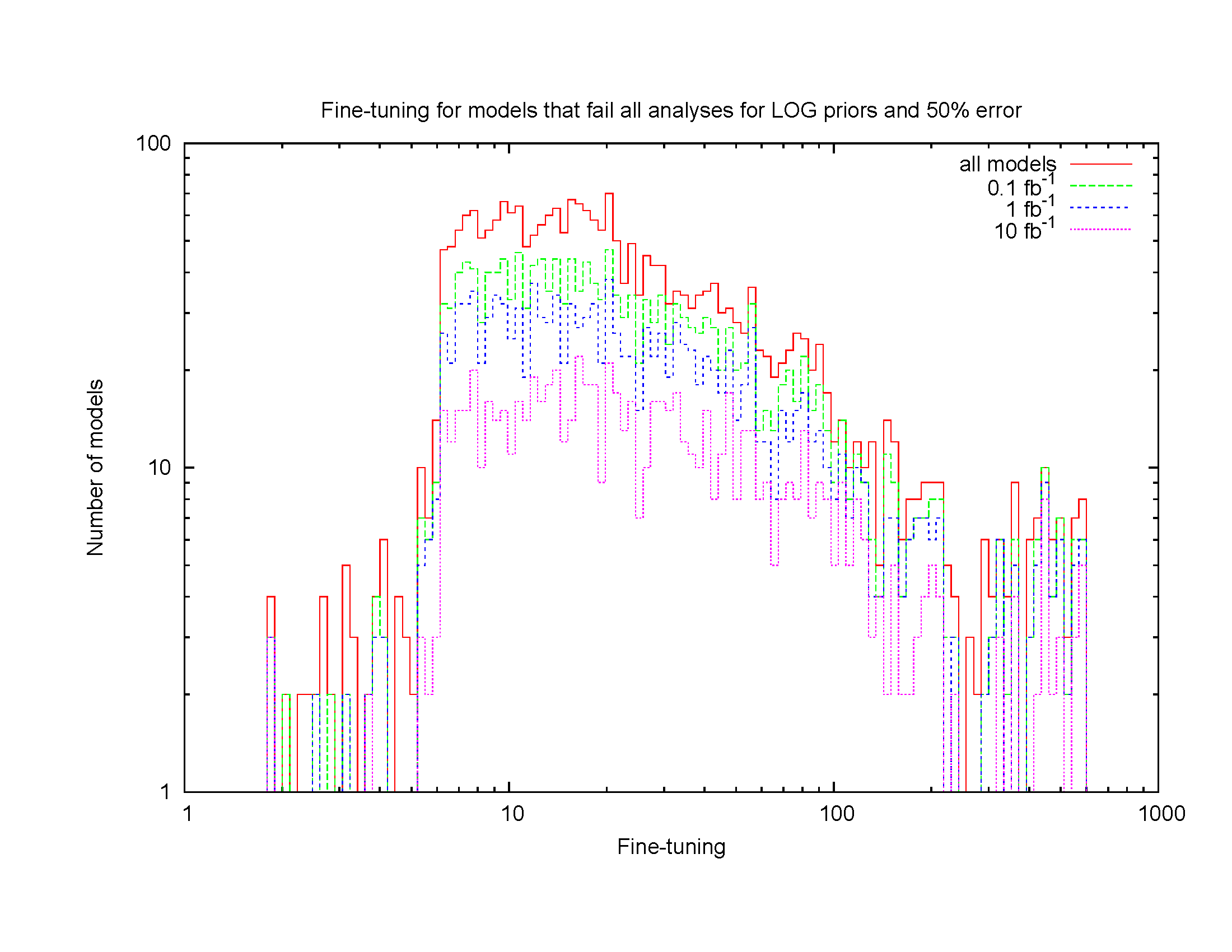}}
%\vspace*{0.1cm}
%    \includegraphics[width=0.70\columnwidth]{figs/ft-compare-FLAT-50.png} \\
%    \includegraphics[width=0.70\columnwidth]{figs/ft-compare-FLAT-50-log.png}
    \caption{Two projections of the fine-tuning distributions for models in our flat (top panels) and 
log (bottom panels) prior sets. The top red histogram in all panels 
     shows the result for the full model set while the subsequently lower histograms correspond to models not observed by the ATLAS \MET\ search analyses for various 
     values of the integrated luminosity as indicated, assuming a background uncertainty of 50\%.}
    \label{fineflat}
\end{figure}

Figure~\ref{fineflat} display the results of this analysis assuming a background systematic error of 50\% for both the flat and 
log prior model samples. Clearly, in the flat prior case, one sees that as the integrated luminosity increases and more models can be discovered by 
ATLAS, those remaining yet {\it undiscovered} tend to be more fine-tuned as expected.  In other words, the fractional loss of models from the full distribution 
occurs more rapidly with increasing luminosity 
for models with smaller amounts of fine-tuning. This is not too surprising as, overall, models with less tuning tend to have lighter 
SUSY sparticle spectra and are thus more easily discovered at the LHC. On the other hand, the results from the log prior model set appear to be affected 
somewhat differently in 
that the overall shape of the fine-tuning distribution does not appear to change very much by removing models that should have been already been discovered by 
ATLAS as the luminosity increases. In this case, we see that there is not much of an increase in the amount of fine-tuning as the set of undiscovered models shrinks. 
This represents one of the few apparent differences between these two different model sets. This can be explained by the fact that while the log prior model set tend to 
have light sparticle spectra (though they extend out to larger mass values than do those for the flat prior models) and are thus less fine-tuned to begin with, these same 
spectra are generally compressed making these models more difficult to discover at the LHC as was discussed above. This would imply that the models missed by the 
ATLAS \MET\ analyses in the log prior case are generally {\it not} much more fine-tuned than those appearing in the originally generated model set. Thus we find that 
the amount of fine-tuning that remains in the LHC-undiscovered pMSSM model sample can depend upon the prior used to generate the original model set.

\clearpage

\subsection{Implications of pMSSM Searches for the Linear Collider}

If there are no clear SUSY signals as the 7 TeV LHC integrates more luminosity, the question arises whether the production (and study) of charged sparticles  
remains viable at the proposed 500 GeV Linear Collider (LC). Based on mSUGRA/CMSSM model coverage projections from both the ATLAS \cite{ATLAS:1278474} and 
CMS \cite{CMSnote1} Collaborations at 7 TeV with a 1 fb$^{-1}$ integrated luminosity (as well as their results from the 2010 SUSY searches), it would seem very 
unlikely that either light sleptons or gauginos (other 
than perhaps the LSP itself) will remain kinematically accessible at a 500 GeV LC if nothing is observed.  Here, we address the question whether 
this expectation also remains true for 
our pMSSM model sets. To this end we examine the set of flat and log prior models which are not detected by {\it any} of the ATLAS \MET\ search analyses for 
assumed values of both the integrated luminosity and background systematic error and then determine the part of the sparticle spectrum within these models which is 
kinematically accessible at a 500 GeV LC.\footnote{Note that we have not performed any analysis here to ascertain whether or not a given kinematically 
accessible sparticle is actually {\it observable} at such a LC. See, however, the work in Refs.~\cite{Berger:2007ut,Berger:2007yu}.}  A similar analysis can also be 
performed for a 1 TeV LC. We remind the reader that, e.g., in the flat prior case we have no sparticles heavier than $\sim 1$ TeV.

We present the results of this analysis in various ways. Tables~\ref{lc1} and~\ref{lc2} show the number of sparticles of various species that are 
kinematically accessible at a 500 GeV or 1 TeV LC within the subset of ATLAS-undetected models assuming 
$\lum=1(10)$\; fb$^{-1}$ with $\delta B=50(20)\%$, respectively. These 
two cases represent both a possibly conservative and a more optimistic performance for the LHC over the 2011-12 running period. Here we see several things: ($i$) In 
addition to the gauginos and sleptons, sparticles such as stops, sbottoms and other squarks are potentially almost as likely to also be kinematically accessible 
at the LC \cite{Rizzo:2010xn}. ($ii$) The number of models with kinematically accessible sparticles and their variety is significantly greater in the log prior 
sample as these models are more likely to have a lighter and more compressed sparticle spectrum ($iii$) The difference 
between the two cases presented in these Tables is quite significant; in particular, we see that for the flat prior model set there is a huge depletion in the number 
of unobserved models at higher luminosity and with lower background systematics. There are extremely few flat prior models remaining at high luminosity with any 
accessible sparticles at a 500 GeV LC. ($iv$) Going from a 500 GeV to a 1 TeV LC {\it substantially} increases the number of models with kinematically accessible 
sparticles, especially in the flat prior case.   It is clear that, at least for the flat prior model sample with luminosities 
in excess of 1 fb$^{-1}$ at the 
LHC, that the 500 GeV LC does not seem to be a good place to study our pMSSM models if no signal for SUSY is found at the LHC in 2011-12.

\begin{table}
\centering
\begin{tabular}{|l|r|r|r|r|} \hline\hline
 & \multicolumn{2}{c|}{$\sqrt s=500$ GeV} & \multicolumn{2}{c|}{$\sqrt s=1$ TeV} \\ \hline
Sparticle & Flat & Log &  Flat & Log \\ \hline
$\tilde e_L$ & 107 & 101 & 3052 & 347 \\
$\tilde e_R$ & 260 & 209 & 3938 & 565 \\
$\tilde \tau_1$ & 730 & 381 & 7431 & 869 \\
$\tilde \tau_2$ & 30 & 36 & 1288 & 207 \\
$\tilde\nu_e$ & 151 & 117 & 3168 & 356 \\
$\tilde\nu_\tau$ & 386 & 236 & 4366 & 553 \\
$\tilde\chi_1^0$ & 5487 & 1312 & 14,510 & 1539 \\
$\tilde\chi_2^0$ & 2738 & 1035 & 10,714 & 1395 \\
$\tilde\chi_3^0$ & 429 & 352 & 5667 & 903 \\
$\tilde\chi_4^0$ & 10 & 18 & 1267 & 202 \\
$\tilde\chi_1^\pm$ & 4856 & 1208 & 13,561 & 1495 \\
$\tilde\chi_2^\pm$ & 94 & 54 & 3412 & 456 \\
$\tilde g$ & 0 & 0 & 1088 & 65 \\
$\tilde d_L$ & 35 & 11 & 2459 & 117 \\
$\tilde d_R$ & 220 & 96 & 3630 & 526 \\
$\tilde u_L$ & 52 & 16 & 2545 & 123 \\
$\tilde u_R$ & 124 & 64 & 3581 & 273 \\
$\tilde b_1$ & 289 & 75 & 5553 & 590 \\
$\tilde b_2$ & 1 & 0 & 409 & 21 \\
$\tilde t_1$ & 93 & 9 & 3727 & 217 \\
$\tilde t_2$ & 0 & 0 & 2 & 0 \\
\hline\hline
\end{tabular}
\caption{Number of kinematically accessible sparticles from our set of 14623(1546) flat(log) prior pMSSM models that are unobservable 
by the ATLAS \MET\ searches assuming 
$\lum=1$ fb$^{-1}$ with $\delta B=50\%$ for both a 500 GeV and 1 TeV LC.}
\label{lc1}
\end{table}

\begin{table}
\centering
\begin{tabular}{|l|r|r|r|r|} \hline\hline
 & \multicolumn{2}{c|}{$\sqrt s=500$ GeV} & \multicolumn{2}{c|}{$\sqrt s=1$ TeV} \\ \hline
Sparticle & Flat & Log &  Flat & Log \\ \hline
$\tilde e_L$ &  0 & 37 & 63 & 142 \\
$\tilde e_R$ & 0 & 72 & 53 & 223 \\
$\tilde \tau_1$ & 2 & 142 & 165 & 338 \\
$\tilde \tau_2$ & 0 & 11 & 9 & 69 \\
$\tilde\nu_e$ & 0 & 42 & 64 & 146 \\
$\tilde\nu_\tau$ & 0 & 85 & 81 & 236 \\
$\tilde\chi_1^0$ & 26 & 507 & 587 & 626 \\
$\tilde\chi_2^0$ & 4 & 397 & 352 & 557 \\
$\tilde\chi_3^0$ & 0 & 136 & 57 & 357 \\
$\tilde\chi_4^0$ & 0 & 5 & 5 & 66 \\
$\tilde\chi_1^\pm$ & 25 & 467 & 505 & 608 \\
$\tilde\chi_2^\pm$ & 0 & 17 & 16 & 170 \\
$\tilde g$ & 0 & 0 & 27 & 5 \\
$\tilde d_L$ & 0 & 3 & 73 & 24 \\
$\tilde d_R$ & 1 & 18 & 63 & 157 \\
$\tilde u_L$ & 0 & 5 & 81 & 24 \\
$\tilde u_R$ & 0 & 14 & 86 & 79 \\
$\tilde b_1$ & 0 & 20 & 103 & 189 \\
$\tilde b_2$ & 0 & 0 & 3 & 4 \\
$\tilde t_1$ & 1 & 2 & 94 & 58 \\
$\tilde t_2$ & 0 & 0 & 0 & 0 \\
\hline\hline
\end{tabular}
\caption{Same as the previous Table but now corresponding to the 672(663) undetected flat(log) prior models assuming $\lum=10$ fb$^{-1}$ with $\delta B=20\%$.}
\label{lc2}
\end{table}

\begin{figure}
\centerline{
\includegraphics[width=5.5cm,angle=90]{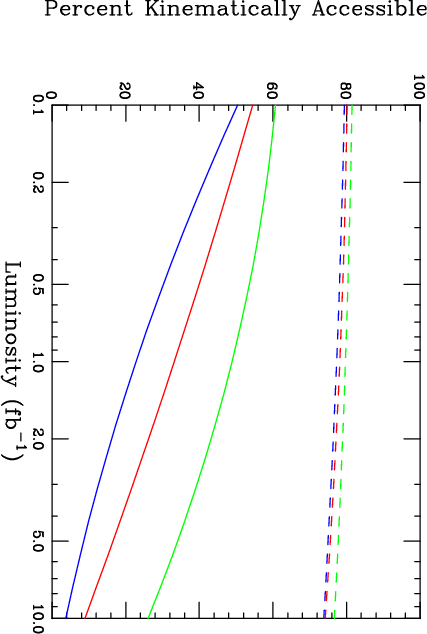}
\hspace*{0.4cm}
\includegraphics[width=5.5cm,angle=90]{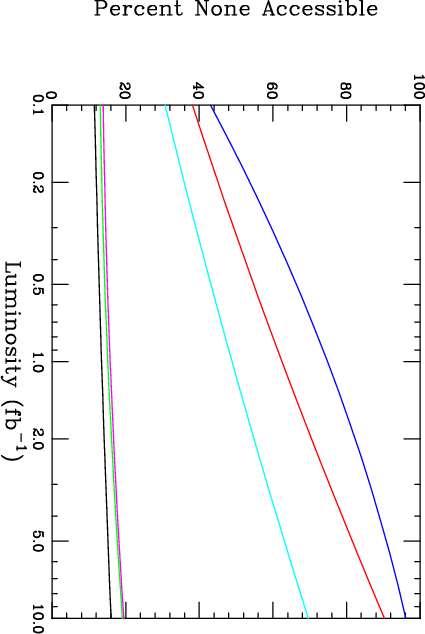}}
\vspace*{0.5cm}
\centerline{
\includegraphics[width=5.5cm,angle=90]{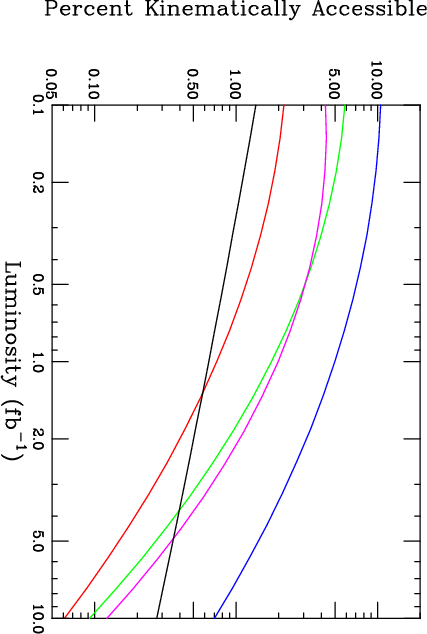}
\hspace*{0.4cm}
\includegraphics[width=5.5cm,angle=90]{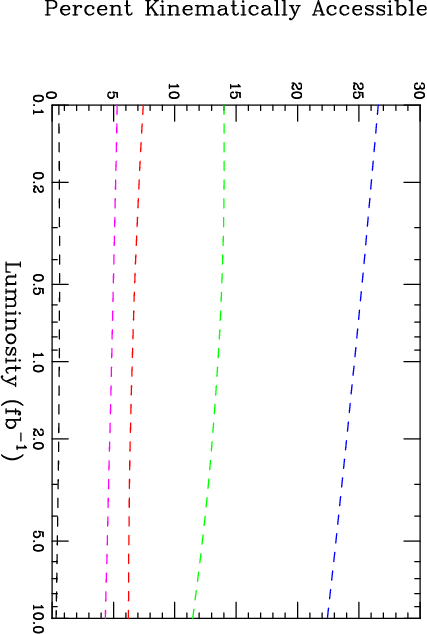}}
\vspace*{0.9cm}

    \caption{Top left: Fractional number of undetected models with kinematically accessible $\tilde \chi_1^\pm$ at a 500 GeV LC as a function of the LHC integrated 
    luminosity for 
    flat(solid) and log(dashed) prior models. The red(green, blue) curves correspond to background systematic uncertainties of 50(100,20)\%, respectively.
    Bottom left: Same as the previous panel but now for flat prior models only with $\delta B=50\%$ for the $\tilde \tau_1$(blue), $\tilde e_R$(green), 
    $\tilde e_L$(red), $\tilde t_1$(solid) and $\tilde b_1$(magenta). Bottom right: Same as the previous panel but now for log prior models. Top right: Fraction 
    of undetected models with {\it no} sparticles kinematically accessible. The magenta(green,blue) curves are for log prior models with $\delta B=20(50,100)\%$ while 
    the blue(red,cyan) curves are the corresponding results for the flat prior set.}
 \label{lcfin1}
\end{figure}

\begin{figure}
\centerline{
\includegraphics[width=5.5cm,angle=90]{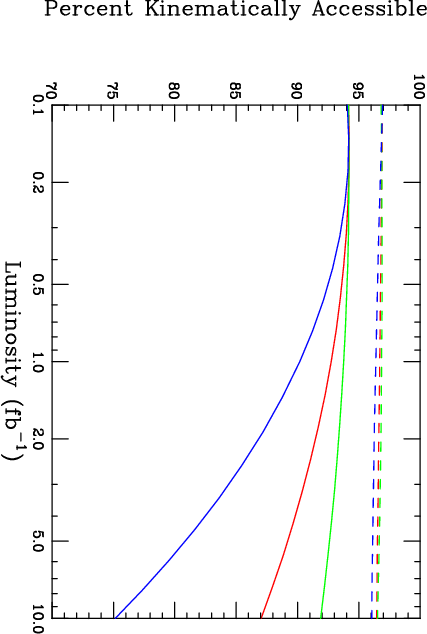}
\hspace*{0.4cm}
\includegraphics[width=5.5cm,angle=90]{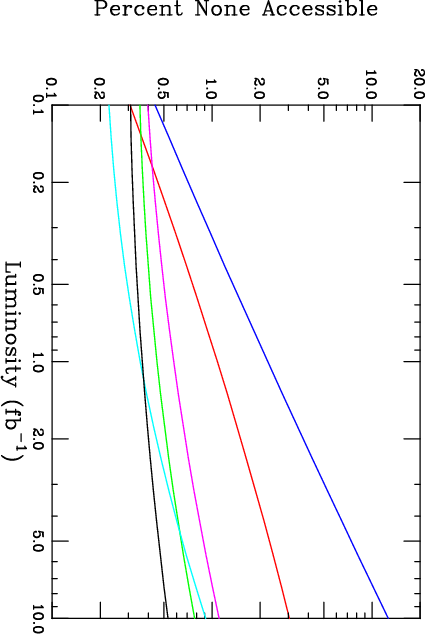}}
\vspace*{0.5cm}
\centerline{
\includegraphics[width=5.5cm,angle=90]{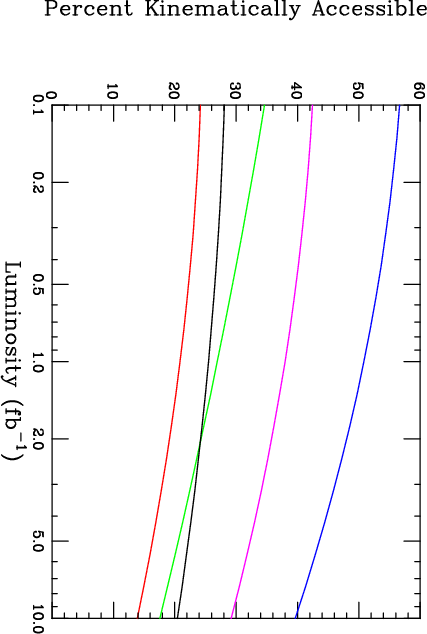}
\hspace*{0.4cm}
\includegraphics[width=5.5cm,angle=90]{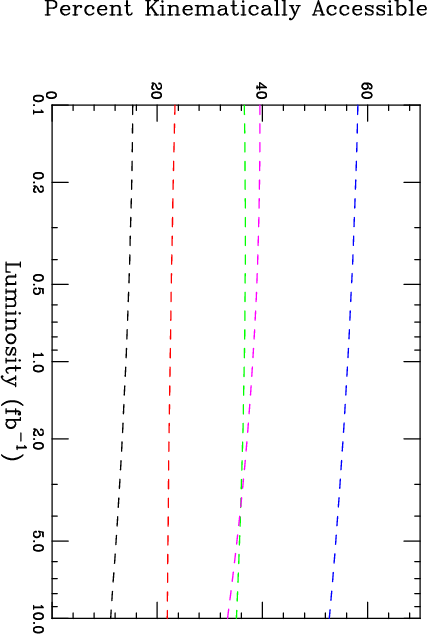}}
\vspace*{0.9cm}
    \caption{Same as in the previous Figure but now for a 1 TeV LC.}
 \label{lcfin2}
\end{figure}

In order to study these LC results in more detail we examine their dependence on the LHC integrated luminosity; this 
is shown for the 500 GeV LC in Fig.~\ref{lcfin1} and for the 1 TeV LC in Fig.~\ref{lcfin2}. Here we display the {\it fraction} of the unobserved set of 
models that have a kinematically accessible sparticle of a particular variety. At a 500 GeV LC, this fraction for charginos, stops, sbottoms, selectrons (or smuons) 
and staus in the flat prior model set is seen to decrease significantly as the LHC accumulates integrated luminosity without observation of  a signal for SUSY.
However, note that for the corresponding log prior model set, while the number of surviving models decreases with any 
corresponding increase in the LHC integrated 
luminosity (or with any decrease in the SM background uncertainty) as one would expect, the {\it fraction} of the surviving models with a kinematically accessible 
sparticle changes very little, if at all! For charginos at a 500 GeV LC this fraction is found to be quite large, $\sim 75\%$, but it is found to be somewhat smaller 
for the other sparticles, e.g., $\sim 25\%$ for $\tilde \tau_1$ and $\sim 7\%$ for $\tilde e_L$.

One possible explanation of this unexpected behavior in the log prior sample is as follows: 
As we saw in the previous discussion of fine-tuning, in the log prior case, the removal of 
pMSSM models from the log prior set (as they are `discovered' by ATLAS) must affect the various sparticle mass distributions in a roughly 
uniform manner.  Otherwise the observed amount of fine-tuning would necessarily increase. However, in the flat prior case, models with lighter sparticles are 
preferentially `discovered' by ATLAS searches. This hypothesis can qualitatively explain why there is no significant reduction in the {\it fraction} of the log prior 
models with kinematically accessible sparticles at a 500 GeV (or 1 TeV) LC (especially in the case of non-colored sparticles as we have seen above).  It also 
simultaneously explains why the apparent amount of fine-tuning does not change appreciably 
as the LHC covers more of the log scan parameter space. Of course, as can be seen from these figures, at a 1 TeV LC a substantial portion of both the log and flat 
prior model sets which remain undiscovered at the LHC have sparticles which are kinematically accessible even at high LHC integrated luminosities. We further note 
that the fraction of LHC-unobserved models where {\it no} SUSY sparticles whatsoever are accessible at a LC (not even the LSP) is quite large for the flat model set at a 
500 GeV LC, but is only in the 12-20\% range for the corresponding log prior model set. On the other hand, at a 1 TeV LC,  this value is seen to 
lie below $\sim 1\%$ for the log prior models, while for the flat prior set it remains below $\sim 13\%$. 
Clearly, a 1 TeV LC will be far better at accessing the sparticles in our pMSSM model sets.

\begin{figure}
\centerline{
\includegraphics[width=8.0cm]{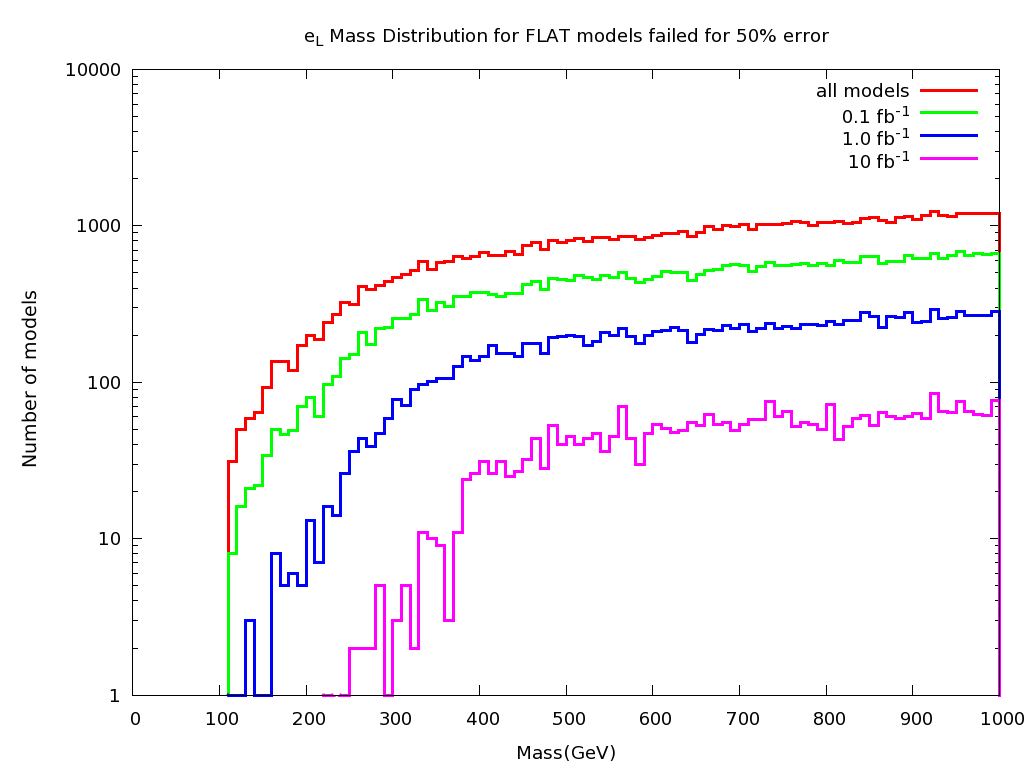}
\hspace*{0.4cm}
\includegraphics[width=8.0cm]{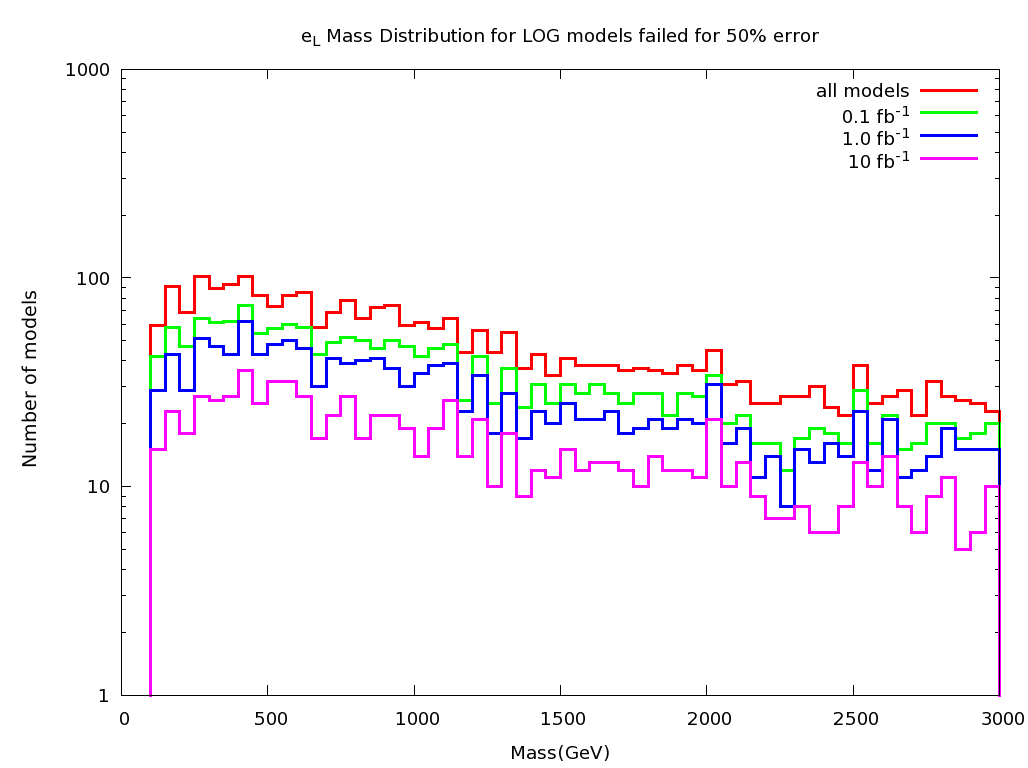}}
\vspace*{1.0cm}
\centerline{
\includegraphics[width=8.0cm]{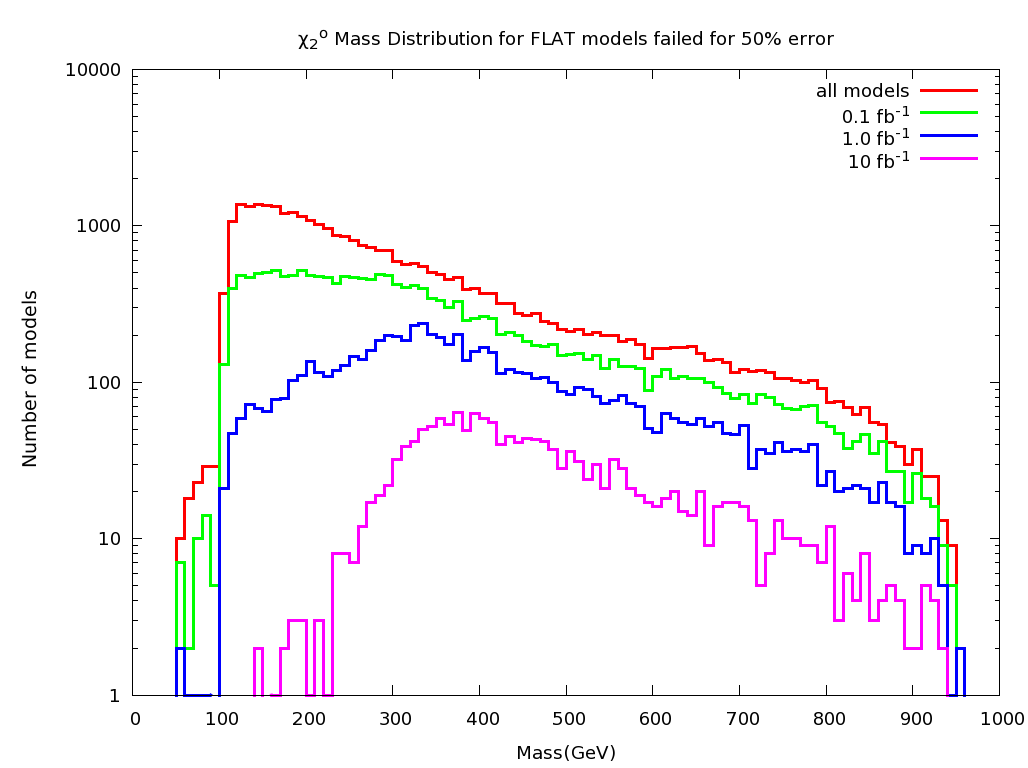}
\hspace*{0.4cm}
\includegraphics[width=8.0cm]{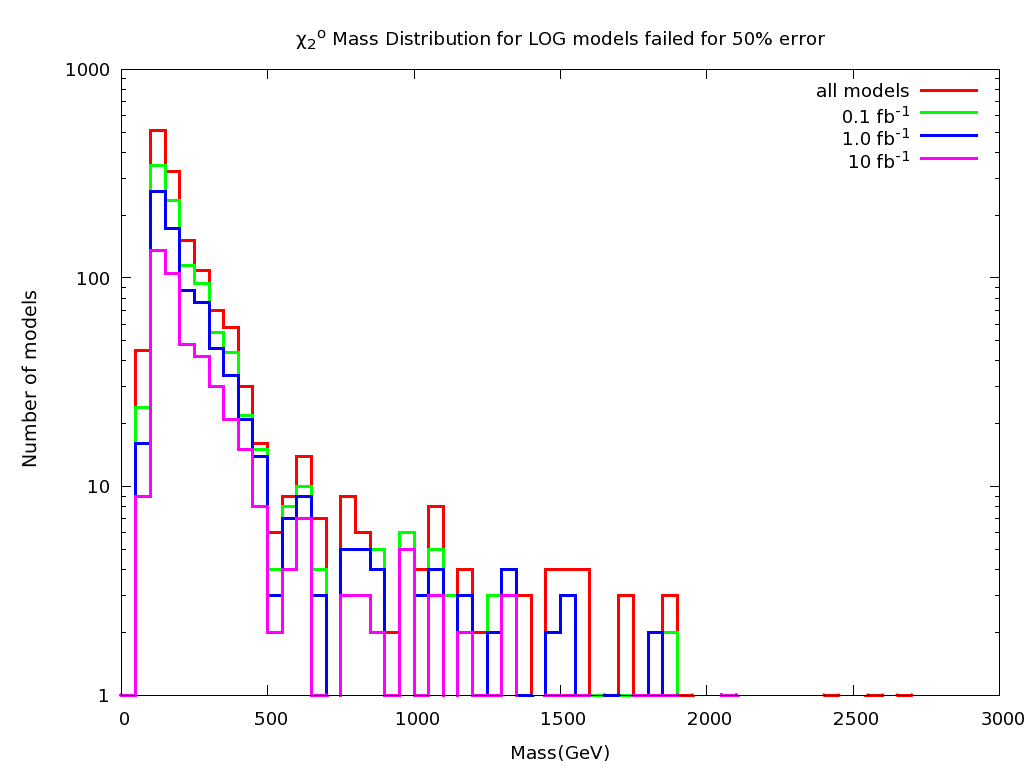}}
\vspace*{0.9cm}
    \caption{Mass distributions for $\tilde e_L$(top) and $\tilde \chi_2^0$(bottom) for the ATLAS-undetected flat(left) and log(right) prior models assuming 
$\delta B=50\%$ for different values of the LHC integrated luminosity as indicated. The top red histogram in each case corresponds to the original 
model sets before any of the ATLAS analyses are considered while the subsequently lower histograms correspond to those subsets of models undetected by the ATLAS 
\MET\ analyses at fixed integrated luminosities.}
 \label{massloss}
\end{figure}

\begin{figure}
\centerline{
\includegraphics[width=8.0cm]{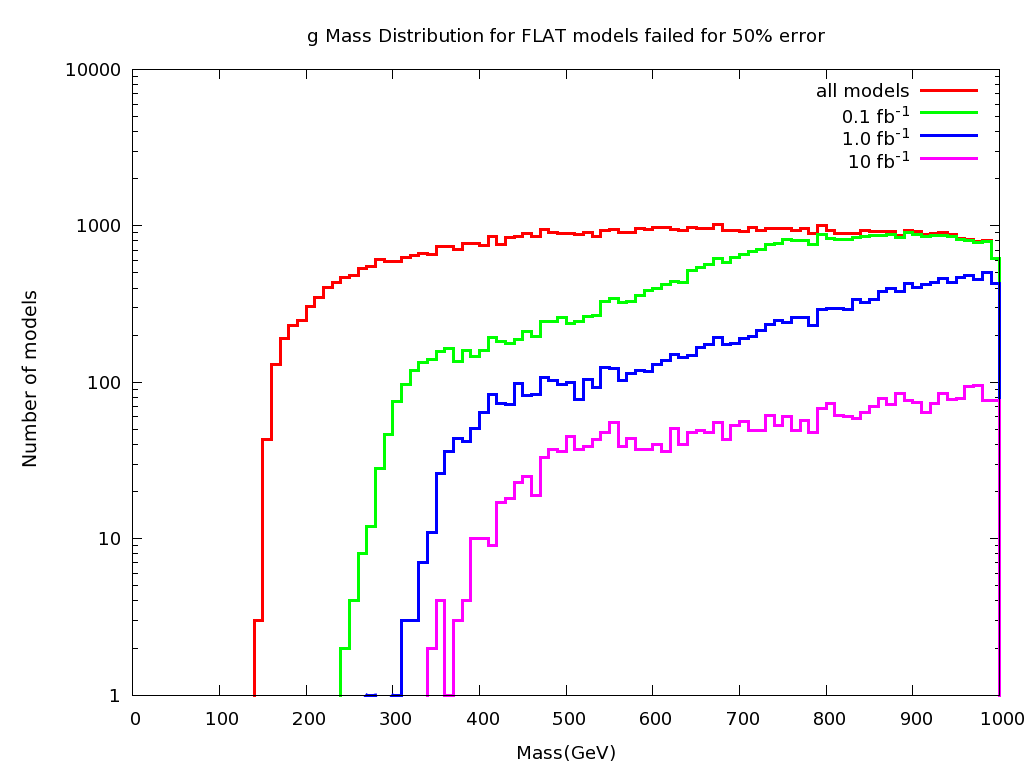}
\hspace*{0.4cm}
\includegraphics[width=8.0cm]{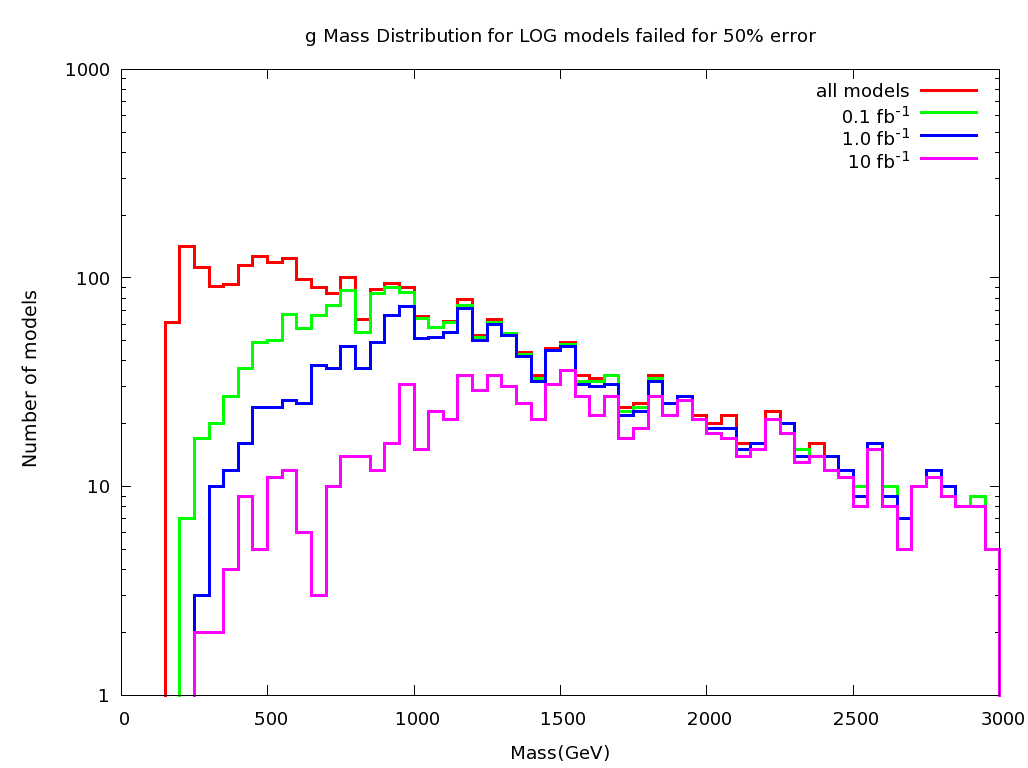}}
\vspace*{1.0cm}
\centerline{
\includegraphics[width=8.0cm]{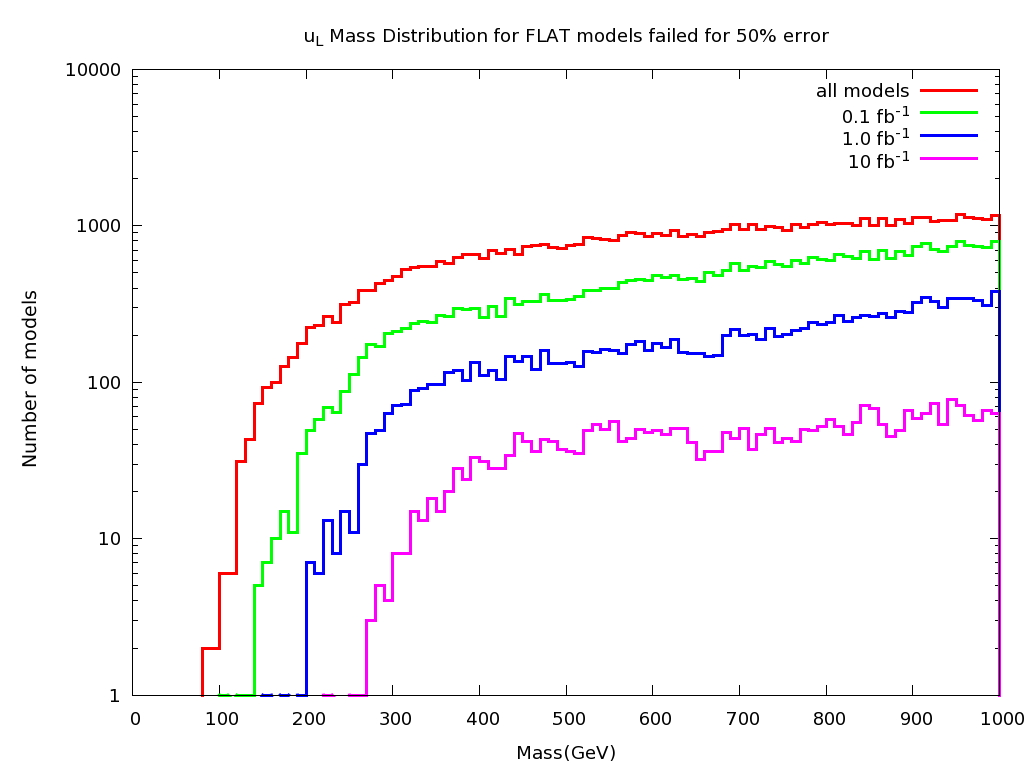}
\hspace*{0.4cm}
\includegraphics[width=8.0cm]{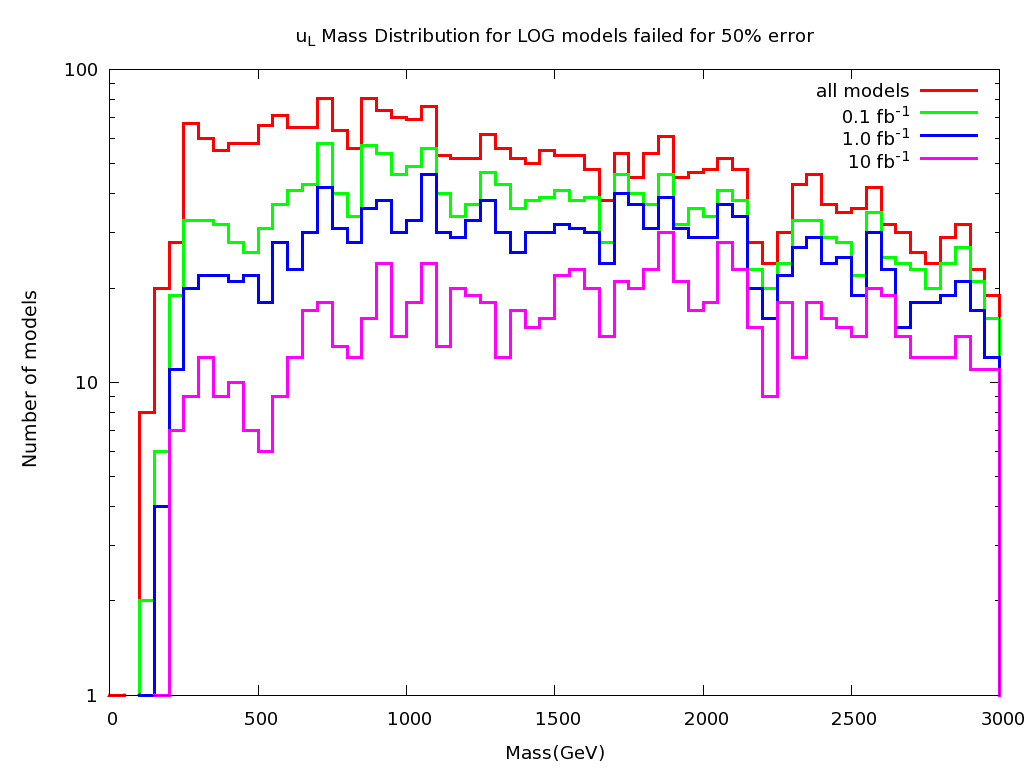}}
\vspace*{0.9cm}
    \caption{Same as the previous Figure but now for gluinos and $\tilde u_L$.}
 \label{massloss2}
\end{figure}

In order to test this hypothesis, we show the mass distributions for the $\tilde e_L$ and $\tilde \chi_2^0$ in both the flat and log prior model sets in 
Fig.~\ref{massloss}.\footnote {Note that there is nothing special about the choice of these two particular non-colored sparticles and the features that we will now 
describe are also found in the mass distributions of other sparticles.}  Indeed, we see that this hypothesis is true for non-colored states: in the flat 
prior set, models being observed by the ATLAS analyses mostly correspond to those with lower sparticle masses. On the other hand, for the log prior case, we see 
that the mass distributions for non-colored sparticles approximately maintain their overall shape as the models are observed, showing not much preference for the 
lighter sparticle masses. Fig.~\ref{massloss2} shows, however, that in the case of colored sparticles, here for the gluinos and (one of the) squarks, which are most 
directly sensitive to most of the ATLAS \MET\ analyses, this same effect is somewhat less significant. In particular for the gluinos we see that even in the log 
prior case there is a significant loss in the fraction of models with lighter masses as the LHC integrated luminosity is increased.

\clearpage

%\section{Discussion of Difficult Models}

% stable particles section
%\input{results-stable.tex}

%\section{Summary}

% conclusions
\section{Summary and Conclusions}

In this paper we have analyzed the capability of the ATLAS \MET-based SUSY analyses to discover Supersymmetry in a model
independent fashion at the 7 TeV LHC.  To this end, we tested these search channels on a large set of model points, $\sim 71$k,
in the 19-dimensional parameter space of the pMSSM.  This model sample contains a wide variety of properties and
characteristics and provides a framework to explore the breadth of possible SUSY signatures at colliders and elsewhere.
These models were generated in a previous work and comply with a set of minimal theoretical assumptions as well as the global precision
electroweak, heavy flavor, collider searches, and astrophysical data sets.  We simulated ten ATLAS \MET\ search channels, which
were designed in the context of mSUGRA-based SUSY, and employed the SM backgrounds as provided directly by the ATLAS SUSY working
group.  We first checked that our analyses were in agreement with ATLAS results for the mSUGRA benchmark point that the collaboration 
had previously simulated.

We passed our model sample through the ATLAS analysis chain and computed the significance of the signal for each
model in each search channel. A significance of $S\geq 5$ was used as the criteria for discovery in each channel; we employed 
the same numerical technique that ATLAS does for calculating this value.  We found
that the systematic error due to uncertainties in the size of the expected background made a substantial impact on model
discovery for the range of expected integrated luminosities.  In fact, some channels become systematics dominated at luminosities
of order $5-10$ fb$^{-1}$.  Overall, for 1(10) fb$^{-1}$ of integrated luminosity roughly 80(95)\% of the flat prior model sample
is discoverable, assuming a 50\% background systematic error.  Larger (or smaller) systematic errors greatly reduce (or increase)
this model coverage.  We found that the 4j0l channel is the most powerful in terms of observing a signal, whereas the leptonic
channels had a much reduced model coverage.
This is due to the suppression of leptonic cascade decays appearing in our model sample compared to expectations from \eg, mSUGRA.
The model coverage was worse for the log prior sample due to kinematic reasons.

We explored the characteristics that caused a model to not be observed in these search channels.  While production cross
section values as related to the sparticle mass obviously plays a role, it does not tell the whole story.  There are cases with
low mass gluinos and/or squarks with large cross sections that are missed by these search analyses, while models with heavy
masses and small cross sections are sometimes observed.  We found that the mass splitting between the gluino/squarks and the
LSP plays an important role in detecting models, and that this can sometimes be compensated by very large production rates
or ISR.  We also saw that subtleties in the sparticle spectrum can conspire to render a model to not be detected.  A fraction
of our model set contains detector-stable sparticles which appear at the end of their cascade decay chains and hence are not
detected by the \MET-based searches.  We studied the effectiveness of the planned stable charged particle searches in these
cases and found that some, but not all, of these models will be discovered.

We briefly considered potential modifications to the ATLAS kinematic cuts in these \MET\ analyses that would improve their
discovery potential.  We studied the optimal cut on $M_{eff}$ as well as for the $p_T$ of the leading jet and overall
\MET.  Our results indicate that the cuts for both the leading jet transverse momentum and the \MET\ could be increased
from their nominal value without seriously impacting model coverage.

Lastly, we studied the implications of a null result from the 7 TeV LHC run.  We found that the degree of fine-tuning that
would be imposed on the pMSSM depended on the choice of priors which generated the model sample, but overall
would not be as large as in the case of mSUGRA.  However, the expectations for sparticle production at a high energy Linear Collider 
would be greatly impacted if Supersymmetry is not discovered during this LHC run.  Basic kinematics would essentially exclude
sparticle production at a 500 GeV Linear Collider, and would point towards the need for a higher energy machine in order to
study Supersymmetry.

In summary, we find that the mSUGRA motivated \MET-based searches for Supersymmetry perform well over a larger and more
complicated SUSY parameter space such as the pMSSM.  However, there are some exceptions and coverage is not perfect.  The
details of the full sparticle spectrum play a very important role in the observability of a model. There are no blanket
statements regarding the potential for discovery, or in setting a mass limit, that that can honestly be made.

We anxiously await the discovery of Supersymmetry in the near future.

\noindent{\Large\bf Acknowledgments}

This project would not have been possible without the assistance and input 
from many people. 

The authors would like to thank the members of the ATLAS SUSY group, in 
particular, S.~Caron, P.~ de Jong, G.~Polesello and G.~Redlinger, for 
discussions and, most importantly, for providing us with additional 
information detailing the ATLAS-generated SM background distributions for 
their SUSY studies and the corresponding results for their mSUGRA benchmark 
analyses.

We would also like to thank the SLAC ATLAS group for assistance with computing 
resources.

We would like to thank T.~Plehn for his help overcoming the special problems 
we had with the implementation of PROSPINO for our pMSSM model set. We would 
also like to thank A.~Djoudai and J.~Conway for similar help with the 
implementations of SDECAY and PGS, respectively.

We would like to thank L.~Dixon for discussions about the theoretical 
assessment of the size of the systematic errors for the SM backgrounds to 
SUSY signals. 

We would like to thank R.~Cotta for important discussions related to our 
analysis.

We would like to thank J.~Cogan for his assistance with setting up the 
ATLAS analysis code during the earlier part of this work.

Work supported by the Department of Energy, Division of High
Energy Physics, Contracts DE-AC02-76SF00515 DE-AC02-06CH11357, and
DE-FG02-91ER40684, and by the BMBF ``Verbundprojekt HEP-Theorie'' under
contract 05H09PDE.


\begin{thebibliography}{99}
\bibitem{thanks}
We would like to thank the members of the ATLAS SUSY Physics Group for providing us with this information.
\end{thebibliography}
\end{document}